\documentclass[aps,pra,twocolumn,preprintnumbers,longbibliography,floatfix]{revtex4-2}

\usepackage{graphicx}  
\usepackage{multirow}
\usepackage{braket}
\usepackage{float}
\usepackage{hyperref}
\usepackage{amsfonts}  
\usepackage{amsmath}   
\usepackage{amssymb}   
\usepackage{mathtools}
\usepackage{comment}
\usepackage{bm}
\usepackage{color}

\DeclareMathOperator{\sech}{sech}
\DeclareMathOperator{\am}{am}
\DeclareMathOperator{\dn}{dn}
\begin{document}

\title{Beyond the parametric approximation: \\ pump depletion, entanglement and squeezing in macroscopic down-conversion}

\begin{abstract}
We study the dynamics of the pump mode in the down-conversion Hamiltonian using the cumulant expansion method, perturbation theory, and the full numerical simulation of systems with a pump mean photon number of up to one hundred thousand. We particularly focus on the properties of the pump-mode such as depletion, entanglement, and squeezing for an experimentally relevant initial state in which the pump mode is initialized in a coherent state. Through this analysis, we obtain the short-time behaviour of various quantities and derive timescales at which the above-mentioned features, which cannot be understood through the parametric approximation, originate in the system. We also provide an entanglement witness involving moments of bosonic operators that can capture the entanglement of the pump mode. Finally, we study the photon-number statistics of the pump and the signal/idler modes to understand the general behaviour of these modes for experimentally relevant time scales.
\end{abstract}

\author{Karthik Chinni}
\email{karthik.chinni@polymtl.ca}
\affiliation{Department of Engineering Physics, \'Ecole Polytechnique de Montr\'eal, Montr\'eal, Qu\'ebec H3T 1J4, Canada}
\author{Nicol\'as Quesada}
\email{nicolas.quesada@polymtl.ca}
\affiliation{Department of Engineering Physics, \'Ecole Polytechnique de Montr\'eal, Montr\'eal, Qu\'ebec H3T 1J4, Canada}

\maketitle 
\date{\today}

\section{Introduction}
The interaction of light with nonlinear materials is important for the production of nonclassical light including squeezed states, Fock states, entangled states, and non-Gaussian states. These states of light have applications in a variety of fields including quantum computation \cite{Knill2001, Menicucci2006, asavanant2019}, Gaussian boson sampling \cite{Zhong2020, Arrazola2021,hamilton2017gaussian,kruse2019detailed,deshpande2022quantum,grier2022complexity}, metrology \cite{Pezze2008, Aasi2013} and communication \cite{Laudenbach2018}. In particular, non-Gaussian states have been of interest in recent works because of their importance for error correction procedures in the context of continuous-variable quantum computing \cite{Niset2009}. We analyze the nature of the states, which are expected to be non-Gaussian, that are produced in the down-conversion process beyond the parametric approximation. The down-conversion Hamiltonian, sometimes referred to as the trilinear Hamiltonian, is realized when a beam of light, referred to as the pump mode, passes through a material with a second-order nonlinear response, resulting in the production of photons in the signal and idler beams. The down-conversion process has been studied since the 1960s \cite{Louisell1961quantum, Mollow1967_1, Mollow1967_2, Walls1969topics_dissertation, Walls1970quantum, Walls1972nonlinear, Tucker1969quantum, Agrawal1974dynamics, Gambini1977parametric, Drobny1992quantum,Drobny1997quantum,Coelho2009three, Bandilla2000parametric,Xing2023pump}, particularly in the context of squeezing \cite{Scharf1984effect, Milburn1981production, Hillery1984path, Crouch1988limitations}. This Hamiltonian also describes several other processes such as the interaction of $N$ two-level atoms with a mode of the electromagnetic field in the Tavis-Cummings model \cite{Drobny1992quantum}, coupling of two optical fields with an acoustic wave, a process known as Brillouin-Mandelstam scattering \cite{Cryer2023}, zero-dimensional model of Hawking radiation \cite{Nation2010} and the modelling of a quantum absorption refrigerator \cite{Levy2012,Correa2014}. The strength of the nonlinearity in the down-conversion process through the use of bulk materials is typically very weak; this issue can be tackled by increasing the interaction strength or increasing the intensity of the pump mode. The former is typically achieved by using integrated photonic structures such as waveguides and resonators. This Hamiltonian has also been realized on a superconducting platform where high nonlinearity strengths can be easily achieved in the microwave regime \cite{Abdo2013nondegenerate}. 

The state dynamics under the down-conversion process is typically understood through the parametric approximation \cite{Drummond2014quantum} where the number of pump photons is assumed to be constant and the resulting process is modeled by replacing the pump-mode destruction operator with the amplitude of the pump coherent state at the initial time. This approximation yields an accurate description of the dynamics of the system at short times, but the resulting state deviates from this prediction as the number of pump photons decreases with time. The depleted pump region has also been explored experimentally \cite{Allevi2014, Pevrina2016, Florez2020pump, Ding2018}, and theoretically in a number of studies \cite{Nation2010, Xing2023pump, Birrittella2020phase,Scharf1984effect, Bandilla2000parametric}. However, the pump mode features, particularly their behaviour as a function of initial mean pump-mode population ($|\alpha_{0}|^{2}$) for large values, has not been fully explored. Here, we study various pump-mode features of the system that cannot be understood through the parametric approximation such as depletion time, entanglement with the rest of the system, squeezing, and the evolution of the zero-delay autocorrelation function associated with the pump mode, using various analytical tools and the full numerical simulation of these features for pump coherent states containing on average up to  $|\alpha_{0}|^{2}=10^{5}$ photons. In particular, we analyze these properties using the cumulant expansion method, perturbation theory, and photon number statistics and analytically identify the timescales at which these features originate in the system for an experimentally relevant initial state, where the pump-mode is in coherent state, the signal and idler modes are in vacuum. In addition to studying pump-mode entanglement with the other modes, we also provide an explicit witness expressed in terms of bosonic moments based on the positive partial transpose (PPT) criterion that can capture this entanglement. Finally, we supplement this analysis with the study of photon-number statistics of the pump and signal mode to provide a general picture of the behaviour of the state for experimentally relevant times.

The rest of the manuscript is organized as follows. In Sec. \ref{sec:Macroscopic pump}, we review the parametric approximation and provide general background material. In Sec. \ref{sec:dynamics}A, we numerically study the system using the cumulant expansion method and obtain an analytic solution to the equations of motion in the second order which allows us to extract the pump-depletion time analytically. To understand other important features of the system, in Sec. \ref{sec:dynamics}B, we perform a transformation on the down-conversion Hamiltonian using the second-order cumulant solution and then apply the time-dependent perturbation theory using the transformed Hamiltonian. Through this, we obtain perturbation theory expressions for various important quantities that help explain their short-time behaviour. In Sec. \ref{sec:dynamics}C, we analyze the photon-number statistics. Following this, in Sec. \ref{sec:entanglement}, we study the entanglement between the pump and the rest of the modes using the purity of the reduced density operator and introduce a witness dependent on fourth-order moments of the quadratures that can well capture the entanglement dynamics seen using the purity.

\section{Evolution Under a Macroscopic Pump State}
\label{sec:Macroscopic pump}
The interaction picture Hamiltonian describing the down-conversion process on resonance ($\omega_p = \omega_s+\omega_i$ where $\omega_k$ is the frequency of mode $k$) is given by 
\begin{align}\label{eq:Int Hamiltonian}
H&=i\hbar \chi\bigl(a_{p}a_{s}^{\dagger}a_{i}^{\dagger}-\;a_{p}^{\dagger} a_{s}a_{i}\bigr),
\end{align}
where $a_\mu (a_\mu^\dagger)$ is the destruction (creation) operator for mode $\mu \in \{p,s,i\}$ satisfying canonical bosonic commutation relations and $\chi$ is a quantity that is real (cf. Appendix \ref{sec:Gauge freedom}), with the units of frequency, describing the coupling between the modes. Here subscripts $p,s,i$ refer to the pump, signal, and idler modes respectively. This Hamiltonian has two conserved quantities: $N_{p}+N_{s}$ and $N_{p}+N_{i}$ where $N_{\mu}=a_{\mu}^{\dagger}a_{\mu}$ is the number operator of mode $\mu$. As a consequence of these conserved quantities, the dynamics of the state is constrained to a subspace of the full Hilbert space if the initial state is an eigenstate of $N_{p}+N_{s}$ and $N_{p}+N_{i}$ \cite{Mollow1967_1}. These conserved quantities map to the conservation of angular momentum and excitation number in the case of the Tavis-Cummings model \cite{Drobny1992quantum}. In this work, we are interested in an experimentally relevant initial state given by a coherent state in the pump mode, vacuum in the signal and idler modes, 
\begin{align}\label{eq:InitialKet}
\ket{\psi(0)} &= |\alpha_{0}\rangle|0\rangle |0\rangle = \exp\left(\alpha_0 [ a_p^\dagger - a_p]\right) |0\rangle|0\rangle |0\rangle  \\
&=\sum_{N=0}^{\infty} \sqrt{p_N}  |N\rangle|0\rangle |0\rangle \text{ with } p_N = e^{-\alpha_0^2} \frac{\alpha_0^{2N}}{N!} 
\end{align}
{where} the first ket corresponds to the pump mode, the second to the signal mode, and the third to the idler mode. In the second line we used the well known-expansion of a coherent state in the Fock basis and assumed without loss of generality (cf. Appendix \ref{sec:Gauge freedom}) that the amplitude of the coherent state is real, allowing us to write amplitudes directly in terms of square roots of Poisson distribution probabilities with mean $\bar{N} = \alpha_0^2$. Note that the state in Eq.~\eqref{eq:InitialKet}, as well as the Hamiltonian in Eq.~\eqref{eq:Int Hamiltonian}, is invariant under the permutation of signal and idler modes. Thus any single mode quantity pertaining to the idler mode will have exactly the same value as its corresponding quantity for the signal mode. The time-evolved state under the \emph{parametric approximation}, where the pump mode is assumed to remain in the initial coherent state, is then given by \cite{Agarwal2012quantum}
\begin{align}\label{eq:TMSV}
|\psi(t)\rangle =e^{r (a_{s}^{\dagger}a_{i}^{\dagger}- a_{s}a_{i})}|\alpha_{0}\rangle |0\rangle |0\rangle =|\alpha_{0}\rangle |\text{TMSV}(r )\rangle,
\end{align}
where TMSV refers to two-mode squeezed vacuum, $|\text{TMSV}(y)\rangle \equiv \sum_{n=0}^{\infty} \frac{\tanh^{n}(y)}{\cosh(y)}|nn\rangle$, $\tau=\chi t$ is a reduced time, which will be used throughout the rest of this manuscript and finally $r = \alpha_0 \tau$ is the squeezing parameter, which is the single parameter that specifies the dynamics of the down-conversion problem in the parametric approximation. As we will see later, this single-parameter dependence no longer holds true beyond the parametric approximation. The average number of signal photons in this state is given by $\langle N_{s}\rangle=\sinh^{2}(\alpha_{0} \tau)$, which grows exponentially in time. This exponential growth results from the assumption that the pump state remains undepleted. Since the state in Eq. (\ref{eq:TMSV}) accurately describes the system only for short times, the prediction associated with exponential growth is only reliable for these initial times as shown by the dashed line corresponding to parametric approximation in Fig. \ref{fig:fig1}(a) when the population of the pump mode remains close to being undepleted. Note that $\langle N_{p}\rangle=\alpha_{0}^{2}-\langle N_{s}\rangle$ from the conservation laws, so the nonzero values of $\langle N_{s}\rangle$ correspond to the pump-depletion region in Fig. \ref{fig:fig1}(a).

The exact evolution of the state $|\alpha\rangle |0\rangle |0\rangle$ under the down-conversion Hamiltonian can be better understood by considering the case where the pump mode is initialized in a state with $N$ photons, and vacuum in both the signal and idler modes, $|\psi_N(0)\rangle=|N\rangle|0\rangle|0\rangle$. In this $N$-photon case, due to the presence of symmetries in the system, the Hilbert space explored by the time-evolved state is spanned only by $N+1$ states, which are given by $\{|N-k\rangle|k\rangle|k\rangle\}$ where $k=\{0,1,2,...,N\}$ \cite{Walls1969topics_dissertation, Walls1970quantum}. Therefore, the state at an arbitrary time can be expressed as 
\begin{align}
    |\psi_N(\tau)\rangle=e^{-i\frac{H}{\hbar \chi}\tau}|N\rangle|0\rangle|0\rangle=\sum_{k=0}^{N}c_{k}(\tau)|N-k\rangle|k\rangle|k\rangle.
\end{align}
Substituting the above state into the Schr\"{o}dinger equation $i\hbar \partial_{\tau} |\psi_N(\tau)\rangle=\frac{H}{\chi}|\psi_N(\tau)\rangle$, we get 
\begin{align}
    \partial_{\tau}\bm{c}(\tau) &= \bm{M} \bm{c}(\tau),
\end{align}
where $\bm{c}(\tau)=\{c_{0}(\tau),c_{1}(\tau),...,c_{N}(\tau)\}$ is a real vector with $c_{k}(\tau)=\langle N-k|\langle k|\langle k|\psi(\tau)\rangle$,  and $\bm{M}$ is an $(N+1) \times (N+1)$ dimensional real anti-symmetric matrix with nonzero elements $M_{\nu+1,\nu}=-M_{\nu,\nu+1} = (\nu+1)\sqrt{(N-\nu)}$~ \cite{Walls1969topics_dissertation,Walls1970quantum}. The expansion coefficients of the state $|\psi(\tau)\rangle$ are then obtained by $\bm{c}(\tau)=\exp\left(\tau \bm{M} \right) \bm{c}(0)$ where $\bm{c}(0)=(1,0,0,...,0)^T$. It should be noted that in this specific case, the state is obtained exactly without resorting to any truncation procedure.

Returning to the case of system being initialized in $\ket{\psi} = |\alpha_{0}\rangle|0\rangle |0\rangle$, the time-evolved state is given by
	\begin{subequations}
\begin{align}   
	\label{eq:linear superposition}
    |\psi(\tau)\rangle&=\sum_{N=0}^{\infty}\sqrt{p_N} \ket{\psi_N(\tau)}  
\\&=\sum_{N=0}^{\infty}\sum_{k=0}^{N}\beta_{N-k,k}(\tau)\;|N-k\rangle|k\rangle|k\rangle \label{eq:full state}.
\end{align}
\end{subequations}
The above equation shows that, in this case, the state at an arbitrary time is obtained by superposing the time-evolved states acquired from the evolution of initial states with a fixed number of pump photons, $\{|n\rangle |0\rangle|0\rangle \}$, for $n=\{0,1,2,...\}$ and weighting them appropriately using the Poisson distribution. For our numerical simulations, we limited the range of $n$ from $n_{1}$ to $n_{2}$, which are chosen to satisfy the condition $p_n>10^{-16}$, and used Scipy's~\cite{2020SciPy-NMeth} \verb|expm_multiply| function to efficiently apply the exponential of a sparse matrix on a vector using the methods from Refs.~\cite{Al2011computing, Higham2010computing}. Because of the conserved quantities, only one parameter needs to be truncated for numerical approximation as opposed to three for describing the {state's} support in the Fock basis if the knowledge of conserved quantities was not utilized. Also note that the form of the state (resulting from conserved quantities) shown in Eq. (\ref{eq:linear superposition}) leads many of the moments of the bosonic operators to become zero. For instance, 
\begin{align}\label{eq:momentrule}
\langle A_{p}a_{s}^{l}a_{i}^{m}\rangle=&\sum_{n,n'=0}^{\infty}\sum_{k,k'=0}^{n,n'}\beta_{n'-k',k'}^{*}(\tau)\beta_{n-k,k}(\tau) \\
&  \ \times \;\langle n'-k'|A_p|n-k\rangle\langle k'|a_{s}^{l}|k\rangle \langle k'|a_{i}^{m}|k\rangle  \nonumber 
\\
    =&f(A_{p},l) \; \delta_{l,m},
\end{align}
where $A_{p}$ is an arbitrary operator involving the pump mode. Likewise, $\langle A_{p}\bigl(a_{s}^{\dagger}\bigr)^{l}\bigl(a_{i}^{\dagger}\bigr)^{m}\rangle =g(A_{p},l) \; \delta_{l,m}$ and $\langle A_{p}\bigl(a_{s}^{\dagger}\bigr)^{l} a_{i}^{m}\rangle =h(A_{p}) \; \delta_{l,0}\delta_{m,0}$. These relations will be  used in Sec. \ref{sec:entanglement} to show that the quadrature covariance matrix of the three modes does not have any cross terms between the pump and the signal or idler modes.

\section{Dynamics of the System}
\label{sec:dynamics}
In this section, we present our analysis of the dynamics of the system with a focus on understanding the pump mode's behaviour beyond the parametric approximation by resorting to cumulant expansions, perturbation theory and the analysis of photon-number statistics.  
\subsection{Cumulant Expansion Method}
\label{sec:cumulant method}
The dynamics of the expectation values of operators can be obtained using the Heisenberg equations of motion $\bigl(\partial_{\tau}\langle A\rangle= \frac{i}{\hbar \chi}\langle[H,A]\rangle\bigr)$. The equations of motion associated with the single-moment variables under the down-conversion Hamiltonian are given by 
\begin{align}
\partial_{\tau}\langle a_{p} \rangle &= -\langle a_{s} a_{i} \rangle, \\
\partial_{\tau} \langle a_{s} \rangle&= \langle a_{p}a_{i}^{\dagger} \rangle, \\
\partial_{\tau} \langle a_{i}^{\dagger} \rangle &=\langle a_{p}^{\dagger}a_{s}\rangle.
\end{align}
The derivatives of the first order moments couple to second-order moments, whose differential equations are further coupled to third order moments. In general, the derivatives of $n^{\text{th}}$-order moments are coupled to $(n+1)^{\text{th}}$-order moments leading to an infinite number of equations that needs to be truncated using some approximation to be solved numerically. In contrast, for a system that is initialized in a state with a fixed number of pump photons ($N_{p}$) and vacuum in the signal and idler modes, the state at an arbitrary time always has certain higher-order moments equal to zero (without any approximation) leading to only $N_{p}+1$ differential equations as opposed to the infinite number of differential equations in the general case. This is consistent with the number of differential equations that would have to be solved when using the Schr\"{o}dinger equation for such an initial state.

We resort to the cumulant expansion method \cite{Kubo1962} that allows us to systematically neglect higher-order correlations furnishing a way to truncate the system of differential equations that would arise for an initial state with the pump mode in a coherent state, and signal, idler modes in vacuum. By neglecting higher-order correlations, the memory requirements and the time required to solve the dynamics of such a system can be reduced considerably, at least for lower-order solutions. 

According to this method, all higher-order correlations above a certain order are ignored by setting the cumulants of that particular order to zero, $\langle\langle A_{1}A_{2}...A_{n}\rangle\rangle=0$. This allows the $n^{\text{th}}$ order moments to be expressed in terms of lower order moments and therefore obtain a closed set of differential equations. The $n^{\text{th}}$ order cumulant ~\cite{fisher1932derivation} (also called Ursell~\cite{ursell_1927} functions, truncated correlation functions~\cite{duneau1973decrease} or cluster functions~\cite{duneau1973decrease})  of a product of $n$ operators is defined as 
\begin{align}
\langle\langle O_{1}O_{2}...O_{n} \rangle\rangle \equiv \sum_{p \in P(I) }(|p|-1)!(-1)^{|p|-1} \prod_{B \in p} \left\langle \prod_{i \in B }O_{i}\right \rangle,
\end{align}
where $I=\{O_{1},O_{2},...,O_{n} \}$, $P(I)$ is the set of all partitions of $I$, $|p|$ is the length of the partition $p$, $B$ goes over each block of the partition and $i$ goes through every element in the block of partition while maintaining the order of the operators \cite{Plankensteiner2022,Kubo1962}. For instance, in the case of $n=3$, we have $\langle\langle O_{1}O_{2}O_{3} \rangle\rangle = \langle O_{1}O_{2}O_{3} \rangle - \langle O_{1}O_{2} \rangle \langle O_{3} \rangle - \langle O_{1} \rangle \langle O_{2}O_{3} \rangle - \langle O_{1}O_{3} \rangle \langle O_{2} \rangle+ 2 \langle O_{1} \rangle \langle O_{2} \rangle \langle O_{3} \rangle$. In the general case, the $|p|=1$ partition in the set $P(I)$ corresponds to $\langle O_{1}O_{2}...O_{n} \rangle$, so setting the $n^{\text{th}}$ order cumulant to zero allows us to express the $n^{\text{th}}$ order moment as
\begin{align}
\langle O_{1}O_{2}...O_{n} \rangle = \sum_{p \in P'(I)}(|p|-1)!(-1)^{|p|} \prod_{B \in p} \left\langle \prod_{i \in B }O_{i}\right \rangle ,
\end{align}
where $P'(I)$ is the set of all partitions excluding the $|p|=1$ partition. {Hence, in the $n^{th}$ order cumulant expansion method, all the cumulants of order $n+1$ are assumed to be negligible.}
\subsubsection{First Order Cumulant Expansion}
\begin{figure*}[t]
\centering
\includegraphics[width=\textwidth]{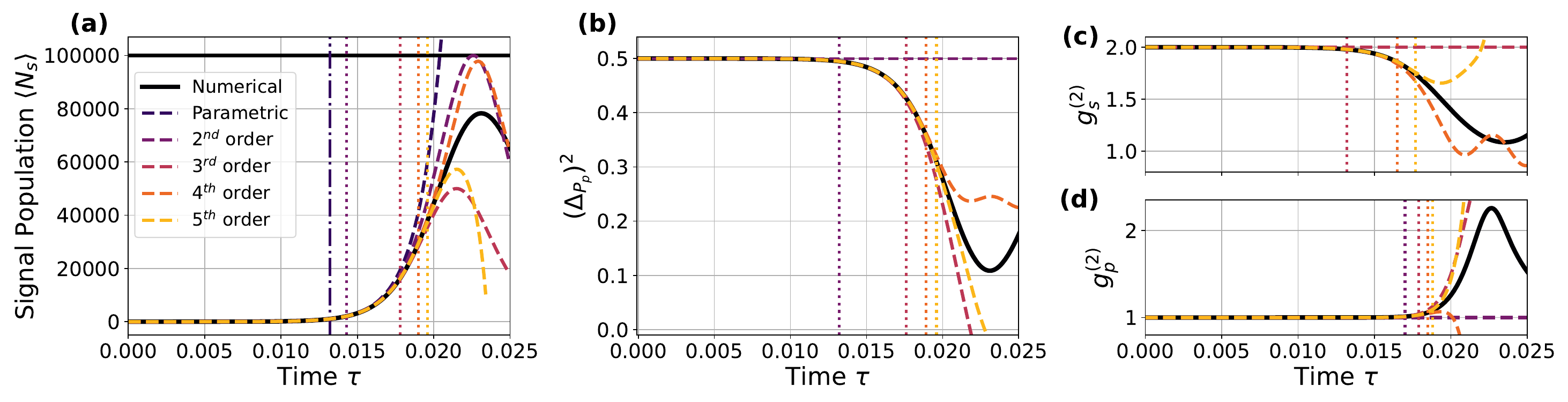}
\caption{Signal-mode population \textbf{(a)}, pump-mode momentum variance \textbf{(b)}, zero-delay autocorrelation functions for both signal mode \textbf{(c)} and pump mode \textbf{(d)} are plotted here as a function of time for $\alpha_{0}^{2}=100,000$ in columns one, two and three respectively. In all of these plots, the data obtained from the numerical simulation is plotted using a thick line while the solution from the cumulant-expansion method is plotted using different colored dashed lines as shown by the label in \textbf{(a)}. Also, different colored vertical dotted lines identify the time at which various quantities ($\langle N_{s}\rangle$, $\left(\Delta_{P_{p}}\right)^{2}$, $g_{s}^{(2)}$ and $g_{p}^{(2)}$) obtained from the $n^{\text{th}}$ order cumulant expansion method differ from the full simulation of the down-conversion process (numerical) value by $1 \%$ relatively. It can be seen from the above plots that the higher order approximation of the cumulant expansion method describes the system dynamics longer than the lower order approximations. \textbf{(a)} The second-order cumulant solution explains the dynamics of the system longer than the parametric approximation. The fifth order cumulant solution explains the signal-mode population dynamics up to $37\%$ of $\alpha_{0}^{2}$ with $1\%$ relative error, and up to $54 \%$ of $\alpha_{0}^{2}$ if a relative error of $5\%$ is accepted. Also, note that $\langle N_{s}\rangle(\tau)=\langle N_{i}\rangle(\tau)=\alpha_{0}^{2}-\langle N_{p}\rangle(\tau)$. \textbf{(b)} It can be seen from here that the pump-mode becomes squeezed and can be identified only from the solution of third and higher order cumulant expansion method. \textbf{(c)} As the signal-idler mode state evolves from being a TSMV state at short times, the $g_{s}^{(2)}$ of the signal mode changes its value, and this change is identified by the solution of the third and higher order cumulant expansion method. Note that the second-order $g_{s}^{(2)}$ is not shown in this plot because it overlaps with the prediction obtained from the third-order cumulant expansion method. \textbf{(d)} The value of $g_{p}^{(2)}$ changes from being one at short times as the pump-mode evolves to a state different from a coherent state, and this change is identified by the solution of the third and higher order cumulant expansion method.}
\label{fig:fig1}
\end{figure*}
This case is referred to as the mean-field limit because setting the second order cumulants to zero is equivalent to ignoring correlations between operators, $\langle A B\rangle =\langle A\rangle \langle B\rangle$. Under this truncation scheme, and the assumption that the pump amplitude $\alpha_0$ is real, we obtain the following equations of motion
\begin{align}
\partial_{\tau}\left\langle a_p\right\rangle &=-\langle a_s \rangle \langle a_i \rangle,\\
\partial_{\tau} \left\langle a_s\right\rangle &=\langle a_i \rangle \langle a_p \rangle, \\
\partial_{\tau} \left\langle a_i\right\rangle &=\langle  a_s\rangle \langle a_p \rangle.
\end{align}
The fixed points of the above set of differential equations are those for which at least two of three expectation values of the destruction operators are zero. The initial state of interest has both $\langle a_{s}\rangle=0$ and $\langle a_{i}\rangle=0$, so the variables in equations of motion do not evolve and the signal-mode population in this approximation is given by
\begin{align}
    \langle N_{s}\rangle^{(1)}(\tau)=0.
\end{align}
A superscript has been added to the signal mode population ($\langle N_{s}\rangle^{(1)}$) obtained from the first-order solution to highlight the difference between this quantity and the signal mode population obtained from the full simulation of the down-conversion process ($\langle N_{s}\rangle$). In the above solution, as expected, the mean-field equations cannot predict the production of a pair of photons in the signal and the idler mode from vacuum. An expression for the pump and signal mode population in terms of Jacobi elliptic functions~\cite{NIST:DLMF} for arbitrary initial conditions has been derived in \cite{Armstrong1962interactions} for the classical case.
\subsubsection{Second Order Cumulant Expansion}
Here we set the third order cumulants to zero, which is equivalent to approximating third-order moments as $\langle ABC \rangle=\langle AB \rangle \langle C \rangle+\langle A \rangle\langle BC \rangle +\langle AC \rangle \langle B \rangle -2 \langle A\rangle\langle B\rangle\langle C\rangle$. In this case, we have $15$ equations of motion ($13$ independent equations due to conserved quantities), which have been obtained using the QuantumCumulants package \cite{Plankensteiner2022} and are shown in Appendix \ref{sec:Gaussian limit}. For our particular initial state of interest, $|\psi(0)\rangle=|\alpha_0\rangle |0\rangle | 0\rangle$, the $15$ equations of motion reduce to $3$ independent equations because other variables are either uncoupled from the relevant equations of motion, or they do not evolve in time and remain at zero for all times. The equations associated with the dynamical variables for real $\alpha_{0}$ are given by 
    \begin{align}
    \label{eq: second order eqn1}
        \partial_{\tau}\left\langle a_p\right\rangle&=-\left\langle a_s a_i \right\rangle,  \\
        \partial_{\tau}\left\langle a_s a_i \right\rangle &=\left\langle a_p\right\rangle \bigl(1 +  2 \left\langle a_s^{\dagger } a_s\right\rangle \bigr),\\
        \partial_{\tau}\left\langle a_s^{\dagger } a_s\right\rangle &=2\left\langle a_p\right\rangle \left\langle a_s a_i\right\rangle.   \label{eq: second order eqn3}
    \end{align}
As expected, $\partial_{\tau}\langle a_{i}^{\dagger}a_{i}\rangle=\partial_{\tau}\langle a_{s}^{\dagger}a_{s} \rangle=-\partial_{\tau}\langle a_{p}^{\dagger}a_{p}\rangle$, so the equations of motion associated with $\partial_{\tau}\langle a_{p}^{\dagger}a_{p}\rangle$ and $\partial_{\tau}\langle a_{i}^{\dagger}a_{i}\rangle$ are not shown. The above set of equations can be solved analytically (see Appendix \ref{sec:Gaussian limit} for more details), and the signal-mode population in this approximation is given by
\begin{align}
\label{eq:second order population}
    \langle N_{s}\rangle^{(2)}(\tau)&\equiv \langle a_{s}^{\dagger}a_{s}\rangle=\sinh^{2}\left(\eta(\tau)\right),\\
\eta(\tau)&=\Im\left( \am\left(i\tau\alpha_{0},-1/\alpha_{0}^{2} \right)\right),
\end{align} 
where $\am(z)$ is the Jacobi amplitude function and $\Im(z)$ is the imaginary part of the complex number $z$. As in the first-order solution, a superscript has been added to the signal mode population ($\langle N_{s}\rangle^{(2)}$) obtained from the second-order solution.
In the above solution, $\am\left(i\tau\alpha_{0},-1/\alpha_{0}^{2} \right)$ is purely imaginary for the times, $\tau$, we are interested in analyzing in this work, but it has a real part that is a multiple of $2\pi$ for longer times. As shown in Fig. \ref{fig:fig1}a, the above expression for the signal mode population is bounded in time unlike the solution obtained from the parametric approximation and approximates the value of $\langle N_{s}\rangle$ for longer times. This can be seen for the case of $\alpha_{0}^{2}= 10^{5}$ in Fig. \ref{fig:fig1}(a) by comparing the location of the dash-dotted vertical line with that of the first dotted line. In this figure, the signal mode population obtained from different orders of the cumulant expansion method are plotted with different colored dashed lines, and the corresponding vertical dotted lines show the value of time until which the colored dashed lines approximate the numerically obtained signal-mode population (thick line) with a relative error smaller than $1\%$. Also, see Ref. \cite{Scharf1984effect} for an exact expression for the mean signal photon number when the pump-mode starts in a number state, and the corresponding approximation for the case when the pump mode is instead initialized in a coherent state, and see Ref. \cite{Nation2010} for a different expression for the population, which is accurate at short times and closely follows the expression given in Eq. \eqref{eq:second order population}. 

\begin{figure}[t]
\includegraphics[width=\linewidth]{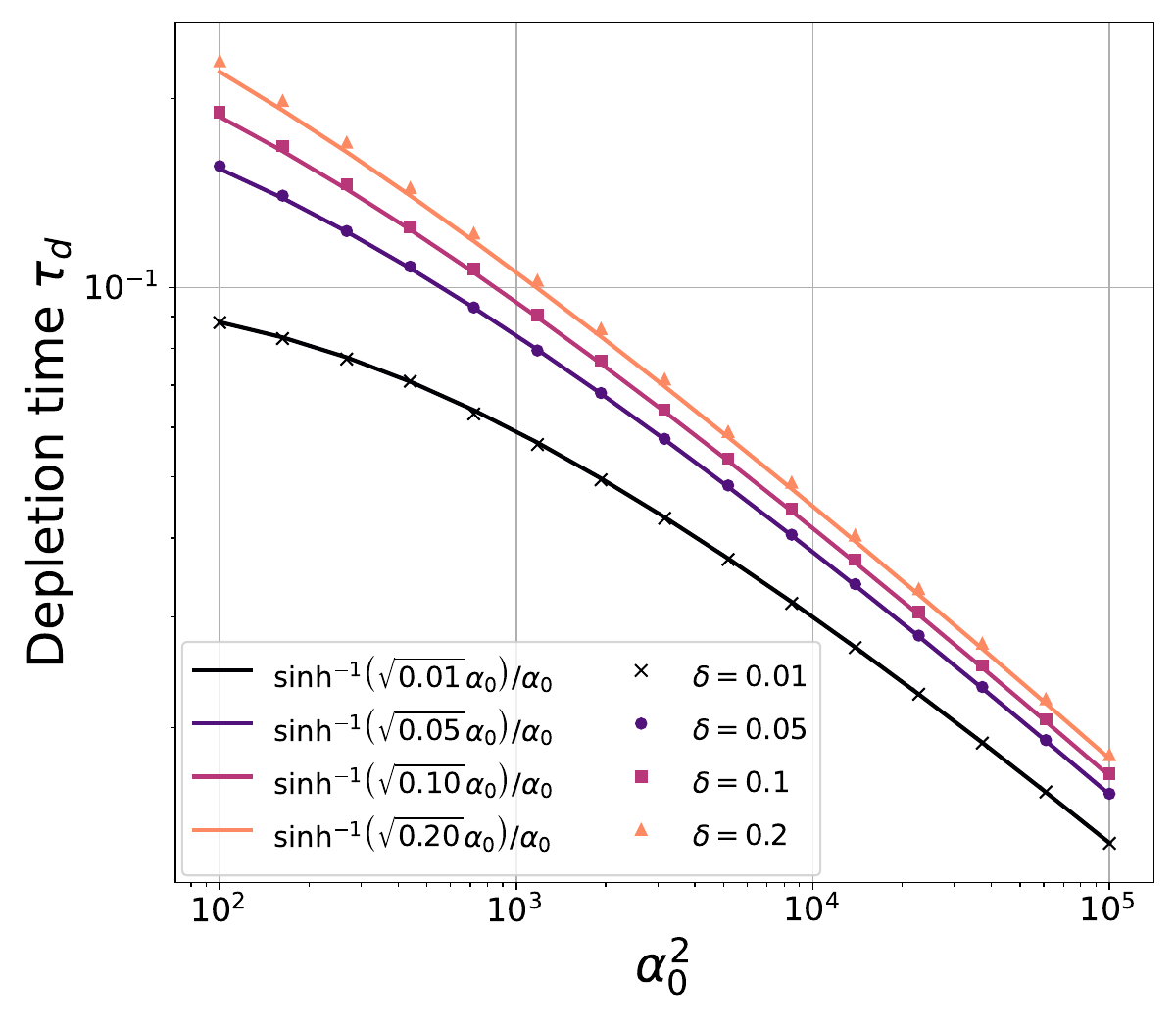}
\caption{{Depletion times obtained numerically are plotted as a function of initial mean population $\alpha_{0}^{2}$. The depletion time is identified as the time when the pump-mode population decreases relatively from $\alpha_{0}^{2}$ by $\delta$. The plots associated with various values of $\delta$ are shown above. The numerical data is well approximated by $\frac{1}{\alpha_{0}}\sinh^{-1}\left( \sqrt{\delta}\alpha_{0} \right)$ (shown by the thick lines) when $\delta$ is small. See Eq. \eqref{eq:approx inverse Jacobi} for more details about the analytic expression describing the pump-depletion time.}}
\label{fig:fig2}
\end{figure}

Note that the problem described by the second order cumulant expansion is identical to extending the parametric approximation to the case where the amplitude of the pump mode is allowed to change in time. That is, the state of the system is given by $|\psi(\tau)\rangle = |\alpha(\tau)\rangle |\chi(\tau)\rangle$, and the equation governing pump-mode amplitude $\alpha(\tau)\equiv \langle \psi(\tau)| a_{p}|\psi(\tau)\rangle$ is obtained using the Heisenberg equation of motion in Eq. \eqref{eq: second order eqn1}. In this equation, $\langle a_{s} a_{i}\rangle$ can be expressed in terms of $\alpha(\tau)$ by solving for $|\psi(\tau)\rangle$. To achieve this, note that
\begin{align}
|\psi(m\Delta \tau\rangle &= e^{ \eta (m \Delta \tau) (a_{s}^{\dagger}a_{i}^{\dagger}-a_{s}a_{i})}|\alpha_{0}\rangle |00\rangle,
\end{align}
where $a_{p}$ has been replaced by $\alpha(\tau)$ and $\eta(m\Delta \tau)=\sum_{z=0}^{m-1} \Delta \tau \; \alpha_{p}(z \Delta \tau)$. In the limit $\Delta \tau \rightarrow 0$, we have 
\begin{align}
\label{eq:coupled TMSV}
    |\psi(\tau)\rangle &= |\alpha(\tau)\rangle \sum_{n=0}^{\infty}\frac{\tanh^{n}(\eta(\tau))}{\cosh(\eta(\tau))}|nn\rangle,
\end{align}
where $\eta(\tau)=\int_{0}^{\tau} d\tau'' \alpha(\tau'')$. Using the above state, we obtain the following system of differential equations
\begin{subequations}
\label{eq:alphatau1}
\begin{align}
    \partial_{\tau} \alpha(\tau)&=-\frac{\sinh(2 \eta(\tau))}{2}, \label{eq: alpha(tau)}\\
    \partial_{\tau} \eta(\tau)&=\alpha(\tau), \label{eq: eta(tau)}
\end{align}
\end{subequations}
which are equivalent to the set of equations given by Eq. \eqref{eq: second order eqn1} - \eqref{eq: second order eqn3} (cf. Appendix for \ref{sec:Gaussian limit}), and whose solutions for $\alpha(0)=\alpha_{0}$ and $\eta(0)=0$ are given by 
    \begin{subequations}
    	    \label{eq:alphatau2}
    	    \begin{align}
    \eta(\tau)&=\Im\left(\am \left(i\tau\alpha_{0},-1/\alpha_{0}^{2} \right)\right), \label{eq:eta}\\
    \alpha(\tau)&=\alpha_{0} \;  \dn \left(i\tau\alpha_{0},-1/\alpha_{0}^{2} \right), \label{eq:alpha}
\end{align}
    \end{subequations}
where $\am(z)$ is the Jacobi amplitude function, $\dn(z)$ is the Jacobi delta amplitude function. Since the solution for $\alpha_{0}^{2}$ in the second-order expansion accurately describes the pump-mode population of the full down-conversion process for short times, it can be used to compute the pump depletion time. The time at which the pump-mode population decreases relatively from $\alpha_{0}^{2}$ by $\delta$ (i.e., $\frac{\alpha_{0}^{2}-\alpha^{2}(\tau_{d})}{\alpha_{0}^{2}}=\delta$) is given by 
\begin{align}
\label{eq:approx inverse Jacobi}
    \tau_{d}& \approx \frac{1}{\alpha_{0}} \sinh^{-1}\left(\sqrt{\delta}\alpha_{0}\right).
\end{align}
For more details on the exact solution and the derivation of the above solution from it, see Appendix \ref{sec:system-size effects}. As can be seen in Fig. \ref{fig:fig2}, this expression accurately identifies the pump depletion time. Moreover, using the second-order solution, we can also approximately estimate the time at which the population of the signal mode reaches its first local maximum, equivalently the time at which $\langle a_{p}\rangle$ or $\langle a_{p}^{\dagger}a_{p}\rangle$ goes to zero in this approximation. It is given by (cf. Appendix \ref{sec:system-size effects}) 
\begin{align}
\label{eq:tmax approx}
    \tau_{\rm{max}} &\approx \frac{1}{\alpha_{0}} \ln(4\alpha_{0}).
\end{align} 
The time value $\tau_{\rm{max}}$ is shown by the first vertical dotted-line in Fig. \ref{fig:fig6} (Appendix \ref{sec:system-size effects}) where it can be seen that it identifies the maximum of $\sinh^{2}(\eta(\tau))$ accurately and provides us with a reasonable estimate of the time when the signal-mode population reaches local maximum and $\langle a_{p}\rangle$ reaches zero in the full numerically simulated down-conversion process. 

We now turn our attention to moments of the quadrature operators defined as
\begin{align}
X_{\mu} = \tfrac{1}{\sqrt{2}}(a_{\mu}+a_{\mu}^{\dagger}) \text{ and } P_{\mu} = \tfrac{1}{\sqrt{2} i }(a_{\mu}-a_{\mu}^{\dagger}).
\end{align}
In the second-order cumulant approximation it is easy to see that the only quadrature with a nonzero expectation is $\braket{X_P} = \sqrt{2} \alpha(\tau)$ and that the covariance matrix of the quadratures $\bm{R} = (X_P,P_P,X_s,P_s,X_i,P_i)$ generally defined as
\begin{align}
V_{i,j} = \frac{\braket{R_i R_j + R_j R_i}}{2} - \braket{R_i} \braket{R_j}
\end{align}
has the form
\begin{align}
\bm{V}^{(2)} = \tfrac{1}{2} \begin{pmatrix} 1 & 0 \\ 0 & 1 \end{pmatrix} \oplus \begin{pmatrix} c & 0 & s &0 \\
	0 & c & 0 & -s\\
	s & 0 & c & 0\\
	0 & -s & 0 & c
 \end{pmatrix}
\end{align}
where $c = \cosh 2 \eta(\tau)$ and $s = \sinh 2 \eta(\tau)$ in the second-order cumulant approximation. The diagonal elements of the covariance matrix contain the quadrature variances given by
\begin{align}
	 \Delta_{X_{\mu}}^{2}&=\langle X_{\mu}^{2}\rangle -\langle X_{\mu}\rangle^{2}, \ 
	 \Delta_{P_{\mu}}^{2}=\langle P_{\mu}^{2}\rangle -\langle P_{\mu}\rangle^{2}.
\end{align}

While the exact form of the covariance will change when it is calculated for the exact state (beyond any cumulant expansion), it is easy to verify, using the moment selection rules described in Eq.~\eqref{eq:momentrule}, that the elements that are zero in the second-order covariance matrix will remain zero for the exact solution. In particular the direct sum structure between pump and signal-idler will hold true, implying that the only possible entanglement between the pump and the rest of the modes is of non-Gaussian type. {The variance of the pump mode, $\Delta_{P_{p}}^{2}$, under the second-order cumulant approximation is plotted in Fig. \ref{fig:fig1}(b), and it remains at $1/2$ for all times as expected, since the pump-mode in a  coherent state (see Eq. \eqref{eq:coupled TMSV}) }
\subsubsection{Third and Higher Order Cumulant Expansions}
Now we set the fourth order cumulants to zero and approximate fourth order moments in terms of third order moments. In this case, the dynamics is described by $43$ differential equations, but most of the variables are either uncoupled from the evolution of $\langle a_{s}^{\dagger}a_{s}\rangle$ or do not evolve in time provided they start at zero, which is the case for our initial state of interest.  The relevant equations of motion are given by  
\begin{align}
\partial_{\tau}\langle a_{p}\rangle =&- \langle a_{s}a_{i}\rangle, \\
\partial_{\tau}\langle a_{s}a_{i}\rangle =& \langle a_{p}\rangle + 2\langle a_{p}a_{s}^{\dagger}a_{s}\rangle ,\\
\partial_{\tau}\langle a_{p}^{2}\rangle =& - 2\langle a_{p}a_{s}a_{i}\rangle, \\
\partial_{\tau}\langle a_{s}^{\dagger}a_{s}\rangle =&2\langle a_{p}a_{s}^{\dagger}a_{i}^{\dagger}\rangle, \\
\partial_{\tau}\langle a_{p}a_{s}a_{i}\rangle =& -2\langle a_{s}a_{i}\rangle^{2}  - 4\langle a_{p}\rangle^{2}\langle a_{s}^{\dagger}a_{s}\rangle + \langle a_{p}^{2}\rangle\\
& + 4\langle a_{p}\rangle \langle a_{p}a_{s}^{\dagger}a_{s}\rangle +2 \langle a_{p}^{2}\rangle\langle a_{s}^{\dagger}a_{s}\rangle, \nonumber
\\
\partial_{\tau}\langle a_{p}a_{s}^{\dagger}a_{s}\rangle =& - 4 \langle a_{p}\rangle^2\langle a_{s}a_{i}\rangle+\langle a_{p}\rangle\langle a_{p}a_{s}a_{i}\rangle \nonumber \\
& + 3\langle a_{p}\rangle \langle a_{p}a_{s}^{\dagger}a_{i}^{\dagger}\rangle - 2\langle a_{s}^{\dagger}a_{s}\rangle \langle a_{s}a_{i}\rangle  \\
& +\langle a_{p}^{\dagger}a_{p}\rangle \langle a_{s}a_{i}\rangle+\langle a_{p}^{2}\rangle \langle a_{s}a_{i}\rangle, \nonumber
\\
\partial_{\tau}\langle a_{p}a_{s}^{\dagger}a_{i}^{\dagger}\rangle =& \langle a_{p}^{\dagger}a_{p}\rangle - \langle a_{s}^{\dagger}a_{s}\rangle^{2} - \langle a_{s}a_{i}\rangle^{2} +2\langle a_{p}^{\dagger}a_{p}\rangle \langle a_{s}^{\dagger}a_{s}\rangle \nonumber \\
& +4\langle a_{p}\rangle \langle a_{p}a_{s}^{\dagger}a_{s}\rangle  - 4\langle a_{p}\rangle^{2} \langle a_{s}^{\dagger}a_{s}\rangle,
\end{align}
where we have used the fact that  all the variables are real and $\langle a_{p}a_{s}^{\dagger}a_{s} \rangle=\langle a_{p}a_{i}^{\dagger}a_{i} \rangle$. Also, the relationships $\partial_{\tau} \langle a_{i}^{\dagger}a_{i}\rangle=\partial_{\tau}\langle a_{s}^{\dagger}a_{s}\rangle=-\partial_{\tau}\langle a_{p}^{\dagger}a_{p}\rangle$ hold true, so the equations of motion associated with $\langle a_{p}^{\dagger}a_{p}\rangle$ and $\langle a_{i}^{\dagger}a_{i}\rangle$ are not shown. This also means that $\langle a_{s}^{\dagger}a_{s} \rangle=\langle a_{i}^{\dagger}a_{i} \rangle=\alpha_{0}^{2}-\langle a_{p}^{\dagger}a_{p} \rangle$. The above set of differential equations become unstable for longer times, where the number of signal/idler photons predicted by these differential equations become negative and thus unphysical. However, the solution to these system of equations for shorter times can still be used to obtain a better description of the system compared to lower-order solution and study important features of the system while avoiding the complexity associated with the straightforward numerical simulation of the full system. For instance, in Fig. \ref{fig:fig1}a, we can see that the third-order cumulant solution explains the signal-mode population with less than $1\%$ relative error until the pump-mode becomes depleted by $15\%$ while the fourth and the fifth-order cumulant solutions explain it until the time the pump gets depleted by $28 \%$ and $37\%$, respectively. The dynamics can be approximated even longer if the required precision for the signal-mode population obtained from the cumulant-expansion method is lowered. 

Using these higher-order expansions we can obtain the variance in the momentum of the pump mode as shown in Fig. \ref{fig:fig1}(b) in a manner similar to \ref{fig:fig1}(a), where the values obtained from the different order cumulant expansion methods are plotted with different coloured-dashed lines. As in Fig. \ref{fig:fig1}(a), the vertical dotted lines of different colours show how long each of the coloured dashed lines can approximate the value obtained from the numerical simulation of the system with  less than $1\%$ relative error. As seen from this figure, the third order cumulant expansion method identifies squeezing in the pump mode, and approximates the numerical curve (black curve) up to $\tau = 0.0176$ ($15 \%$ pump depletion) with $1 \%$ relative error. Also, the fourth and the fifth order solutions approximate pump-mode squeezing until $\tau=0.0189$ ($28\%$ pump depletion) and $\tau=0.0196$ ($38\%$ pump depletion), respectively. Finally, see Fig. \ref{fig:fig1}(c-d) for a comparison of the zero-delay second-order correlation functions obtained from the numerical simulation with the values obtained from the cumulant expansion method. From Fig. \ref{fig:fig1}(a-d), it can be clearly seen that the $n^{\text{th}}$ order cumulant expansion explains the dynamics of the system longer than the $(n-1)^{\text{th}}$ order cumulant expansion, as expected. However, as the order of the approximation is increased, the information that is gained from seeking the next-order solution decreases. That is, the length of the time interval over which the $n^{\text{th}}$-order solution accurately describes the system compared to the $(n-1)^{\text{st}}$-order dynamics decreases as $n$ is increased. Therefore, this method provides its biggest advantage over the full numerical simulation of the Hamiltonian for short-times. For instance, as seen in this section,  all the pump-mode properties ($\langle N_{p}\rangle$, $\Delta^{2}_{X_{P}}$ and $g^{(2)}_{p}$) until the time pump is depleted by $15 \%$ can be explained by the solution from the third order cumulant expansion method, which is described by only seven equations. This solution can be used to obtain important information about various properties of the system (such as the time when the pump-mode becomes squeezed, or the time when $g_{p}^{(2)}$ deviates from one) without having to simulate the full down-conversion dynamics for $\alpha_{0}^{2}=10^{5}$.

\begin{figure*}[t]
\centering
\includegraphics[width=\textwidth]{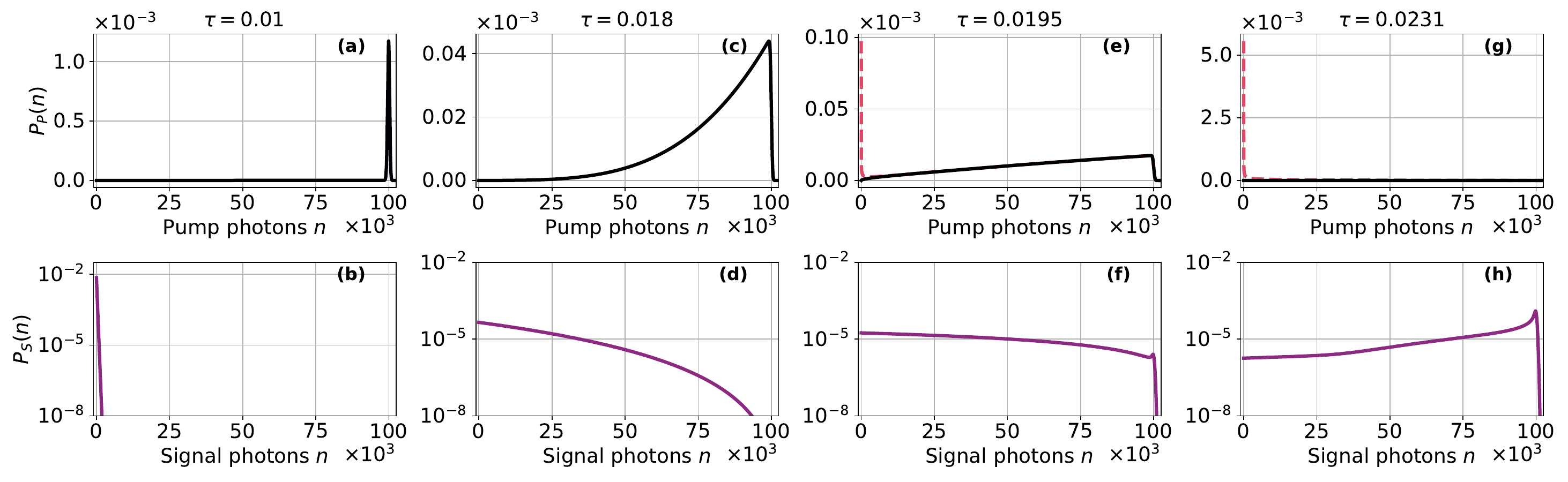}
\caption{The photon-number statistics of the pump mode, $\langle n|\rho_{p}|n\rangle$, and the signal mode $\langle n|\rho_{s}|n\rangle$ are shown here for $\tau=0.01$, $\tau=0.018$, $\tau=0.0195$ and $\tau=0.0231$ in each column for $\alpha_{0}^{2}=10^{5}$. \textbf{(a, b)} For short times, the pump-mode remains in a coherent state and the signal/idler modes evolve into a two-mode squeezed vacuum state. Note that the photon-number statistics of the signal mode are plotted on a semi-log scale. \textbf{(c, d)} As the time progresses, the marginal state of the pump mode becomes skewed towards the $n=0$ Fock state while the marginal state of the signal mode starts to develop support over larger number Fock states. {The time at which the statistics of the pump mode starts to become skewed is also the time at which the pump-mode first becomes squeezed.}  \textbf{(e, f)} When the pump-mode marginal state develops support over the Fock states in the vicinity of $n=0$ state, the photon number statistics of the odd and the even numbered Fock states develop two different distributions. The distribution associated with the odd and even states are shown here by the thick and dashed curves respectively. On the other hand, note that the photon-number statistics of the signal mode becomes flatter for larger times \textbf{(g, h)} This instant corresponds to the time when the signal (pump)-mode population shown in Fig. \ref{fig:fig1} reaches local maxima (minima). At this point, the photon statistics of the pump mode has a large support on even states with a small support on the odd states, as can be seen in the figure above. The signal mode, on the other hand, develops a peak around $\alpha_{0}^{2}=10^{5}$ at this instant.}
\label{fig:fig3}
\end{figure*}
\subsection{Perturbation theory}
In this section, we perform a transformation on the Hamiltonian in Eq. \eqref{eq:Int Hamiltonian} that allows us to further understand the behaviour of various quantities analyzed in the previous subsection. We do this by isolating the non-Gaussian dynamics of the system \cite{Yanagimoto2022onset, Yanagimoto:24}. To achieve this, we substitute the ansatz 
\begin{align}
|\psi(\tau)\rangle=D(\alpha(\tau))S(\eta(\tau))|\phi(\tau)\rangle
\end{align}
 into the Schr\"{o}dinger equation where $D(\alpha(\tau))=\exp[\alpha(\tau)(a_{p}^{\dagger}-a_{p})]$, $S(\eta(\tau))=\exp[\eta(\tau)( a_{s}^{\dagger}a_{i}^{\dagger}-a_{s}a_{i})]$ and solve for the dynamics of $|\phi(\tau)\rangle$. Here $\alpha(\tau)$ and $\eta(\tau)$ have initial conditions $\alpha(0)=\alpha_{0}$ and $\eta(0)=0$, and satisfy Eq.~\eqref{eq:alphatau1}-\eqref{eq:alphatau2}. As a result, $|\phi(\tau)\rangle$ evolves under the effective Hamiltonian
\begin{align}
\label{eq:effective Hamiltonian main}
    \frac{\tilde{H}_{\text{eff}}(\tau)}{i \hbar} \equiv \frac{H_{\rm{eff}}}{ i \hbar \chi}=& \sinh^{2}(\eta(\tau))(a_{p} a_{s}a_{i}-a_{p}^{\dagger}a_{s}^{\dagger}a_{i}^{\dagger}) \\
    & +\cosh^{2}(\eta(\tau))(a_{p}a_{s}^{\dagger}a_{i}^{\dagger}-a_{p}^{\dagger}a_{s}a_{i})  \nonumber \\
    & + \frac{\sinh(2\eta(\tau))}{2}(a_{p}-a_{p}^{\dagger})(a_{s}^{\dagger}a_{s}+a_{i}^{\dagger}a_{i}),  \nonumber
\end{align}
with initial state $|\phi(0)\rangle = |0\rangle|0\rangle|0\rangle$. In order to gain insight into the short-time dynamics of the system, we apply time-dependent perturbation theory and obtain expressions (cf. Appendix \ref{sec:perturbation theory}) for quantities analyzed in the previous subsection. The pump-mode population at short times can be described as
\begin{align}
    \langle \psi(\tau)&|N_{s}|\psi(\tau)\rangle  - \sinh^{2}(\eta(\tau))\\
    & = -\frac{7\alpha_0^4}{45}  \tau^6- \frac{\left(48 \alpha _0^6-77 \alpha_0^4\right)}{630}  \tau ^8+\mathcal{O}(\tau^{9}). \nonumber
\end{align}
where we neglected terms of ninth-order and above in $\tau$. It is interesting to note from the above expression that the second-order cumulant expansion correctly accounts for the signal-mode population up to fifth order in time and it overestimates the value of the signal mode population beyond that time, which is the time identified by the first vertical dotted line in Fig. \ref{fig:fig1}(a). This is consistent with the fact that $\sinh^{2}(\eta(\tau))$ reaches $\alpha_{0}^{2}$, while $\langle N_{s}\rangle$ has a local maximum at about $78\%$ of $\alpha_{0}^{2}$ in the case of $\alpha_{0}^{2}=10^{5}$. Also, we confirm the accuracy of the equation in Fig. \ref{fig:fig5}(a-d), where the sixth-order expression is compared with $\langle N_{s}\rangle -\sinh^{2}(\eta(\tau))$, where $\langle N_{s}\rangle$ is obtained from the full numerical simulation of the down-conversion process for small times. The perturbation theory expression for the variance in the momentum of the pump mode is given by 
\begin{align}
\label{eq:pert pump variance}
    \Delta_{P_{p}}^{2}&= \frac{1}{2}-\frac{1}{6} \alpha_0^2 \tau^4+\left(\frac{\alpha
   _0^2}{18}-\frac{\alpha _0^4}{45}\right) \tau
   ^6 \\
   & +\frac{\left(-4 \alpha _0^6+144 \alpha _0^4-19 \alpha
   _0^2\right)}{2520}\tau ^8 + \mathcal{O}(\tau^{9}). \nonumber
\end{align}
Note that $1/2$ variance is expected at the second-order cumulant expansion since the pump-mode state predicted in this approximation is a coherent state, and the value of the variance in the pump-mode momentum deviates from the second-order cumulant expansion value at fourth-order in time. We compare the fourth-order term with the numerical values of $\Delta_{P_{p}}^{2}-1/2$ obtained from the full simulation of the down-conversion process in Fig. \ref{fig:fig5}(e-h), and the numerical data behaves in a manner predicted by the above expression for short times. Also, since the variance remains at $1/2$ for short times, this quantity has a threshold behaviour, so it has a value close to $1/2$ before a certain time $\tau_{\rm{sqz}}$, and then it decreases relatively quickly beyond this time (see Fig. \ref{fig:fig1}(b) and \ref{fig:fig6}(e-h)). The squeezing threshold time $\tau_{\rm{sqz}}$ has been plotted as a function of $\alpha_{0}^{2}$ in Fig. \ref{fig:fig8}(a) (Appendix \ref{sec:system-size effects}), where it can be seen that $\tau_{\rm{sqz}} \propto \alpha_{0}^{-0.653}$. Here we define $\tau_{\rm{sqz}}$ as the time at which the pump variance decreases relatively by $1 \%$ from $1/2$. In Fig. \ref{fig:fig8}(a), we also compare the threshold time obtained numerically with the estimates obtained using the sixth and eight order expressions from Eq. \eqref{eq:pert pump variance} (cf. Appendix \ref{sec:system-size effects}). As shown in this figure, both the sixth and the eighth order expressions explain the behaviour of threshold time reasonably well. Moreover, an analytic estimate obtained using the sixth order expression in $\tau$ shows that $\tau_{\rm{sqz}} \propto \alpha_{0}^{-2/3}$ for $\alpha_{0} \gg 1$, and this scaling behaviour is close to the one observed using the numerical data. The perturbation theory expressions for  $g^{(2)}_{p}$ and $g^{(2)}_{s}$ are given by 
\begin{subequations}
\begin{align}
\label{eq:pert gp}
    g^{(2)}_{p}=&1+\frac{1}{3}\tau^4 + \frac{2}{45} \left(11 \alpha _0^2+5\right) \tau^6 \\
    & +\frac{1}{630} \left(114 \alpha _0^4-128 \alpha_0^{2} +97\right) \tau ^8 +\mathcal{O}(\tau^{9}), \nonumber \\
\label{eq:pert gs}
    g^{(2)}_{s}=& 2-\frac{2}{3} \tau ^2+\frac{1}{810}\left(45-432 \alpha _0^2\right) \tau ^4 \\
    &+\frac{\left(198 \alpha _0^4-459 \alpha _0^2+69\right)}{405} \tau^6 \nonumber \\
    & +\frac{\left(-888 \alpha _0^6+1935 \alpha _0^4-2838 \alpha _0^2+415\right)}{4050}\tau ^8 +\mathcal{O}(\tau^{9}). \nonumber 
\end{align}
\end{subequations}
The lowest-order corrections to $g_{p}^{(2)}$ and $g_{s}^{(2)}$ in time are at fourth and second order, respectively. Similar to the case of pump-mode squeezing, these $g^{(2)}$ functions also show threshold behaviour (see Fig. \ref{fig:fig1}(c-d)), and the threshold time has been plotted as a function of $\alpha_{0}^{2}$ in Fig. \ref{fig:fig8}(c-d). As seen in this figure, the threshold times for $g_{p}^{(2)}$ and $g_{s}^{(2)}$ scale as $\tau_{p} \propto \alpha_{0}^{-0.633}$ and $\tau_{s} \propto \alpha_{0}^{-0.713}$ respectively. {The eighth order expressions shown in Eq. \eqref{eq:pert gp} and Eq. \eqref{eq:pert gs} cannot accurately predict this threshold time, particularly $g^{(2)}_{p}$ as shown in Fig. \ref{fig:fig8}(c). This results from the fact that the $g^{(2)}_{p}$ changes from one at longer times, so the threshold time is longer compared to other threshold times (See Fig. \ref{fig:fig1}). Higher-order corrections need to be obtained for a more accurate estimate, but a rough estimate of these times could be still obtained from the eight-order expressions.} Also see Ref. \cite{Xing2023pump} for perturbation theory expressions derived by taking the pump depletion into account but with the interaction term as the perturbation parameter, and see Ref. \cite{Nation2010} for the state obtained through Baker–Campbell–Hausdorff formula up to second order in time.

\subsection{Photon-Number Statistics}
In this subsection, we study the marginal pump-mode and signal/idler-mode photon number statistics in the Fock basis to obtain a general understanding of the behaviour of the state. Using the state of the system given in Eq.~\eqref{eq:linear superposition}, the density operators corresponding to the pump and signal modes after tracing out the other modes can be expressed as
\begin{align}   
\label{eq:rho pump}
\rho_{p}&=\sum_{n,n'=0}^{\infty}\sum_{k=0}^{n}\beta_{n-k,k} \beta^{*}_{n'-k,k}\;|n-k\rangle\langle n'-k|, \\
\rho_{s}&=\sum_{n=0}^{\infty}\sum_{k=0}^{n}|\beta_{n-k,k}|^{2}\;|k\rangle\langle k|.\label{eq:rho signal}
\end{align}
At the initial time, $\langle n|\rho_{p}|n\rangle$, is a Poisson distribution since the pump mode is in a coherent state. For short times, the pump-mode statistics continues to remain Poissonian with a mean shifting towards $n=0$ (Fig. (\ref{fig:fig3}a)), while the signal and idler modes evolve into a TMSV, as expected from the second-order cumulant solution. The thermal state statistics of the signal mode can be noticed in Fig. \ref{fig:fig3}(b) where the distribution associated with the photon number statistics traces out a straight line with negative slope when plotted on a semi-log scale. For longer times, the statistics of the pump mode start to become skewed towards $n=0$ as shown in Fig. \ref{fig:fig3}(c), suggesting the role of third-order correlations. The signal mode on the other hand starts to develop support over the Fock states with larger number of photons (Fig. \ref{fig:fig3}(d)). As time progresses, when the pump-mode photon-number statistics start to develop support on the Fock states in the vicinity of $n=0$, the statistics corresponding to odd and even terms develop two different distributions as seen in Fig. \ref{fig:fig3}(e). The distribution associated with the statistics in the even Fock states is shown by a dashed curve whereas the distribution associated with the odd Fock states is shown by a thick curve. Also, the distribution associated with the signal mode statistics at this point becomes flatter (Fig. \ref{fig:fig3}(f)) suggesting a low degree of purity in the marginal signal state. As time increases further, the support on odd (even) Fock states decreases (increases) with time until $\tau=0.0231$, which is the instant where the signal-mode (pump-mode) population has a local maxima (minima) (see Fig. \ref{fig:fig1}(a)). At this instant, the marginal pump-mode state (Fig. \ref{fig:fig3}(g)) has a very small support on odd Fock states. Note that this is also the time when $\langle a_{p}\rangle$ has a value close to zero and the variance of momentum in the pump-mode becomes minimum (note the location of the black dotted line in Fig. \ref{fig:fig6}). A rough time estimate of this point as a function of $\alpha_{0}$ is given by Eq. \eqref{eq:tmax approx}. {For the study of long-time dynamics beyond this point, see Appendix \ref{sec:long-time evolution}}
\section{Entanglement between Modes}
\label{sec:entanglement}
\begin{figure*}[t]
\centering
\includegraphics[width=\textwidth]{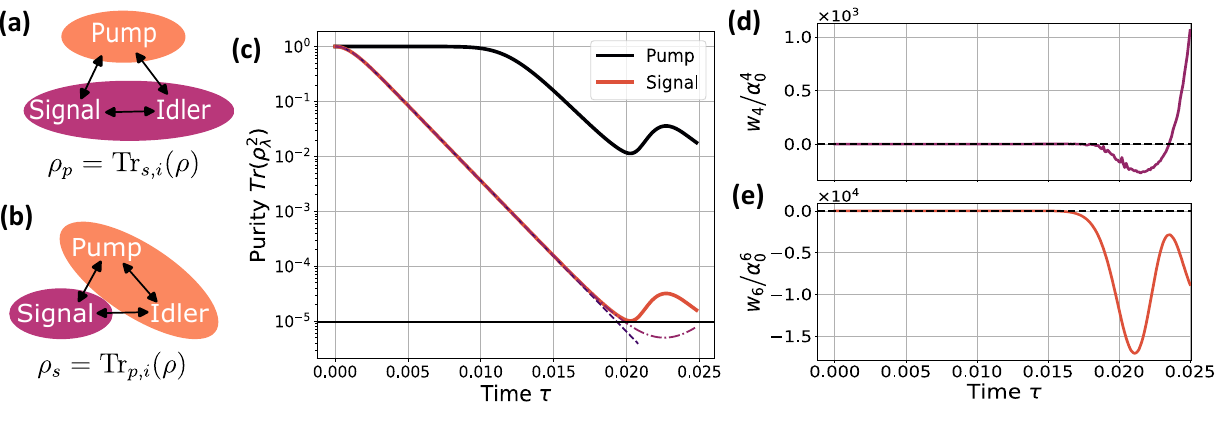}
\caption{\textbf{(a-b)} The bipartite entanglement between pump mode or signal mode and the remaining part of the system is of interest to us in this work, and both the cases are schematically shown here. \textbf{(c)} The purities of the pump and the signal mode obtained after tracing out remaining modes are shown here for $\alpha_{0}^{2}=10^{5}$. The signal mode purity is compared with the purity obtained from the parametric approximation (dashed line) and the second order cumulant expansion method (dash-dotted line). \textbf{(d)} Two entanglement witnesses for the pump-mode corresponding to the fourth and sixth order moments are shown here for $\alpha_{0}^{2}=10^{5}$. As can be seen, the sixth order witness captures the finer details of pump purity as opposed to the fourth-order witness. Note that the rescaling of the witness does not affect where the witness is positive/ negative, which is the essential feature in identifying the entanglement of the state. A separable state has $w_{4}$ and $w_{6}$ positive, so the negativity of these quantities imply that the state is entangled.}
\label{fig:fig4}
\end{figure*}

In this section, we are mainly interested in characterizing the entanglement of the pump mode with the rest of the system and comparing it with that of signal/idler-mode entanglement. As schematically shown in Fig. \ref{fig:fig4}(a) and \ref{fig:fig4}(b), it should be emphasized that we are focused here on identifying the bipartite entanglement between the two parts of the system, unlike the trimodal problem considered here. The entanglement of the pump mode for small $\alpha_{0}^{2} \leq 9$ has been studied using the mutual information in the context of Hawking radiation \cite{Nation2010}, but an analysis identifying a witness for the detection of this entanglement or the time scale of its origin particularly for larger system sizes ($\alpha_{0}^{2}$) has not been given to the best of our knowledge. Here, we aim to analyze the behaviour of pump-mode entanglement and seek a witness that can be expressed in terms of moments of the system. We have seen that the signal and idler modes become entangled even in the parametric approximation where these modes evolve into a two-mode squeezed vacuum state. 
The purities of the above density operators $\rho_{p}$ in Eq. \eqref{eq:rho pump} and $\rho_{s}$ in Eq. \eqref{eq:rho signal}, given by $\text{Tr}(\rho_{p}^{2})$ and $\text{Tr}(\rho_{s}^{2})$, are both shown in Fig. \ref{fig:fig4}c. As seen here, the purity associated with the pump mode {deviates from one} as a function of time indicating that it becomes entangled with the remaining system. Note that for small times the purity of the pump mode is {one}, as this region corresponds to the time when the parametric approximation and the second order cumulant solution accurately describes the system, and the pump mode remains unentangled. Even though the pump mode becomes depleted in the second order solution, it remains unentangled with other modes indicating that pump depletion is not equivalent to pump-mode entanglement in this system. Using perturbation theory, we also show that the purity of the pump mode is given by
\begin{align}
    \gamma(\tau)&=1-\frac{2 \alpha_{0}^4}{9}\tau ^6+\left(\frac{2 \alpha_{0}^4}{9}-\frac{4 \alpha_{0}^6}{45}\right) \tau ^8 +\mathcal{O}(\tau^{9}),
\end{align}
which shows that the purity at short times deviates from one as $\tau^{6}$, and we also confirm this numerically in Fig. \ref{fig:fig5}(i-l) (cf. Appendix \ref{sec:perturbation theory}). Similar to other quantities, the purity of the pump-mode also shows threshold behaviour as shown in Fig. \ref{fig:fig4}(c), where it is one for times smaller than $\tau_{\text{ent}}$ and deviates from one relatively rapidly for $\tau>\tau_{\text{ent}}$. Based on numerical data, we find that the time required for pump-mode purity to decrease from one relatively by $1 \%$ of its possible range, $\left(1/(n_{2}+1)\right) \leq \gamma(t) \leq 1$ scales as $\tau_{\rm{ent}} \propto \alpha_{0}^{-0.768}$ as shown in Fig. \ref{fig:fig8}(b) (Appendix \ref{sec:system-size effects}). As shown in the figure, the eighth-order perturbation theory expression accurately identifies this threshold time. Moreover, we also obtain an analytic expression for threshold time as a function of $\alpha_{0}$ using this expression: $\tau_{\rm{ent}} \propto \alpha_{0}^{-3/4}$ for $\alpha_{0} \gg 1$ (cf. Appendix \ref{sec:system-size effects}), and this agrees with the behaviour of the numerical data. On the other hand, the purity of the thermal state, which is obtained after tracing out idler mode from the TMSV in Eq. (\ref{eq:TMSV}), is given by
\begin{align}
\label{eq:thermal purity}
\gamma(t) &=\sech\left(2 r(\tau)\right),
\end{align}
where $r(\tau)=\alpha_{0} \tau$. This expression, as shown by the dashed line in Fig. \ref{fig:fig4}c, agrees with the value of the signal mode purity obtained numerically at short times. However, the quantity in Eq. (\ref{eq:thermal purity}) is unbounded with $r(\tau)=\alpha_{0} \tau$, as it is obtained from the parametric approximation. The purity obtained from the second-order cumulant expansion method is also given by Eq. \eqref{eq:thermal purity} with $r(\tau)=\eta(\tau)$ where $\eta(\tau)$ is defined in terms of Jacobi amplitude function (Eq. \eqref{eq:eta}). This expression is bounded and better approximates the signal purity as shown by the dashed-dotted line in Fig. \ref{fig:fig4}(c). Moreover, the numerical value of the signal-mode purity reaches close to the minimum that can be observed in the system given by $1/(n_{2}+1)$, where $n_{2}$ is the upper cutoff used for numerical simulations as defined in Sec. \ref{sec:Macroscopic pump}.

As mentioned before, the state of the system in the second order cumulant expansion method is Gaussian at all times because the {third-order cumulants are set to zero} in this case. A sufficient condition for a Gaussian state to be separable is that its associated covariance matrix can be written as a direct sum of pump and signal, idler modes, $\sigma=\sigma_{p} \oplus \sigma_{s,i}$ \cite{Serafini2023quantum}. The off-diagonal block elements between the pump and signal mode are given by 
\begin{align}
\begin{split}
\langle A_{p}A_{s}\rangle =\sum_{n,n'=0}^{\infty}\sum_{k=0}^{n}&\beta_{n'-k,k}^{*}(\tau)\beta_{n-k,k}(\tau)\; \\ 
&  \times \langle n'-k|A_{p}|n-k\rangle \langle k|A_{s}|k\rangle, 
\end{split}
\end{align}
where $A_{p}$ and $A_{s}$ are local operators acting only on the pump and signal mode, respectively. The expectation value $\langle k|A_{s}|k\rangle_{s}$ is zero if $A_{s}=X_{s}$ or $A_{s}=P_{s}$. Hence, $\langle X_{p}X_{s} \rangle=\langle P_{p}X_{s} \rangle= \langle X_{p}P_{s} \rangle=\langle P_{p}P_{s} \rangle=0$. Likewise, for the idler mode, it can be shown that $\langle X_{p}X_{i} \rangle=\langle P_{p}X_{i} \rangle= \langle X_{p}P_{i} \rangle=\langle P_{p}P_{i} \rangle=0$. Therefore, the covariance matrix for all times (without any approximation) is of the form $\sigma=\sigma_{p} \oplus \sigma_{s,i}$. This further corroborates the fact that the pump mode is not entangled with the other modes for short times, where the second order cumulant expansion is accurate. 

The block-diagonal form of $\sigma$ suggests the need for higher-order witnesses for probing the pump-mode entanglement. In fact, the form of effective Hamiltonian in Eq. \eqref{eq:effective Hamiltonian main} provides us with a hint of the type of witness that could be helpful. In particular, the first term in the above effective Hamiltonian is identical to the Hamiltonian described in \cite{Agusti2020tripartite} where a witness is prescribed for identifying tripartite entanglement. The witness is given by 
\begin{align}
|\langle a_{p}a_{s}a_{i}\rangle_{\phi(\tau)}| \leq \sqrt{\langle N_{p}\rangle_{\phi(\tau)} \langle N_{s}N_{i}\rangle_{\phi(\tau)}},
\end{align}
where we emphasize that expectation values should be taken with respect to the state 
\begin{align}
\ket{\phi(\tau)} &= D(-\alpha(\tau))S(-\eta(\tau)) \ket{\psi(\tau)} \\
&\approx \ket{0}\ket{0}\ket{0} - \frac{\alpha_0^2 \tau^3}{3} \ket{1}\ket{1}\ket{1}
\end{align}
 Our analysis shows that this witness also detects entanglement of the pump mode for the system described here, but it only identifies entanglement for a small-time interval and is not a reliable witness for larger system sizes. This suggests that a witness involving fourth-order moments is likely required to probe this entanglement. Also note that since the operations performed on the down-conversion Hamiltonian to obtain the effective Hamiltonian are local unitaries with respect to pump and signal-idler subsystems, they do not alter the entanglement between the pump and the signal-idler modes.

\subsection{Higher-order Witness}
In this subsection, we identify witnesses involving fourth order and sixth order moments that detect the pump-mode entanglement through inequalities provided by the positive partial transpose (PPT) criterion \cite{Shchukin2005inseparability}. The problem of identifying entanglement through the PPT criterion is typically addressed by checking if the state of the system violates an inequality that is satisfied by all separable states: a violation of this type of inequality for a particular state successfully provides a witness that detects entanglement. 
According to Shchukin and Vogel~\cite{Shchukin2005inseparability}, any separable state in mode $a$ and mode $b$ has all the principal minors of the following matrix nonnegative.
\begin{align} M_{j,k} &= \big\langle (a^{\dagger})^{k_{2}} a^{k_{1}}(a^{\dagger})^{j_{1}}a^{j_{2}}(b^{\dagger})^{j_{4}}b^{j_{3}}(b^{\dagger})^{k_{3}}b^{k_{4}}\big\rangle,
\end{align}
where $j\equiv \{j_{1},j_{2},j_{3},j_{4}\}$ and $k \equiv \{k_{1},k_{2},k_{3},k_{4}\}$ correspond to $j^{\text{th}}$ and $k^{\text{th}}$ index in some ordered sequence as described in \cite{Shchukin2005inseparability}. 

In our case, we identify mode $a$ with the pump mode ($a\equiv a_{p}$) and mode $b$ with a linear combination of signal and idler ($b\equiv a_{s} \cos\theta+a_{i} \sin\theta$). Upon inspection, we were able to find that the principal minors involving fourth order moments can detect entanglement, but only for a short window of time. For instance, a witness for $\theta=\pi/4$ is given by 
\begin{align}
    w_{4}&=\begin{vmatrix}
    \langle a_{p}^{\dagger}a_{p}b^{\dagger}b\rangle & \langle (a_{p}^{\dagger})^{2}b^{\dagger}b\rangle & \langle (a_{p}^{\dagger})^{2}b^{2}\rangle  \\
    \langle a_{p}^{2}b^{\dagger}b\rangle &  \langle a_{p}a_{p}^{\dagger}b^{\dagger}b\rangle & \langle a_{p}a_{p}^{\dagger}b^{2}\rangle \\
    \langle a_{p}^{2}(b^{\dagger})^{2}\rangle & \langle a_{p}a_{p}^{\dagger}(b^{\dagger})^{2}\rangle & \langle a_{p} a_{p}^{\dagger}bb^{\dagger} \rangle  \\
\end{vmatrix},
\end{align}
and the full form of the witness in terms of $a_{s}$ and $a_{i}$ is shown in Appendix \ref{sec:witness} for conciseness. This witness is plotted in Fig. \ref{fig:fig4}(d) for $\alpha_{0}^{2}=10^{5}$ where it can be seen that the entanglement is detected only for a small time window, and the witness does not capture the finer details associated with the behaviour of the purity. Note that the above witness rescaled by $\alpha_{0}^{4}$ has been plotted in Fig. \ref{fig:fig4}(d), but this rescaling does not alter the positivity/negativity of the witness, which is the crucial factor for probing entanglement. 
Upon further analysis, we were also able to identify a sixth-order witness that captures the behaviour of entanglement as characterized by the pump-mode purity. The witness is given by
\begin{align}
w_{6}&= \begin{vmatrix}
1 & \langle a_{p} (b^{\dagger})^{2}\rangle & \langle a_{p}^{\dagger}b^{2}\rangle\\ 
 \langle a_{p}^{\dagger}b^{2}\rangle & \langle a_{p}^{\dagger}a_{p} (b^{\dagger})^{2}b^{2}\rangle & \langle (a_{p}^{\dagger})^{2}b^{4}\rangle\\
\langle a_{p} (b^{\dagger})^{2}\rangle & \langle a_{p}^{2}(b^{\dagger})^{4}\rangle & \langle a_{p}a_{p}^{\dagger}b^{2}(b^{\dagger})^{2}\rangle
\end{vmatrix},
\end{align}
where $\theta$ is chosen to be $\pi/4$. The complete form of this expression is shown in Appendix \ref{sec:witness}. The rescaled witness is plotted as a function of time in Fig. \ref{fig:fig4}(e) for $\alpha_{0}^{2}=10^{5}$. As can be seen there, the witness starts to become negative around the time where the entanglement becomes nonzero, and the oscillations in purity correspond well with the witness. 

\section{Summary and Outlook}
We studied the dynamics of the down-conversion process through the cumulant-expansion method, perturbation theory and photon number statistics with a focus on the pump mode features such as its depletion time, squeezing, and entanglement with the rest of the system. Through the cumulant-expansion method, we obtained an analytic solution for the state of the system that explains the system's behaviour for short times. This solution provides an expression for pump depletion time, and also shows that the pump depletion is not equivalent to pump mode entanglement. Moreover, we show that the third order cumulant expansion method (described by seven coupled differential equations) allows us to numerically study features such as the time when the pump-mode squeezing originates in the system and the time when the pump-mode autocorrelation function deviates from the values expected from the parametric approximation. Moreover, we performed a transformation of the Hamiltonian and used perturbation theory to study the behaviour of the system analytically beyond the second-order cumulant expansion method. Through this, we obtained expressions for the time when the pump-mode squeezing and its entanglement with the rest of the system originates in the system. Then, we studied the photon number statistics of the pump and the signal mode to obtain a general understanding of the behaviour of the modes. Finally, we provide an entanglement witness based on the PPT criterion using sixth order moments that can detect the entanglement of the pump mode.

While we leave the study of the properties of the system in the presence of experimental imperfections for future work, we make some general comments about the effects of imperfections on three pump-mode properties studied in this work: depletion, squeezing and entanglement. First, it should be noted that PDC experiments are typically performed with in wavelengths  away from  resonances and thus where the material is transparent and absorption can be neglected~\cite{Steinlechner2013absorption}; However, scattering losses, mode mismatch and detector inefficiencies can lead to loss. Despite these losses, the process of pump depletion can still be observed in experiments by calibrating the losses as was done in \cite{Florez2020pump}. Furthermore, the variance of a particular quadrature of a mode in the presence of losses is given by $V_{\text{meas}}=\eta V_{\text{ide}}+(1-\eta) V_{\text{vac}}$ where $V_{\text{ide}}$, $V_{\text{mes}}$ and $\eta$ are the ideal variance, measured variance and the energy transmission coefficient respectively. From this equation, it can be seen that $\left(V_{\text{meas}}/V_{\text{vac}}\right) < 1$ if $\left(V_{\text{ide}}/V_{\text{vac}}\right) < 1$, thus loss will reduce the amount of measurable noise reduction (squeezing) but will not destroy it. Finally, entanglement being the ultimate quantum correlation is of course the quantity of interest that will be most sensitive to decoherence. The measurement of this quantity will require a more detailed analysis incorporating the particular details of a given experimental setup.

The code that has been used to produce results in this manuscript is available at \url{https://github.com/polyquantique/beyond_parametric}.
\section*{Acknowledgements}
We would like to thank Jeff Lundeen, John Sipe, Milica Banic, Elizabeth Agudelo and Salvador Poveda-Hospital for insightful discussions. We also thank Javier Martínez-Cifuentes for providing feedback on the manuscript. This work was supported by the Ministère de l’Économie et de l’Innovation du Québec and the Natural Sciences and Engineering Research Council of Canada. K.C. acknowledges the support from the Fonds de recherche du Québec-Nature et technologies (FRQNT) under the Programme PBEEE / Bourses de stage postdoctoral. This research was enabled in part by support provided by the Digital Research Alliance of Canada and l’Institut transdisciplinaire d'information quantique (INTRIQ).
\appendix
\begin{widetext}
\section{Gauge Freedom} 
\label{sec:Gauge freedom}
In this section, we show that the coupling between the modes ($\chi e^{i\theta}$) and $\alpha_{0}$ can be assumed to be real without any loss of generality. The Hamiltonian in the general case is given by 
\begin{align}\label{eq:Int Hamiltonian with complex kappa}
H&=i\hbar \chi\bigl(e^{i\theta}a_{p}a_{s}^{\dagger}a_{i}^{\dagger}-e^{-i\theta}a_{p}^{\dagger} a_{s}a_{i}\bigr),
\end{align}
where we assume that the coupling constant ($\chi e^{i\theta}$) can be complex. The state of interest in this work is given by
\begin{align}
    |\psi_{\theta,\alpha_{0}}(\tau)\rangle&= e^{-i\frac{H}{\hbar \chi} \tau}|\alpha_{0}\rangle|0\rangle|0\rangle= e^{\tau(e^{i\theta} a_{p}a_{s}^{\dagger}a_{i}^{\dagger}-e^{-i\theta} a_{p}^{\dagger}a_{s}a_{i})}e^{(\alpha_{0} a^{\dagger}_{p}-\alpha_{0}^{*}a_{p})}|0\rangle|0\rangle|0\rangle \\
    \label{eq:unitary with phase}
    &\equiv U(\tau, \theta,|\alpha_{0}|e^{i\phi_{0}})|0\rangle|0\rangle|0\rangle,
\end{align}
where $\alpha_{0}=|\alpha_{0}|e^{i\phi_{0}}$ and $\tau=\chi t$. Now, we have
\begin{align}
\begin{split}
    |\psi_{\theta,\alpha_{0}}(\tau)&\rangle = U_{1}(\theta,\phi_{0})\left(U_{1}^{\dagger}(\theta,\phi_{0}) e^{\tau(e^{i\theta}a_{p}a_{s}^{\dagger}a_{i}^{\dagger}-e^{-i\theta}a_{p}^{\dagger}a_{s}a_{i})}U_{1}(\theta,\phi_{0})\right)e^{-i(\theta+\phi_{0})a_{s}^{\dagger}a_{s}}\left(e^{-i\phi_{0} a_{p}^{\dagger}a_{p}}e^{|\alpha_{0}|(e^{i\phi_{0}} a_{p}^{\dagger}-e^{-i\phi_{0}}a_{p})}e^{i\phi_{0} a_{p}^{\dagger}a_{p}}\right)\\
    & \qquad\qquad\qquad\qquad\qquad\qquad\qquad\qquad\qquad\qquad\qquad\qquad\qquad\qquad\qquad\qquad\qquad\qquad e^{-i\phi_{0} a_{p}^{\dagger}a_{p}}|0\rangle|0\rangle|0\rangle,
\end{split}
\end{align}
where $U_{1}(\theta,\phi_{0})=e^{i\phi_{0}a_{p}^{\dagger}a_{p}}e^{i(\theta+\phi_{0})a_{s}^{\dagger}a_{s}}$. Now, using the fact that $e^{-i\phi_{\mu} a_{\mu}^{\dagger}a_{\mu}}a_{\mu}e^{i\phi_{\mu} a_{\mu}^{\dagger}a_{\mu}}=a_{\mu}e^{i\phi_{\mu}}$ and $e^{-i (\phi_{0} a_{p}^{\dagger}a_{p}+(\phi_{0}+\theta)a_{s}^{\dagger}a_{s})}|0\rangle|0\rangle|0\rangle=|0\rangle|0\rangle|0\rangle$, we have
\begin{align}
\begin{split}
    |\psi_{\theta,\alpha_{0}}(t)\rangle &= U_{1}(\theta,\phi_{0})e^{\tau(a_{p}a_{s}^{\dagger}a_{i}^{\dagger}-a_{p}^{\dagger}a_{s}a_{i})} e^{|\alpha_{0}|( a_{p}^{\dagger}-a_{p})}|0\rangle|0\rangle|0\rangle
\end{split}\\
&= U_{1}(\theta,\phi_{0})U(\tau,0,|\alpha_{0}|)|0\rangle|0\rangle|0\rangle \\
    & = e^{i (\phi_{0} a_{p}^{\dagger}a_{p}+(\phi_{0}+\theta)a_{s}^{\dagger}a_{s})}|\psi_{0,|\alpha_{0}|}(\tau)\rangle.
\end{align}
The above derivation shows that the state obtained using complex coupling coefficient and complex $\alpha_{0}$ is same as the state obtained using the absolute values ($\chi$,$|\alpha_{0}|$) of the associated coupling constant and $\alpha_{0}$ but with an extra factor $e^{i (\phi_{0} a_{p}^{\dagger}a_{p}+(\phi_{0}+\theta)a_{s}^{\dagger}a_{s})}$ added. The effect of this factor on a general expectation value is given by 
\begin{align}
    \langle (a_{p}^{\dagger})^{c} a_{p}^{d} (a_{s}^{\dagger})^{e} a_{s}^{f}(a_{i}^{\dagger})^{g} a_{i}^{h}\rangle_{\theta,\alpha_{0}} &=\langle \psi_{\theta,\alpha_{0}}(\tau)|(a_{p}^{\dagger})^{c} a_{p}^{d} (a_{s}^{\dagger})^{e} a_{s}^{f}(a_{i}^{\dagger})^{g} a_{i}^{h}|\psi_{\theta,\alpha_{0}}(\tau)\rangle\\
    &= \langle \psi_{|\alpha_{0}|}(\tau)|U_{1}^{\dagger}(\theta,\phi_{0})\left((a_{p}^{\dagger})^{c} a_{p}^{d} (a_{s}^{\dagger})^{e} a_{s}^{f}(a_{i}^{\dagger})^{g} a_{i}^{h} \right) U_{1}(\theta,\phi_{0})|\psi_{|\alpha_{0}|}(\tau)\rangle \\
    &=e^{i\phi_{0}(d-c)}e^{i(\theta+\phi_{0})(f-e)}\langle \psi_{|\alpha_{0}|}(\tau)|(a_{p}^{\dagger})^{c} a_{p}^{d} (a_{s}^{\dagger})^{e} a_{s}^{f}(a_{i}^{\dagger})^{g} a_{i}^{h} |\psi_{|\alpha_{0}|}(\tau)\rangle \\
    \label{eq:alpha dependent expec}
    &= e^{i\phi_{0}(d-c)}e^{i(\theta+\phi_{0})(f-e)}\langle (a_{p}^{\dagger})^{c} a_{p}^{d} (a_{s}^{\dagger})^{e} a_{s}^{f}(a_{i}^{\dagger})^{g} a_{i}^{h} \rangle_{|\alpha_{0}|}.
\end{align}
The effect of complex coupling constant and $\alpha_{0}$ for an expectation value is just an extra phase factor $e^{i\phi_{0}(d-c)}e^{i(\theta+\phi_{0})(f-e)}$ added to the expectation value obtained from the state that is evolved with the absolute value of the coupling constant in the Hamiltonian and initialized in the state with $\alpha_{0}=|\alpha_{0}|$. Note that the same phase factor would result even if we have antinormal ordering of the operators. If the operator over which the expectation value is needed has equal powers of raising and lowering operators, the phase disappears.
\section{Gaussian Limit} 
\label{sec:Gaussian limit}
In this section, we explicitly show all the equations of motion that result from the second order cumulant expansion method and obtain an analytic solution for these differential equations for the initial state of interest. The equations are
\begin{align}
\label{eq:eq1}
\partial_{\tau}\left\langle a_p\right\rangle &=-\left\langle a_s a_i \right\rangle, \\
\partial_{\tau}\left\langle a_s \right\rangle&=\langle  a_p a_i^{\dagger} \rangle,\label{eq:eq2}\\
\partial_{\tau}\left\langle a_i\right\rangle&=\langle  a_p a_s^{\dagger }\rangle, \label{eq:eq3} \\
\begin{split}
\label{eq:eq4}
   \partial_{\tau}\langle a_s a_i\rangle&=\langle a_p\rangle +\langle a_p\rangle  \langle a_i^{\dagger}a_i\rangle +\langle a_i\rangle  \langle a_p a_i^{\dagger} \rangle +\langle a_s\rangle  \langle  a_p a_s^{\dagger}\rangle +\langle a_p\rangle  \langle a_s^{\dagger } a_s\rangle-2 \langle a_i\rangle \langle a_p\rangle  \langle a_i^{\dagger }\rangle +\langle a_p a_i\rangle  \langle a_i^{\dagger }\rangle -2 \langle a_p\rangle  \langle a_s\rangle  \langle a_s^{\dagger }\rangle \\
   &\qquad +\langle a_p a_s \rangle  \langle a_s^{\dagger }\rangle,
\end{split}
\\
\begin{split}
\label{eq:eq5}
   \partial_{\tau}\langle a_i^{\dagger } a_i\rangle &=\langle a_s\rangle  \langle a_p^{\dagger } a_i\rangle
    +\langle a_i\rangle  \langle a_p^{\dagger} a_s\rangle +\langle a_p\rangle \langle a_s^{\dagger} a_i^{\dagger }\rangle +\langle a_p a_s^{\dagger } \rangle  \langle a_i^{\dagger }\rangle -2 \langle a_i\rangle \langle a_s\rangle  \langle a_p^{\dagger }\rangle +\langle a_s a_i\rangle  \langle a_p^{\dagger }\rangle+\langle a_p a_i^{\dagger } \rangle  \langle a_s^{\dagger }\rangle \\
    & \qquad -2\langle a_p\rangle  \langle a_i^{\dagger}\rangle  \langle a_s^{\dagger }\rangle, 
\end{split}
 \\
 \begin{split}
 \label{eq:eq6}
\partial_{\tau} \langle a_s^{\dagger } a_s\rangle &=\langle a_s\rangle  \langle a_p^{\dagger } a_i\rangle
+\langle a_i\rangle  \langle a_p^{\dagger} a_s\rangle +\langle a_p\rangle  \langle a_s^{\dagger} a_i^{\dagger }\rangle +\langle a_p a_s^{\dagger } \rangle  \langle a_i^{\dagger }\rangle -2\langle a_i\rangle  \langle a_s\rangle  \langle a_p^{\dagger }\rangle +\langle a_s a_i\rangle  \langle a_p^{\dagger}\rangle +\langle  a_p a_i^{\dagger }\rangle  \langle a_s^{\dagger}\rangle\\
&\qquad -2 \langle a_p\rangle  \langle a_i^{\dagger}\rangle  \langle a_s^{\dagger }\rangle,
 \end{split} \\
 \begin{split}
 \label{eq:eq7}
\partial_{\tau} \langle a_p^{\dagger} a_p\rangle &=-\langle a_s\rangle \langle a_p^{\dagger} a_i\rangle
-\langle a_i\rangle  \langle a_p^{\dagger} a_s\rangle -\langle a_p\rangle  \langle a_s^{\dagger} a_i^{\dagger}\rangle -\langle a_p a_s^{\dagger} \rangle  \langle a_i^{\dagger}\rangle +2 \langle a_i\rangle  \langle a_s\rangle  \langle a_p^{\dagger}\rangle -\langle a_s a_i\rangle  \langle a_p^{\dagger}\rangle-\langle a_p a_i^{\dagger} \rangle  \langle a_s^{\dagger}\rangle\\
&\qquad +2 \langle a_p\rangle  \langle a_i^{\dagger}\rangle  \langle a_s^{\dagger}\rangle,
 \end{split}
 \\
\begin{split}
\label{eq:eq8}
\partial_{\tau} \langle a_p^2\rangle &=4 \langle a_i\rangle  \langle a_p\rangle  \langle a_s\rangle
-2 \langle a_s\rangle  \langle a_p a_i\rangle -2\langle a_p\rangle \langle a_s a_i\rangle
-2 \langle a_i\rangle \langle a_p a_s \rangle,
\end{split}
\\
\begin{split}
\label{eq:eq9}
    \partial_{\tau} \langle a_p a_i^{\dagger}\rangle &=\langle a_s\rangle \bigl(\langle a_p^{\dagger} a_p\rangle-\langle a_i^{\dagger} a_i\rangle -2 |\langle a_p\rangle |^{2}  \bigr)  -\langle a_i\rangle\bigl(  \langle a_s a_i^{\dagger} \rangle -2 \langle a_s\rangle  \langle a_i^{\dagger}\rangle\bigr) -\langle a_s a_i\rangle  \langle a_i^{\dagger }\rangle +\langle a_p\rangle  \langle a_p^{\dagger} a_s\rangle  +\langle a_p a_s \rangle \langle a_p^{\dagger}\rangle,
\end{split}
\\
\begin{split}
\label{eq:eq10}
   \partial_{\tau}\langle a_p a_s^{\dagger} \rangle &= \langle a_i\rangle \left(\langle a_p^{\dagger } a_p \rangle -  \langle a_s^{\dagger}a_s\rangle   +2  |\langle a_s\rangle|^{2}  \right)-2  |\langle a_p\rangle|^{2}\langle a_{i}\rangle-\langle a_s^{\dagger}\rangle \langle a_s a_i\rangle
  -\langle a_s\rangle  \langle a_s^{\dagger} a_i\rangle +\langle a_p\rangle  \langle a_p^{\dagger } a_i\rangle +\langle a_p a_i\rangle  \langle a_p^{\dagger}\rangle,  
\end{split}
 \\
 \begin{split}
 \label{eq:eq11}
\partial_{\tau}\langle a_p a_i \rangle &=\langle a_s\rangle \bigl(2\langle a_i\rangle^{2}  -\langle a_i^{2}\rangle\bigr) +\langle a_s^{\dagger}\rangle\bigl(\langle a_p^{2}\rangle -2 \langle a_p\rangle^{2} \langle a_s^{\dagger}\rangle \bigr)-2 \langle a_i\rangle \langle a_s a_i\rangle +2 \langle a_p\rangle \langle a_p a_s^{\dagger} \rangle, 
 \end{split}
 \\
 \begin{split}
 \label{eq:eq12}
\partial_{\tau} \langle a_p a_s\rangle &=\langle a_i\rangle  \bigl(2 \langle a_s\rangle^{2}-
\langle a_s^{2}\rangle\bigr) -2\langle a_s\rangle \langle a_s a_i\rangle+\langle a_i^{\dagger}\rangle\bigl(\langle a_p^{2}\rangle -2 \langle a_p\rangle^{2} \bigr) +2 \langle a_p\rangle \langle a_p a_i^{\dagger} \rangle, 
 \end{split}
\\
\begin{split}
\label{eq:eq13}
\partial_{\tau} \langle a_s a_i^{\dagger}\rangle &=2 \langle a_s\rangle \bigl(\langle a_p^{\dagger } a_s\rangle-\langle a_s\rangle
 \langle a_p^{\dagger}\rangle \bigr)
+2 \langle a_i^{\dagger}\rangle \bigl(\langle a_p a_i^{\dagger} \rangle  - \langle a_p\rangle \langle
a_i^{\dagger}\rangle \bigr)+\langle a_p\rangle  \langle \bigl(a_i^{\dagger}\bigr)^{2}\rangle  +\langle a_s^2\rangle \langle a_p^{\dagger}\rangle,
\end{split}
 \\
 \begin{split}
 \label{eq:eq14}
\partial_{\tau} \langle a_i^2\rangle &=2 \langle
a_i\rangle \langle a_p a_s^{\dagger} \rangle  +2   \langle a_s^{\dagger}\rangle \bigl(\langle a_p a_i\rangle -2 \langle a_i\rangle \langle a_p\rangle \bigr) +2\langle a_p\rangle  \langle a_s^{\dagger} a_i\rangle,
 \end{split}
\\
\begin{split}
\label{eq:eq15}
\partial_{\tau} \langle a_s^2\rangle &=2 \langle a_s\rangle  \langle a_p a_i^{\dagger} \rangle +2 \langle a_i^{\dagger }\rangle\bigl(\langle a_p a_s\rangle  -2\langle a_p\rangle \langle a_s\rangle  \bigr)  +2 \langle a_p\rangle  \langle a_s a_{i}^{\dagger} \rangle.
\end{split}
\end{align}
There are 15 equations in total. As we can see from above (consequence of symmetries), $\partial_{\tau}\langle N_{s}\rangle=\partial_{\tau}\langle N_{i}\rangle=-\partial_{\tau}\langle N_{p}\rangle$, which implies $\langle a_i^{\dagger}a_i\rangle(\tau)=\langle a_s^{\dagger}a_s\rangle(\tau)$ for all times and $\langle a_{p}^{\dagger}a_p\rangle(\tau)=\alpha_{0}^{2}-\langle a_s^{\dagger}a_s\rangle(\tau)$. This reduces the number of independent variables to 13.
Note that for the particular initial state of interest $|\psi(0)\rangle=|\alpha_{0}\rangle |0\rangle | 0\rangle$, Eqns. (\ref{eq:eq2}), (\ref{eq:eq3}), (\ref{eq:eq9})-(\ref{eq:eq15}) don't evolve and they all remain at their initial values, so all these variables are zero in this case. Hence, 
the equations of motion reduce to four, which are given by 
\begin{align}
\label{eq:eq16}
   \partial_{\tau} \left\langle a_p\right\rangle &=-\left\langle a_s a_i \right\rangle,  \\
    \partial_{\tau} \left\langle a_s a_i \right\rangle &=\left\langle a_p\right\rangle \bigl(1 + 2  \left\langle a_s^{\dagger } a_s\right\rangle \bigr),\label{eq:eq17}\\
     \partial_{\tau} \left\langle a_s^{\dagger } a_s\right\rangle &=\left\langle a_p\right\rangle  \left\langle a_s^{\dagger} a_i^{\dagger }\right\rangle +\left\langle a_s a_i\right\rangle  \left\langle a_p^{\dagger }\right\rangle, \label{eq:eq18}\\
     \partial_{\tau} \left\langle a_p^2\right\rangle &= -2 \left\langle a_p\right\rangle  \left\langle a_s a_i\right\rangle. 
\end{align}
Note that $\partial_{\tau} \langle a_p^2 \rangle =\partial_{\tau}\langle a_{p}\rangle^{2}$, which allows us to eliminate one of the equations. Furthermore, all the variables are real, so we can further eliminate another differential equation by solving Eq. (\ref{eq:eq17}) and Eq. (\ref{eq:eq18}) leading to
\begin{align}
    2\langle a_s^{\dagger } a_s \rangle &= -1+\sqrt{1+4\langle a_s a_i \rangle^{2}}.
\end{align}
As a result, we have two independent differential equations that need to be solved:
\begin{align}
    \partial_{\tau} \left\langle a_p\right\rangle &=-\langle a_{s}a_{i}\rangle,\\
    \partial_{\tau} \left\langle a_s a_i \right\rangle &=\left\langle a_p\right\rangle \sqrt{1+4\langle a_s a_i \rangle^{2}}.
\end{align}
Defining $\langle a_{s}a_{i}\rangle\equiv \frac{1}{2}\sinh(2\eta (\tau))$, we have 
\begin{align}
    \partial_{\tau} \left\langle a_p\right\rangle &=-\frac{1}{2}\sinh(2\eta (t)), \\
    \partial_{\tau}\eta(t) &=\frac{1}{\sqrt{1+4\langle a_{s}a_{i})\rangle^{2}}}  \partial_{\tau}\langle a_{s}a_{i}\rangle =\langle a_{p}\rangle,
\end{align}
whose solutions are given by 
\begin{align}
    \langle a_{p}\rangle (\tau) &= \alpha_{0} \dn\bigl(i\alpha_{0} \tau,-1/\alpha_{0}^{2}
 \bigr),\\
 \eta(\tau) &=\Im\bigl[\am\bigl(i\alpha_{0} \tau,-1/\alpha_{0}^{2}\bigr)\bigr],
\end{align} 
where $\am(z)$ is the Jacobi amplitude function, $\dn(z)$ is the
Jacobi delta amplitude function, $\Im$ is the imaginary part of the argument. As a result, we have 
 \begin{align}
 \label{eq:sec_order_pop}
     \langle a_{s}^{\dagger}a_{s}\rangle &= \sinh^{2}(\eta(t)), \\
     \langle a_{p}^{2}\rangle &=\langle a_{p}\rangle^{2}= \alpha_{0}^{2} \dn\bigl(i\alpha_{0} \tau,-1/\alpha_{0}^{2}
 \bigr)^{2}.
 \end{align}
\section{Perturbation theory}
\label{sec:perturbation theory}
\begin{figure*}[t]
\centering
\includegraphics[width=\textwidth]{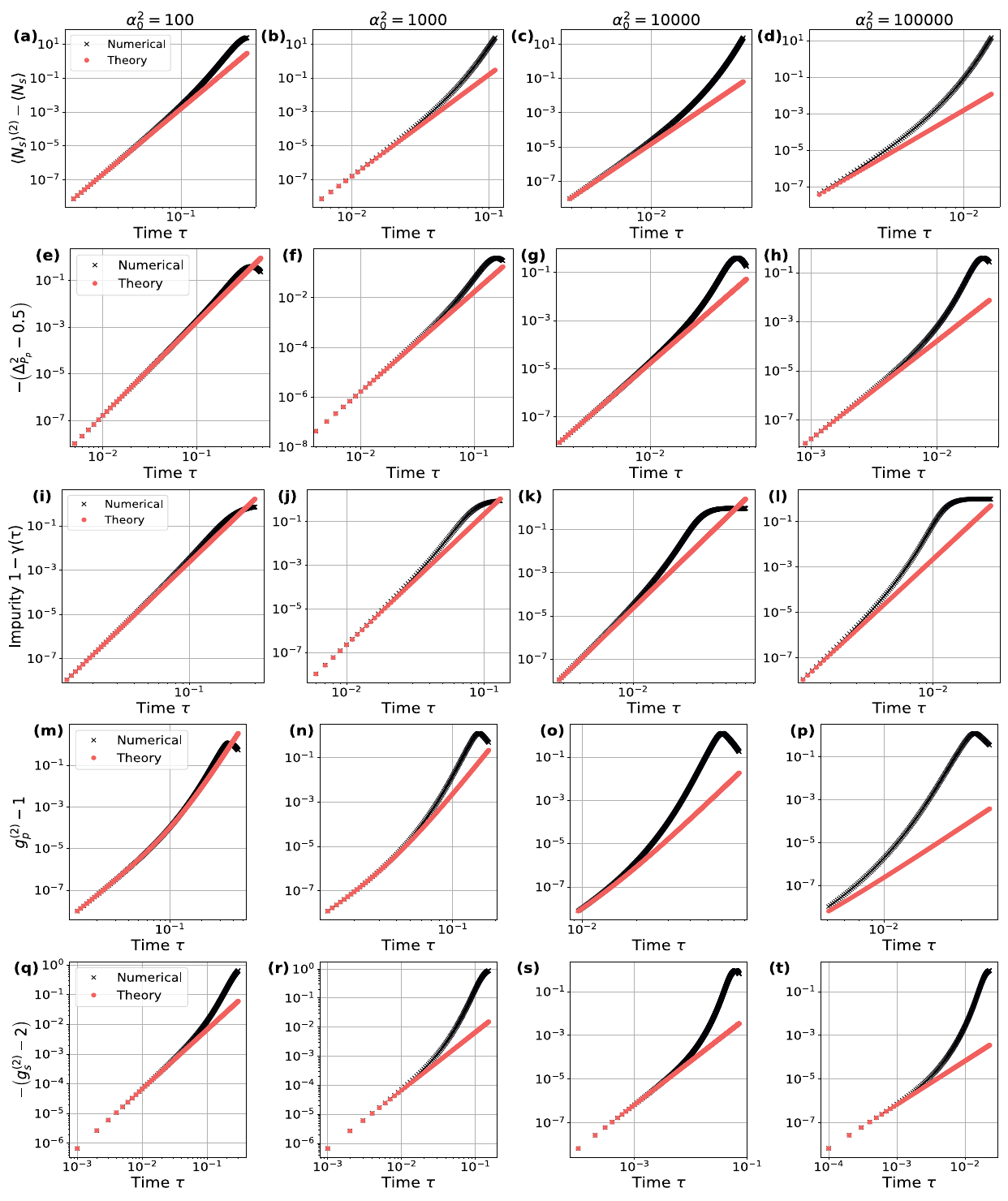}
\caption{The numerical data obtained from the full simulation (crosses) is compared with the corresponding analytic predictions from perturbation theory (dots) for various quantities. Data corresponding to $\alpha_{0}^{2}=\{10^{2},10^{3},10^{4},10^{5}\}$ is plotted in four columns for different quantities in each row. \textbf{(a-d)} These plots confirm the $\tau^{6}$ correction to the signal mode population given in Eq. \eqref{eq:population correction} (note that $\langle N_{s}\rangle^{(2)}-\langle N_{s}\rangle=\langle N_{p}\rangle-\langle N_{p}\rangle^{(2)}$). \textbf{(e-h)} The overlap of the numerical data with the analytic prediction validates the predicted $\tau^{4}$ growth in squeezing in Eq. \eqref{eq:variance correction}. \textbf{(i-l)} These plots substantiate the claim in Eq. \ref{eq:purity correction} that impurity grows as $\tau^{6}$ for short times. \textbf{(m-p)} These plots confirm the prediction in \eqref{eq:gp2 perturbation}. Note that the eighth order expression  in \eqref{eq:gp2 perturbation} is compared with the numerical data here, since the lower order corrections of perturbation theory are too small in this case, especially for the times relevant to the $\alpha_{0}^{2}=10^{5}$ case. \textbf{(q-t)} These plots corroborate the $\tau^{2}$ correction predicted by Eq. \eqref{eq:gs2 perturbation}.}
\label{fig:fig5}
\end{figure*}
In this section, we perform a transformation on the Hamiltonian in Eq. \eqref{eq:Int Hamiltonian} to understand the behaviour of the quantities analyzed in the main text. To achieve this, we substitute an ansatz into the Schr\"{o}dinger equation that the state of the system is given by $|\psi(\tau)\rangle=D(\alpha(\tau))S(\eta(\tau))|\phi(\tau)\rangle$, where $D(\alpha(\tau))=\exp[\alpha(\tau)(a_{p}^{\dagger}-a_{p})]$, $S(\eta(\tau))=\exp[\eta(\tau)( a_{s}^{\dagger}a_{i}^{\dagger}-a_{s}a_{i})]$, and both the variables $\alpha(\tau)$ and $\eta(\tau)$ are assumed to be real. As a result, we have
\begin{align}
\label{eq:effective Hami derivation1}
i \hbar \partial_{\tau}|\phi(\tau)\rangle &=\left(S^{\dagger}(\eta(\tau))D^{\dagger}(\alpha(\tau))\tilde{H} D(\alpha(\tau))S(\eta(\tau))-i\hbar  D^{\dagger}(\alpha(\tau))\partial_{\tau}D(\alpha(\tau)) - i\hbar S^{\dagger}(\eta(\tau)) \partial_{\tau}S(\eta(\tau))\right)|\phi(\tau)\rangle \\
&\equiv \left(\frac{H_{\text{eff}}(\tau)}{\chi}\right)|\phi(\tau)\rangle,
\end{align}
where $\tilde{H}\equiv \frac{H}{\chi}$ (in Eq. (\ref{eq:effective Hami derivation1})) is the Hamiltonian from Eq. \eqref{eq:Int Hamiltonian} up to a factor of $\chi$. Since $\partial_{\tau}D(\alpha(\tau))= \partial_{\tau}\alpha(\tau) \; D(\alpha(\tau)) (a^{\dagger}-a)$, $\partial_{\tau}S^{\dagger}(\eta(\tau))=\partial_{\tau}\eta(\tau) \; S(\eta(\tau)) (b^{\dagger}c^{\dagger}-bc)$, the effective Hamiltonian, $H_{\rm{eff}}(\tau)$, is given by
\begin{align}
\tilde{H}_{\rm{eff}}(\tau)\equiv \left(\frac{H_{\rm{eff}}(\tau)}{\chi}\right) &= \left(S^{\dagger}(\eta(\tau))D^{\dagger}(\alpha(\tau))\tilde{H}D(\alpha(\tau))S(\eta(\tau))-i\hbar \partial_{\tau}\alpha(\tau) \; (a_{p}^{\dagger}-a_{p})  -i\hbar\partial_{\tau}\eta(\tau) \; (a_{s}^{\dagger}a_{i}^{\dagger}-a_{s}a_{i})\right).
\end{align}
Evaluating the term $S^{\dagger}(\eta(\tau))D^{\dagger}(\alpha(\tau))\tilde{H} S(\alpha(\tau))S(\eta(\tau))$ in the above expression, we have
\begin{align}
\begin{split}
    \frac{\tilde{H}_{\text{eff}}(\tau)}{\hbar} &= i\alpha(\tau)(a_{s}^{\dagger}a_{i}^{\dagger}-a_{s}a_{i})+i\sinh^{2}(\eta(\tau))(a_{p}a_{s}a_{i}-a_{p}^{\dagger}a_{s}^{\dagger}a_{i}^{\dagger})+i\cosh^{2}(\eta(\tau))(a_{p}a_{s}^{\dagger}a_{i}^{\dagger}-a_{p}^{\dagger}a_{s}a_{i}) \\
    &+ i\frac{\sinh(2\eta(\tau))}{2}(a_{p}-a_{p}^{\dagger})(a_{s}^{\dagger}a_{s}+a_{i}^{\dagger}a_{i}+1)-i \partial_{\tau}\alpha(\tau) \; (a_{p}^{\dagger}-a_{p})  -i\partial_{\tau}\eta(\tau) \; (a_{s}^{\dagger}a_{i}^{\dagger}-a_{s}a_{i}).
\end{split}
\end{align}
Setting the variables $\alpha(\tau)$ and $\eta(\tau)$ to satisfy the differential equations, $\partial_{\tau}\alpha(\tau)=-\sinh(2 \eta(\tau))/2$ and $\partial_{\tau}\eta(\tau)=\alpha(\tau)$ with $\alpha(0)=\alpha_{0}$ and $\eta(0)=0$, which are the ones satisfied by those in the second-order cumulant expansion method, we have the following effective Hamiltonian 
\begin{align}
\label{eq:effective Hamiltonian}
    \frac{\tilde{H}_{\text{eff}}(\tau)}{\hbar}&= i\sinh^{2}(\eta(\tau))(a_{p} a_{s}a_{i}-a_{p}^{\dagger}a_{s}^{\dagger}a_{i}^{\dagger}) +i\cosh^{2}(\eta(\tau))(a_{p}a_{s}^{\dagger}a_{i}^{\dagger}-a_{p}^{\dagger}a_{s}a_{i}) +i \frac{\sinh(2\eta(\tau))}{2}(a_{p}-a_{p}^{\dagger})(a_{s}^{\dagger}a_{s}+a_{i}^{\dagger}a_{i}).
\end{align}
Therefore, after the above transformation, the down-conversion process is described by the state $|\phi(0)\rangle = |0\rangle|0\rangle|0\rangle$ evolving under the effective Hamiltonian (\ref{eq:effective Hamiltonian}). Next, in order to gain insight into the short-time dynamics of the system described by the above Hamiltonian, we apply the time-dependent perturbation theory, according to which the unitary acting on the state is given by the standard Dyson series
\begin{align}
\begin{split}
        U_{\rm{eff}}(\tau) &= 1 - \frac{i}{\hbar} \int_{0}^{\tau} d\tau_{1} \tilde{H}_{\rm{eff}}(\tau_{1})+\ldots+\left(-\frac{i}{\hbar}\right)^{n}\int_{0}^{\tau} d\tau_{1} \int_{0}^{\tau_{1}} d\tau_2 \ldots \int_{0}^{\tau_{n-1}} d\tau_{n} \tilde{H}_{\rm{eff}}(\tau_{1}) \ldots \tilde{H}_{\rm{eff}}(\tau_{n})+\ldots . 
\end{split}
\end{align}
Note that $\eta(\tau)=\text{Re}\left(-i\;\am \left(i\tau\alpha_{0},-1/\alpha_{0}^{2} \right)\right)$, and $\am \left(i\tau\alpha_{0},-1/\alpha_{0}^{2} \right)$ is purely imaginary for the first time period, which is also the time domain of interest for us in this analysis. The function $\eta(\tau)$ has a Taylor series expansion given by
\begin{align}
\begin{split}
    \eta(\tau) &=\alpha_0 \tau-\frac{\alpha _0 }{6}\tau^3 -\frac{\alpha_0 \left(4 \alpha _0^2-1\right)}{120}  \tau^{5}-\frac{\alpha _0 \left(16 \alpha_0^4-44 \alpha
   _0^2+1\right) }{5040} \tau^{7} + \mathcal{O}\left(\tau^{9}\right).
\end{split}
\end{align}
Substituting the above expression in the effective Hamiltonian and obtaining $U_{\rm{eff}} |000\rangle$ up to eighth order, we have
\begin{align}
\begin{split}
    &|\phi(\tau)\rangle = U_{\rm{eff}}(\tau)|000\rangle\\
    &\qquad\;\;=|000\rangle  -\frac{\tau
   ^3}{3} \alpha_0^2 \left|111\right\rangle   +\frac{\tau ^4}{504}\left(42 \sqrt{2}
   \alpha_0^2 \left|200\right\rangle -84 \alpha _0^2 \left|022\right\rangle \right) + \frac{\tau ^5}{5040} \biggl(-336\left|111\right\rangle  \alpha _0^4-672 \left|011\right\rangle  \alpha _0^3 \\
   &+672 \sqrt{2}
   \left|211\right\rangle  \alpha _0^3+840 \left|111\right\rangle  \alpha_0^2\biggr) +\frac{\tau ^6}{5040}\biggl(-280
   \left|000\right\rangle  \alpha_0^4-672
   \left|022\right\rangle  \alpha_0^4+336 \sqrt{2}
   \left|200\right\rangle  \alpha_0^4+560 \sqrt{2}
   \left|222\right\rangle  \alpha_0^4\\
   & +112
   \left|100\right\rangle  \alpha_0^3+1008
   \left|122\right\rangle  \alpha_0^3-112 \sqrt{6}
   \left|300\right\rangle  \alpha_0^3+280
   \left|022\right\rangle  \alpha_0^2-140 \sqrt{2}
   \left|200\right\rangle  \alpha_0^2\biggr) +\frac{\tau ^7}{5040}\biggl(-32
   \left|111\right\rangle  \alpha_0^6\\
   &-416
   \left|011\right\rangle  \alpha_0^5 +416 \sqrt{2}
   \left|211\right\rangle  \alpha_0^5+1432
   \left|111\right\rangle  \alpha_0^4+840
   \left|133\right\rangle  \alpha_0^4-412 \sqrt{6}
   \left|311\right\rangle  \alpha_0^4+336
   \left|011\right\rangle  \alpha_0^3+432
   \left|033\right\rangle  \alpha_0^3\\
   &-656 \sqrt{2}
   \left|211\right\rangle  \alpha_0^3-152 \left|111\right\rangle  \alpha_0^2\biggr)+ \frac{\tau ^8}{5040} \biggr(-112 \left|000\right\rangle  \alpha _0^6-232
   \left|022\right\rangle  \alpha _0^6+4 \sqrt{2} \left(29
   \left|200\right\rangle +56 \left|222\right\rangle
   \right) \alpha _0^6\\
   & +304 \left|100\right\rangle  \alpha _0^5+1888
   \left|122\right\rangle  \alpha _0^5-136 \sqrt{6}
   \left|300\right\rangle  \alpha _0^5-448 \sqrt{6} \left|322\right\rangle  \alpha _0^5+175 \left|000\right\rangle  \alpha _0^4+1212 \left|022\right\rangle  \alpha _0^4+420 \left|044\right\rangle  \alpha _0^4\\
   &-354 \sqrt{2} \left|200\right\rangle  \alpha _0^4-1478 \sqrt{2} \left|222\right\rangle  \alpha _0^4+103 \sqrt{6} \left|400\right\rangle  \alpha _0^4-42 \left|100\right\rangle  \alpha _0^3-700
   \left|122\right\rangle  \alpha _0^3+82 \sqrt{6} \left|300\right\rangle  \alpha _0^3\\
   & -38 \left|022\right\rangle  \alpha _0^2 +19 \sqrt{2} \left|200\right\rangle  \alpha _0^2\biggr) +\mathcal{O}(\tau^{9}),
\end{split}
\end{align}
and the above state is normalized up to the eighth order in $\tau$. Note that we use the notation $|x,y,z\rangle \equiv |x\rangle |y\rangle |z\rangle$ for arbitrary $x,y,z$ in the above equation for compactness, and the numbers in the first, second and third kets are associated with the pump, signal and idler modes, respectively. The pump-mode population at short times is given by $\langle \psi(\tau)|N_{p}|\psi(\tau)\rangle =\langle \phi(\tau)|D^{\dagger}(\alpha(\tau))a_{p}^{\dagger}a_{p} D(\alpha(\tau))|\phi(\tau)\rangle$. Using this, we obtain the difference between pump-mode population and the pump-mode population in the second-order cumulant expansion, which is given by 
\begin{align}
    \langle \psi(\tau)|N_{p}|\psi(\tau)\rangle  - \alpha^{2}(\tau)&= \langle \phi(\tau)|\left(a_{p}^{\dagger}a_{p}+\alpha(\tau)a_{p}^{\dagger}+\alpha(\tau)a_{p} \right)|\phi(\tau)\rangle \\
    \label{eq:population correction}
    &=\frac{7\alpha_0^4}{45}  \tau^6+ \frac{\left(48 \alpha _0^6-77 \alpha_0^4\right)}{630}  \tau ^8+\mathcal{O}(\tau^{9}).
\end{align}
It is interesting to note that the second-order cumulant expansion correctly accounts for the pump-mode population up to terms of sixth order in time. 
The variance in the momentum of the pump mode ($P_{p} = \frac{a_{p}-a_{p}^{\dagger}}{\sqrt{2}i}$) is given by 
\begin{align}
    \Delta_{P_{p}}^{2}&=\langle P_{p}^{2}\rangle -\langle P_{p} \rangle^{2} =\frac{1}{2} -\text{Re}\bigl(\langle \psi(\tau)|a_{p}^{2}|\psi(\tau)\rangle - \langle \psi(\tau)|a_{p}|\psi(\tau) \rangle^{2}\bigr) + \bigl(\langle \psi(\tau)
a_{p}^{\dagger}a_{p}|\psi(\tau)\rangle - |\langle \psi(\tau)|a_{p}|\psi(\tau) \rangle|^{2}\bigr)\\
    &=\frac{1}{2}+ \langle \phi(\tau)|a_{p}^{\dagger}a_{p}|\phi(\tau)\rangle - \langle \phi(\tau)|a_{p}^{2}|\phi(\tau)\rangle  \\
    \label{eq:variance correction}
   &= \frac{1}{2}-\frac{1}{6} \alpha_0^2 \tau^4+\left(\frac{\alpha
   _0^2}{18}-\frac{\alpha _0^4}{45}\right) \tau
   ^6 +\frac{\left(-4 \alpha _0^6+144 \alpha _0^4-19 \alpha
   _0^2\right)}{2520}\tau ^8 + \mathcal{O}(\tau^{9}).
\end{align}
Note that a variance of $1/2$ is expected at the second-order cumulant expansion since the pump-mode state predicted in this approximation is a coherent state. Hence, the actual value of the variance in the pump momentum deviates from the value predicted by second-order cumulant expansion at fourth-order in time. Next, we look at the perturbation theory expression for  $g^{(2)}_{p}$ and $g^{(2)}_{s}$:
\begin{align}
    g^{(2)}_{p}&=\frac{\langle \psi(\tau)|(a_{p}^{\dagger})^{2}a_{p}^{2}|\psi(\tau)\rangle}{\left(\langle\psi(\tau)| a_{p}^{\dagger}a_{p}|\psi(\tau)\rangle\right)^{2}} \\
    \label{eq:gp2 perturbation}
    &=1+\frac{1}{3}\tau^4 + \frac{2}{45} \left(11 \alpha _0^2+5\right) \tau^6 +\frac{1}{630} \left(114 \alpha _0^4-128 \alpha_0^{2} +97\right) \tau ^8 +\mathcal{O}(\tau^{9}),
\end{align}
and 
\begin{align}
\begin{split}
    g^{(2)}_{s}&=\frac{\langle \psi(\tau)|(a_{s}^{\dagger})^{2}a_{s}^{2}|\psi(\tau)\rangle}{\left(\langle\psi(\tau)| a_{s}^{\dagger}a_{s}|\psi(\tau)\rangle\right)^{2}}\\
    \label{eq:gs2 perturbation}
   &= 2-\frac{2}{3} \tau ^2+\frac{\left(225-2160 \alpha _0^2\right)}{4050} \tau ^4+\frac{\left(198 \alpha _0^4-459 \alpha _0^2+69\right)}{405} \tau^6+\frac{\left(-888 \alpha _0^6+1935 \alpha _0^4-2838 \alpha _0^2+415\right)}{4050}\tau ^8 +\mathcal{O}(\tau^{9}).
\end{split}
\end{align}
The pump mode density operator after tracing out the signal and idler modes is given by
\begin{align}
    \begin{split}
    \Phi_{p}(\tau) &=|0\rangle \langle 0| + \frac{\tau^4}{12}  \left(\sqrt{2} \alpha_{0}^2 |2\rangle \langle 0|+\sqrt{2} \alpha_{0}^2 |0\rangle
   \langle 2|\right)+\frac{\tau^6 }{180} \biggl(-20 \alpha_{0}^4 |0\rangle \langle 0|+12 \sqrt{2} \alpha_{0}^4 |2\rangle \langle 0|+20 \alpha_{0}^4 |1\rangle \langle 1|\\
   & +12 \sqrt{2} \alpha_{0}^4 |0\rangle \langle 2|+4 \alpha_{0}^3 |1\rangle \langle 0|-4 \sqrt{6} \alpha_{0}^3 |3\rangle \langle 0|+4 \alpha_{0}^3 |0\rangle \langle 1|-4 \sqrt{6} \alpha_{0}^3 |0\rangle \langle 3|-5 \sqrt{2} \alpha_{0}^2 |2\rangle \langle 0|-5 \sqrt{2}
   \alpha_{0}^2 |0\rangle\langle 2| \biggr) \\
   & +\frac{\tau^8}{5040}\biggl(-224 \alpha _0^6 |0\rangle \langle 0|+116 \sqrt{2} \alpha _0^6 |2\rangle \langle
   0|+224 \alpha _0^6 |1\rangle \langle 1|+116 \sqrt{2} \alpha _0^6 |0\rangle \langle
   2|+528 \alpha _0^5 |1\rangle\langle 0|-136 \sqrt{6} \alpha _0^5 |3\rangle \langle 0|\\
   & +528 \alpha _0^5 |0\rangle \langle 1|-224 \sqrt{2} \alpha _0^5 |2\rangle \langle 1|-224 \sqrt{2} \alpha _0^5 |1\rangle \langle 2|-136 \sqrt{6} \alpha _0^5 |0\rangle\langle
   3|+490 \alpha _0^4 |0\rangle \langle 0|-354 \sqrt{2} \alpha _0^4 |2\rangle \langle 0|\\
   & +103 \sqrt{6} \alpha _0^4 |4\rangle \langle 0|-560 \alpha _0^4 |1\rangle\langle 1|-354 \sqrt{2} \alpha _0^4 |0\rangle \langle 2|+70 \alpha _0^4 |2\rangle\langle 2| +103 \sqrt{6} \alpha _0^4 |0\rangle \langle 4|-42 \alpha _0^3 |1\rangle \langle 0| \\
   & +82 \sqrt{6} \alpha_0^3 |3\rangle \langle 0|-42 \alpha _0^3 |0\rangle \langle 1|+82 \sqrt{6} \alpha _0^3 |0\rangle \langle 3|+19 \sqrt{2} \alpha _0^2 |2\rangle \langle
   0|+19 \sqrt{2} \alpha _0^2 |0\rangle \langle 2|\biggr) +\mathcal{O}(\tau^{9}).
   \end{split}
\end{align}
Using the above expression, we get the purity of the pump mode: 
\begin{align}
\label{eq:purity correction}
    \gamma(\tau)&= \text{Tr}(\Phi_{p}^{2}(\tau)) =1-\frac{2 \alpha_{0}^4}{9}\tau ^6+\left(\frac{2 \alpha_{0}^4}{9}-\frac{4 \alpha_{0}^6}{45}\right) \tau ^8 +\mathcal{O}(\tau^{9}).
\end{align}
Finally, we compare the perturbation theory expressions with the numerical data for various system sizes in Fig. \ref{fig:fig5}, which confirms the validity of the analytic expressions derived in this section.
\section{Finite system-size effects}
\label{sec:system-size effects}
\begin{figure*}[t]
\centering
\includegraphics[width=\textwidth]{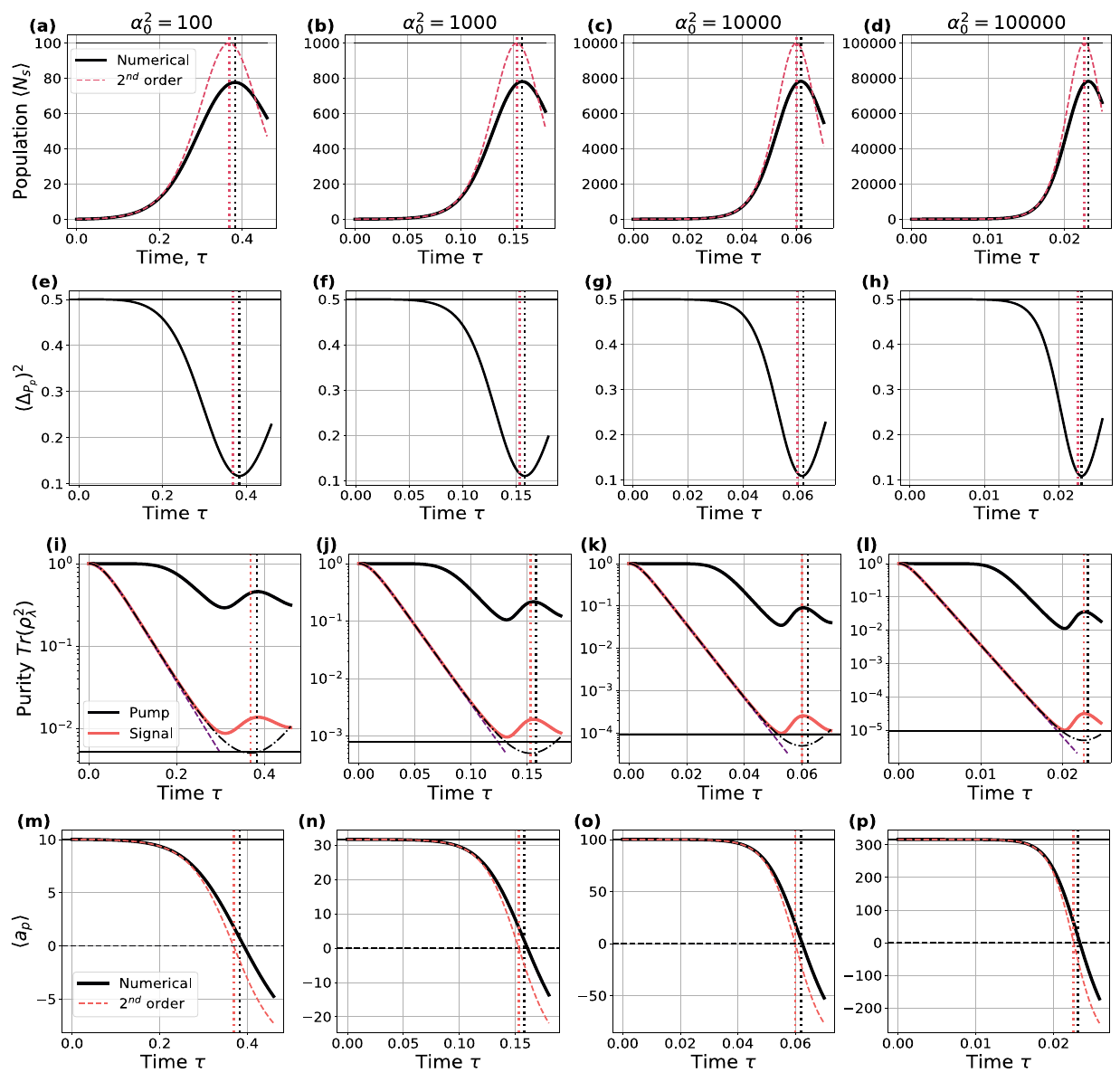}
\caption{Different quantities are plotted here for $\alpha_{0}^{2}=\{10^{2},10^{3},10^{4},10^{5}\}$ in four columns. In all the plots, the location of the black dotted vertical line (the vertical dotted line at later time) corresponds to the time where the signal-mode population($\langle N_{s}\rangle$) reaches local maximum (see thick curves in (a-d)), and the other dotted vertical line indicates the time at which the signal-mode population in the second-order cumulant solution reaches $\alpha_{0}^{2}$ (see dashed curves in (a-d)). \textbf{(a-d)} Signal-mode population from the full numerical simulation (black curve) is plotted here along with the prediction from the second-order cumulant expansion method (dashed curve). \textbf{(e-h)} The pump-mode variance is plotted for different $\alpha_{0}^{2}$. Note that the pump mode is most squeezed at the time when the pump-mode population reaches local minimum (local maximum for signal mode population). The pump-mode variance in momentum shows threshold behaviour: variance is close to $1/2$ for short times and then starts decreasing after a certain time, $\tau_{\text{sqz}}$. \textbf{(i-l)} The purities associated with the pump and the signal modes after tracing out other modes are shown here. Notice that the purity of the pump mode shows threshold behaviour. \textbf{(m-p)} The pump mode amplitude $\langle a_{p}\rangle$ obtained numerically (thick curve) is shown here for different $\alpha_{0}^{2}$ and compared with the amplitude obtained from the second-order cumulant expansion method (dashed curve).}
\label{fig:fig6}
\end{figure*}
\begin{figure*}[t]
\centering
\includegraphics[width=0.4\textwidth]{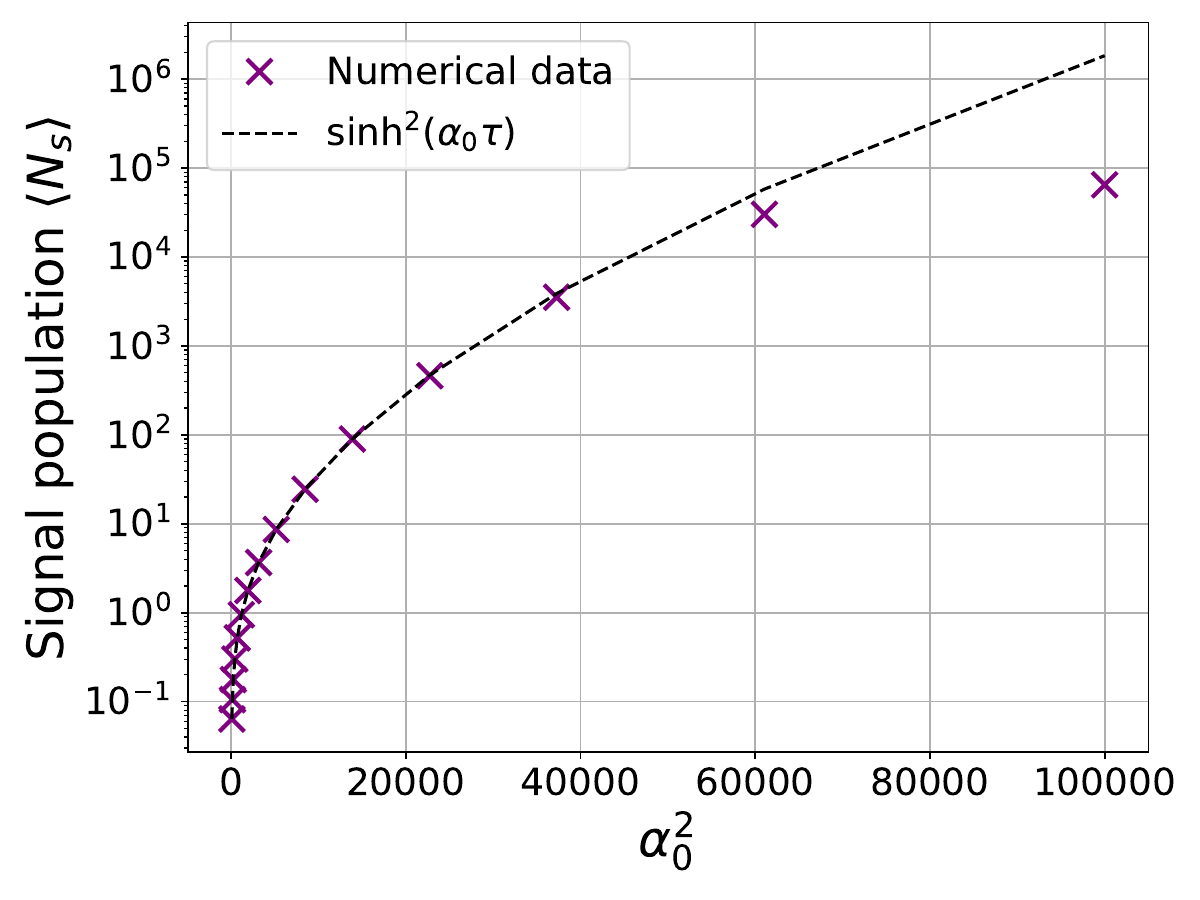}
\caption{The signal-mode population is plotted as a function of $\alpha_{0}^{2}$ for a fixed time, $\tau=0.025$. This type of plot would be relevant for an experimental situation where the pump mode in a coherent state is sent through a crystal with non-linear response whose length is fixed, and the signal-mode population at the output is analyzed for various input coherent states. Note that the signal mode population overlaps with the curve predicted by the parametric approximation for small $\alpha_{0}^{2}$, since the time ($\tau=0.025$) that is being analyzed in this plot corresponds to the undepleted pump region for these values of $\alpha_{0}^{2}$. The numerical data plotted here is consistent with the experimental observation of the signal-mode population shown in \cite{Florez2020pump}.}
\label{fig:fig7}
\end{figure*}
In this section, we analyze the various properties of the system as a function of $\alpha_{0}^{2}$. In Fig. \ref{fig:fig6}, the plots of signal-mode population, pump-mode variance in momentum, purity of the pump and signal modes, and the mean mode amplitude of the pump mode for different values of $\alpha_{0}^{2}$ are shown. In the first row of the figure, the signal-mode population obtained from the full numerical simulation of systems for various values of $\alpha_{0}^{2}$ are shown by thick curves. In these plots, the population obtained from the second-order cumulant expansion method is plotted using dashed lines, and, as it can be seen, they approximate the numerical simulations (black curves) well for short times. Therefore, the solution to the second-order cumulant expansion method can be used to derive the pump depletion time as a function of $\alpha_{0}$:
\begin{align}
\frac{\alpha_{0}^{2}-\alpha^{2}(\tau_{\text{dep}})}{\alpha_{0}^{2}} &=\delta ,
\end{align}
whose solution is given by $\tau_{\text{dep}}=-\frac{i}{\alpha_{0}} \text{arcsn}\left(i\sqrt{\delta} \alpha_{0},-\frac{1}{\alpha_{0}^{2}}\right)$ where arcsn$(z)$ is the inverse of Jacobi elliptic sn$(z)$
function. For more details on Jacobi elliptic functions, see \cite{NIST:DLMF}. Performing an Taylor series expansion in the second argument of arcsn and ignoring terms of the order $\mathcal{O}(1/\alpha_{0}^{2})$ in the expansion, we get $\tau_{\text{dep}}\approx \frac{1}{\alpha_{0}} \sinh^{-1}\left(\sqrt{\delta}\alpha_{0}\right)$. Likewise, we can also obtain an analytic expression for the time at
which the population of the signal mode reaches maximum
in the second-order cumulant expansion method, which is also the time at which $\langle a_{p} \rangle$ or $\langle a_{p}^{\dagger} a_{p}\rangle$ goes to zero in this approximation. It is given by 
\begin{align}
\label{eq:tmax}
    \tau_{\rm{max}} &= \frac{1}{\alpha_{0}}\text{Re}\left(K\left(1+1/\alpha_{0}^{2}\right)\right)\\
    \label{eq:tmax approx appendix}
    &\approx \frac{1}{\alpha_{0}} \ln(4\alpha_{0}),
\end{align} 
where $K$ is the complete elliptic integral of the first kind. The value of $\tau_{\rm{max}}$ shown in Eq. \eqref{eq:tmax approx appendix} is obtained from Eq. \eqref{eq:tmax} by Taylor expanding the function in \eqref{eq:tmax} and ignoring terms of order $\mathcal{O}(1/\alpha_{0}^{2})$. $\tau_{\rm{max}}$ is shown by the first vertical dotted-line in Fig. \ref{fig:fig6} where it can be seen that it identifies the maximum of $\sinh^{2}(\eta(t))$ accurately and provides us with a reasonable estimate of the time when the signal-mode population reaches local maximum and $\langle a_{p}\rangle$ reaches zero in the full numerically simulated down-conversion process. Also, see Fig. \ref{fig:fig7} for a plot of the signal-mode population as a function of $\alpha_{0}^{2}$.

The plot corresponding to the pump-mode momentum variance is shown in the second row of Fig. \ref{fig:fig6}, and it can be seen here that it shows a threshold behaviour. That is, the value of the variance remains close to $1/2$ for small times and then it changes relatively rapidly after a certain value of time, $\tau_{\text{sqz}}$. Since this happens at small times, where we expect the lower order terms of perturbation theory to accurately explain the behaviour of these quantities, we derive an estimate of the threshold time using the results of the previous Appendix. To achieve this, we solve for 
\begin{align}
    \frac{0.5-\Delta_{P_{p}}^{2}(\tau_{\textrm{sqz}})}{0.5} &= \delta.
\end{align}
Using the sixth-order expression in $\tau$ for $\Delta_{P_{p}}^{2}(\tau)$, we have
\begin{align}
\label{eq:squeezing time}
   \left(\frac{\alpha _0^4}{45}-\frac{\alpha_{0}^{2}}{18}\right) \tau_{\text{sqz}}
   ^6 + \frac{\alpha_0^2}{6} \tau_{\text{sqz}}^4-\frac{\delta}{2}=0,
\end{align}
whose solution for $\alpha_{0} \gg 1$ is given by $\tau_{\text{sqz}} \approx \frac{(180 \delta)^{1/6}}{\sqrt{2}} \alpha_{0}^{-2/3}$. A plot of this squeezing threshold time is shown in Fig. \ref{fig:fig8}(a) as a function of $\alpha_{0}^{2}$. For this plot, we numerically obtain the value of time when the variance in the pump mode momentum decreases relatively by $1\%$ percent ($\delta=1/100$) from $1/2$ for different values of $\alpha_{0}$ and then obtain a fit for the curve. The fit shows that the threshold time scales as $\alpha_{0}^{-0.6534}$, which is very close to our analytic estimate.

Similarly, it can be noticed from Fig. \ref{fig:fig6} that the purity of the pump mode also shows threshold behaviour. We can perform a similar analysis as above to obtain the threshold time $\tau_{\text{ent}}$ in this case. Solving for
\begin{align}
    1-\gamma(\tau_{\textrm{ent}}) &= \delta,
\end{align}
where we substitute eight order expression in $\tau$ obtained from perturbation theory for $\gamma(\tau)$ to get
\begin{align}
\label{eq:entanglement time}
    \left(\frac{4 \alpha_{0}^6}{45}-\frac{2\alpha_{0}^{4}}{9} \right)\tau_{\text{ent}} ^8+\frac{2 \alpha_{0}^4}{9}\tau_{\text{ent}} ^6-\delta=0.
\end{align} 
For $\alpha_{0} \gg 1$, the solution to the above equation is given by $\tau_{\text{ent}} \approx (3/2)^{1/4} (5\delta)^{1/8} \alpha_{0}^{-3/4}$. We obtain numerical scaling behaviour in Fig. \ref{fig:fig8}(b),  where we choose $\delta=\frac{1}{100}(\frac{n_{2}}{n_{2}+1})$, which is $1 \%$ of the possible range for purity, $1/(n_{2}+1) \leq \gamma\leq 1$, and for large $\alpha_{0}^{2}$, $\delta \approx \frac{1}{100}$ has no $\alpha_{0}$ dependence. Hence, the observed numerical behaviour $\alpha_{0}^{-0.769}$ agrees with our analytic prediction.
\begin{figure*}[t]
\centering
\includegraphics[width=\textwidth]{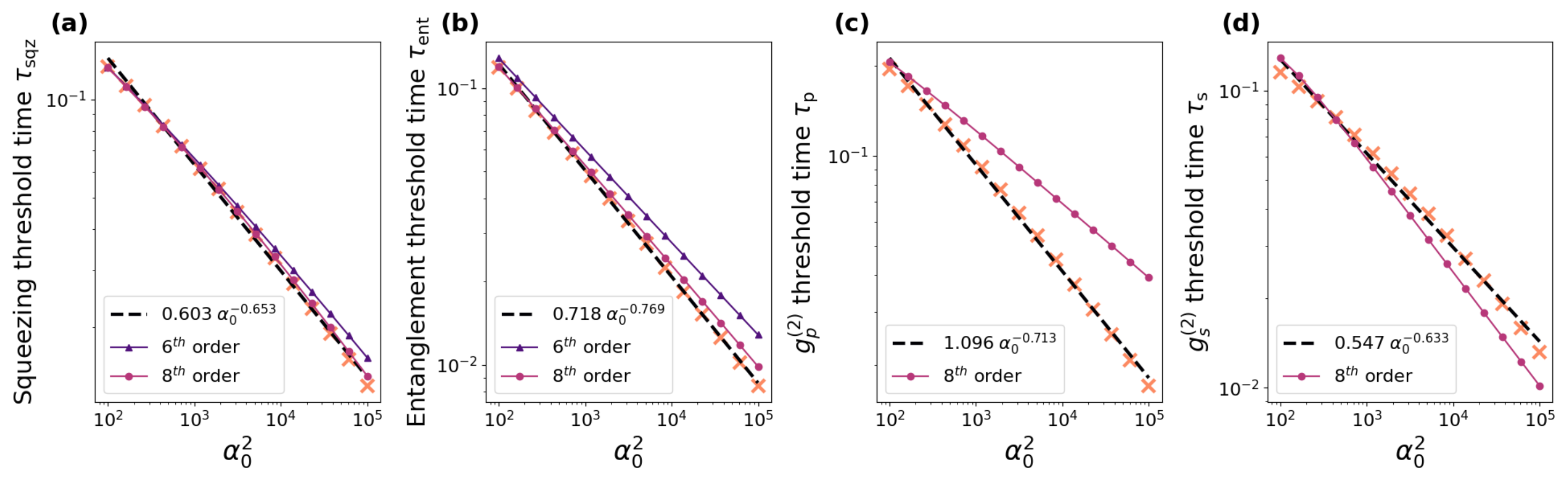}
\caption{The threshold times associated with various quantities are shown here. In each of these plots, the crosses show the numerical data, the dashed line is the fit associated with the numerical data, triangles and dots show the threshold time, $\tau$, obtained using the sixth and eighth order expressions in time derived from perturbation theory. \textbf{(a)} The squeezing threshold time is given by $\tau \sim \alpha^{-0.653}$, which is obtained numerically. The numerical data overlaps well with the eighth order estimate, but, as seen in the plot, even the sixth order expression shown in Eq. \eqref{eq:squeezing time} provides reasonable estimate of the time when squeezing originates in the system. \textbf{(b)} The threshold time of the pump mode purity is plotted as a function of $\alpha_{0}^{2}$, and the numerical data shows that $\tau_{\text{ent}} \sim \alpha_{0}^{-0.769}$. As shown here, the eight-order expression in Eq. \eqref{eq:entanglement time} detects this time reasonably well. \textbf{(c-d)} The threshold times associated with zero time autocorrelation functions of the pump ($\tau_{p}$) and the signal ($\tau_{s}$) mode are plotted here, and their values scale as $\tau_{p} \sim \alpha_{0}^{0.713}$ and $\tau_{s} \sim \alpha_{0}^{0.633}$. As can be seen, the eighth order expressions cannot predict this time accurately, but they can still be used to get a rough estimate of the threshold time.}
\label{fig:fig8}
\end{figure*}
\section{Entanglement Witness}
\label{sec:witness}
In this section, we show the explicit form of the witness mentioned in Sec. \ref{sec:entanglement} in terms of creation and annihilation operators of pump, signal and idler modes. The matrix elements of $w_{4}$ are given by
\begin{align}
    w_{4}&=\begin{vmatrix}
    \langle a_{p}^{\dagger}a_{p}b^{\dagger}b\rangle & \langle (a_{p}^{\dagger})^{2}b^{\dagger}b\rangle & \langle (a_{p}^{\dagger})^{2}b^{2}\rangle  \\
    \langle a_{p}^{2}b^{\dagger}b\rangle &  \langle a_{p}a_{p}^{\dagger}b^{\dagger}b\rangle & \langle a_{p}a_{p}^{\dagger}b^{2}\rangle \\
    \langle a_{p}^{2}(b^{\dagger})^{2}\rangle & \langle a_{p}a_{p}^{\dagger}(b^{\dagger})^{2}\rangle & \langle a_{p} a_{p}^{\dagger}bb^{\dagger} \rangle  \\
\end{vmatrix} \\
&=\begin{vmatrix}
    \frac{1}{2}(\langle a_{p}^{\dagger}a_{p}a_{s}^{\dagger}a_{s}\rangle +\langle a_{p}^{\dagger}a_{p}a_{i}^{\dagger}a_{i}\rangle) & \frac{1}{2}(\langle (a_{p}^{\dagger})^{2}a_{s}^{\dagger}a_{s}\rangle+\langle (a_{p}^{\dagger})^{2}a_{i}^{\dagger}a_{i}\rangle) & \langle (a_{p}^{\dagger})^{2}a_{s}a_{i}\rangle  \\
     \frac{1}{2}(\langle a_{p}^{2}a_{s}^{\dagger}a_{s}\rangle+\langle a_{p}^{2}a_{i}^{\dagger}a_{i}\rangle)  &  \frac{1}{2}(\langle a_{p}^{\dagger}a_{p}a_{s}^{\dagger}a_{s}\rangle+\langle a_{p}^{\dagger}a_{p}a_{i}^{\dagger}a_{i}\rangle  + \langle a_{s}^{\dagger}a_{s}\rangle+ \langle a_{i}^{\dagger}a_{i}\rangle) & \langle a_{p}^{\dagger}a_{p} a_{s}a_{i}\rangle+\langle a_{s}a_{i}\rangle \\
    \langle a_{p}^{2}a_{s}^{\dagger}a_{i}^{\dagger}\rangle  & \langle a_{p}^{\dagger}a_{p} a_{s}^{\dagger}a_{i}^{\dagger}\rangle+\langle a_{s}^{\dagger}a_{i}^{\dagger}\rangle  &  a_{33}  \\
\end{vmatrix},
\end{align}
where $a_{33}=\frac{1}{2}(\langle a_{p}^{\dagger}a_{p}a_{s}^{\dagger}a_{s}\rangle+\langle a_{p}^{\dagger}a_{p}a_{i}^{\dagger}a_{i}\rangle+\langle a_{s}^{\dagger}a_{s}\rangle+\langle a_{i}^{\dagger}a_{i}\rangle)+\langle a_{p}^{\dagger}a_{p} \rangle+1$,
and the matrix elements of $w_{6}$ in terms of $a_{p}$, $a_{s}$ and $a_{i}$ are given by
\begin{align}
w_{6}&=\begin{vmatrix}
1 & \langle a_{p} (b^{\dagger})^{2}\rangle & \langle a_{p}^{\dagger}b^{2}\rangle\\ 
 \langle a_{p}^{\dagger}b^{2}\rangle & \langle a_{p}^{\dagger}a_{p} (b^{\dagger})^{2}b^{2}\rangle & \langle (a_{p}^{\dagger})^{2}b^{4}\rangle\\
\langle a_{p} (b^{\dagger})^{2}\rangle & \langle a_{p}^{2}(b^{\dagger})^{4}\rangle & \langle a_{p}a_{p}^{\dagger}b^{2}(b^{\dagger})^{2}\rangle
\end{vmatrix} 
= \begin{vmatrix}
1 & \langle a_{p} a_{s}^{\dagger}a_{i}^{\dagger}\rangle & \langle a_{p}^{\dagger}a_{s}a_{i}\rangle\\ 
 \langle a_{p}^{\dagger}a_{s}a_{i}\rangle & b_{22}& \frac{3}{2}\langle (a_{p}^{\dagger})^{2}a_{s}^{2}a_{i}^{2}\rangle\\
\langle a_{p} a_{s}^{\dagger}a_{i}^{\dagger}\rangle & \frac{3}{2}\langle a_{p}^{2}(a_{s}^{\dagger})^{2}(a_{i}^{\dagger})^{2}\rangle & b_{33}
\end{vmatrix},
\end{align}
where    
\begin{align}
        b_{22}=&\frac{1}{4}(\langle a_{p}^{\dagger}a_{p} (a_{s}^{\dagger})^{2}a_{s}^{2}\rangle+\langle a_{p}^{\dagger}a_{p} (a_{i}^{\dagger})^{2}a_{i}^{2}\rangle)+ \langle a_{p}^{\dagger}a_{p} a_{s}^{\dagger}a_{s}a_{i}^{\dagger}a_{i}\rangle ,\\
        \begin{split}
        b_{33}  =& \frac{1}{4} \left(\langle a_{p}^{\dagger}a_{p}(a_{s}^{\dagger})^{2}a_{s}^{2}\rangle+4\langle a_{p}^{\dagger}a_{p}a_{s}^{\dagger}a_{s}\rangle +2\langle a_{p}^{\dagger}a_{p}\rangle+\langle (a_{s}^{\dagger})^{2}a_{s}^{2}\rangle +4\langle a_{s}^{\dagger}a_{s}\rangle+2 \right) \\
        & + \frac{1}{4} \left(\langle a_{p}^{\dagger}a_{p}(a_{i}^{\dagger})^{2}a_{i}^{2}\rangle+4\langle a_{p}^{\dagger}a_{p}a_{i}^{\dagger}a_{i}\rangle +2\langle a_{p}^{\dagger}a_{p}\rangle+\langle (a_{i}^{\dagger})^{2}a_{i}^{2}\rangle +4\langle a_{i}^{\dagger}a_{i}\rangle+2\right) \\
        &+ \langle a_{p}^{\dagger}a_{p}a_{s}^{\dagger}a_{s}a_{i}^{\dagger}a_{i}\rangle+\langle a_{p}^{\dagger}a_{p}a_{s}^{\dagger}a_{s}\rangle+\langle a_{p}^{\dagger}a_{p}a_{i}^{\dagger}a_{i}\rangle+\langle a_{s}^{\dagger}a_{s}a_{i}^{\dagger}a_{i}\rangle+\langle a_{p}^{\dagger}a_{p}\rangle+\langle a_{s}^{\dagger}a_{s}\rangle+\langle a_{i}^{\dagger}a_{i}\rangle+1.
    \end{split}
\end{align}
{\section{Long-time evolution of the system}}
\label{sec:long-time evolution}
\begin{figure*}[t]
\centering
\includegraphics[width=\textwidth]{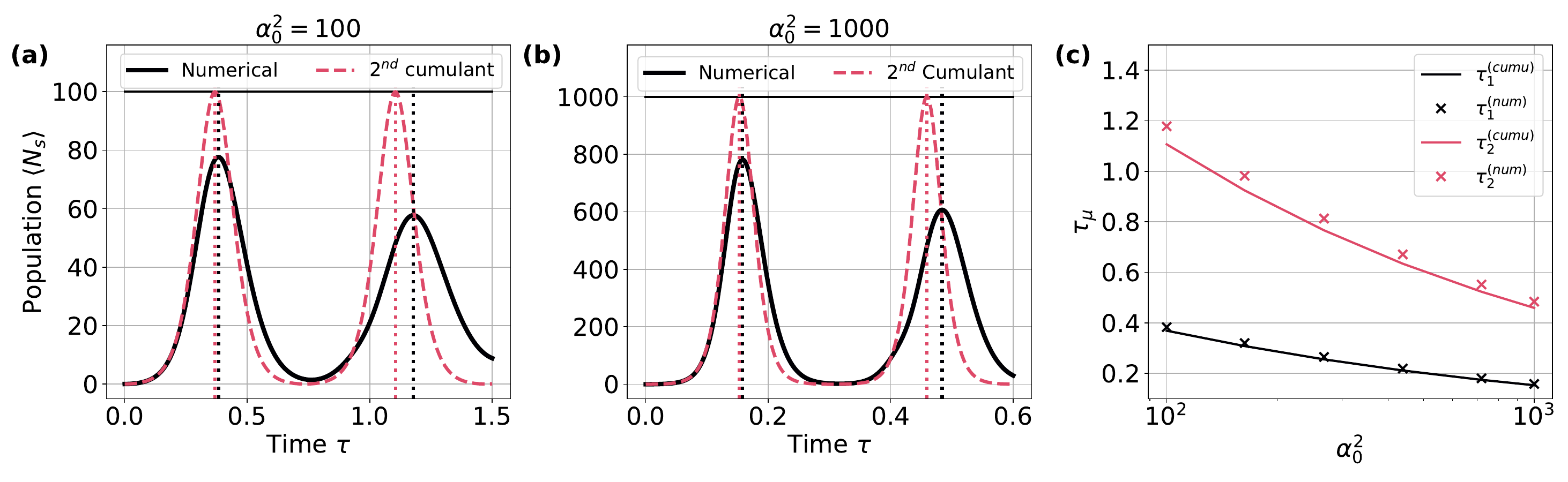}
\caption{ \textbf{(a-b)} The signal-mode population obtained from the full numerical solution (thick curve) is compared with the one obtained from the second-order cumulant solution (dashed curve) for $\alpha_{0}^{2}=100$ and $\alpha_{0}^{2}=1000$. As can be seen in this plot, the position of local maxima in the signal-mode population obtained from the full numerical solution (vertical dotted lines of corresponding color) can be approximated using the period of $\langle N_{s}\rangle$ obtained from the second-order cumulant solution. \textbf{(c)} The location of first and second local maxima obtained from the full numerical simulation, $\tau_{1}^{\text{(num)}}$ and $\tau_{2}^{\text{(num)}}$, are shown by crosses, and they are compared against the position of first and second maxima of the population obtained from the second-order cumulant solution, $\tau_{1}^{\text{(cumu)}}$ and $\tau_{2}^{\text{(cumu)}}$, which are plotted using thick curves. The analytic expressions associated with the second-order cumulant solution are given by $\tau_{1}^{\text{(cumu)}}=\frac{1}{\alpha_{0}}\ln(4\alpha_{0})$ and $\tau_{2}^{\text{(cumu)}}=\frac{3}{\alpha_{0}}\ln(4\alpha_{0})$.}
\label{fig:fig9}
\end{figure*}
In this section, we provide a brief analysis of the dynamics of the system for longer times, beyond the time at which the signal-mode population reaches its first local maximum. In Fig. \ref{fig:fig9}(a) and \ref{fig:fig9}(b), the signal-mode population is plotted both from the full numerical solution and the second-order cumulant solution for $\alpha_{0}^{2}=100$ and $\alpha_{0}^{2}=1000$ respectively. The signal-mode population shows oscillatory, but not periodic, behavior, and the value of the signal-mode population at the local maximum keeps decreasing with time. On the other hand, the population obtained from the second-order cumulant solution, which is expected to approximate the numerically obtained population for short times, can be expressed in terms of Jacobi elliptic function and is therefore periodic in time (see Eq. \eqref{eq:sec_order_pop} and Eq. \eqref{eq:tmax approx}).  While the population obtained from the second-order cumulant expansion method deviates significantly from the full numerical solution for long times, the period of the Jacobi elliptic function can be still used to obtain a rough estimate of the position of the first two local maxima in the signal-mode population obtained from the full numerical solution (compare the position of the vertical dotted lines). More concretely, we compare the position of first two local maxima in signal-mode population obtained from the full numerical solution, labelled by $\tau_{1}^{\text{(num)}}$ and $\tau_{2}^{\text{(num)}}$, with the analytic expressions for the position of first two maxima of $\langle N_{s}\rangle$ obtained from the second-order cumulant solution, labelled by $\tau_{1}^{\text{(cumu)}}$ and $\tau_{2}^{\text{(cumu)}}$ in \ref{fig:fig9}(c). These maxima of the second-order cumulant solution are located at $\tau_{1}^{\text{(cumu)}}=\frac{1}{\alpha_{0}}\ln(4\alpha_{0})$ and $\tau_{2}^{\text{(cumu)}}=\frac{3}{\alpha_{0}}\ln(4\alpha_{0})$. From the plot, it can be seen that the position of first local maximum can be estimated well using the analytic expression $\frac{1}{\alpha_{0}}\ln(4\alpha_{0})$ as already mentioned in Appendix \ref{sec:system-size effects}.

{\section{Analysis of cumulant-expansion method as a function of $\alpha_{0}$}}
\begin{figure*}[t]
\centering
\includegraphics[width=\textwidth]{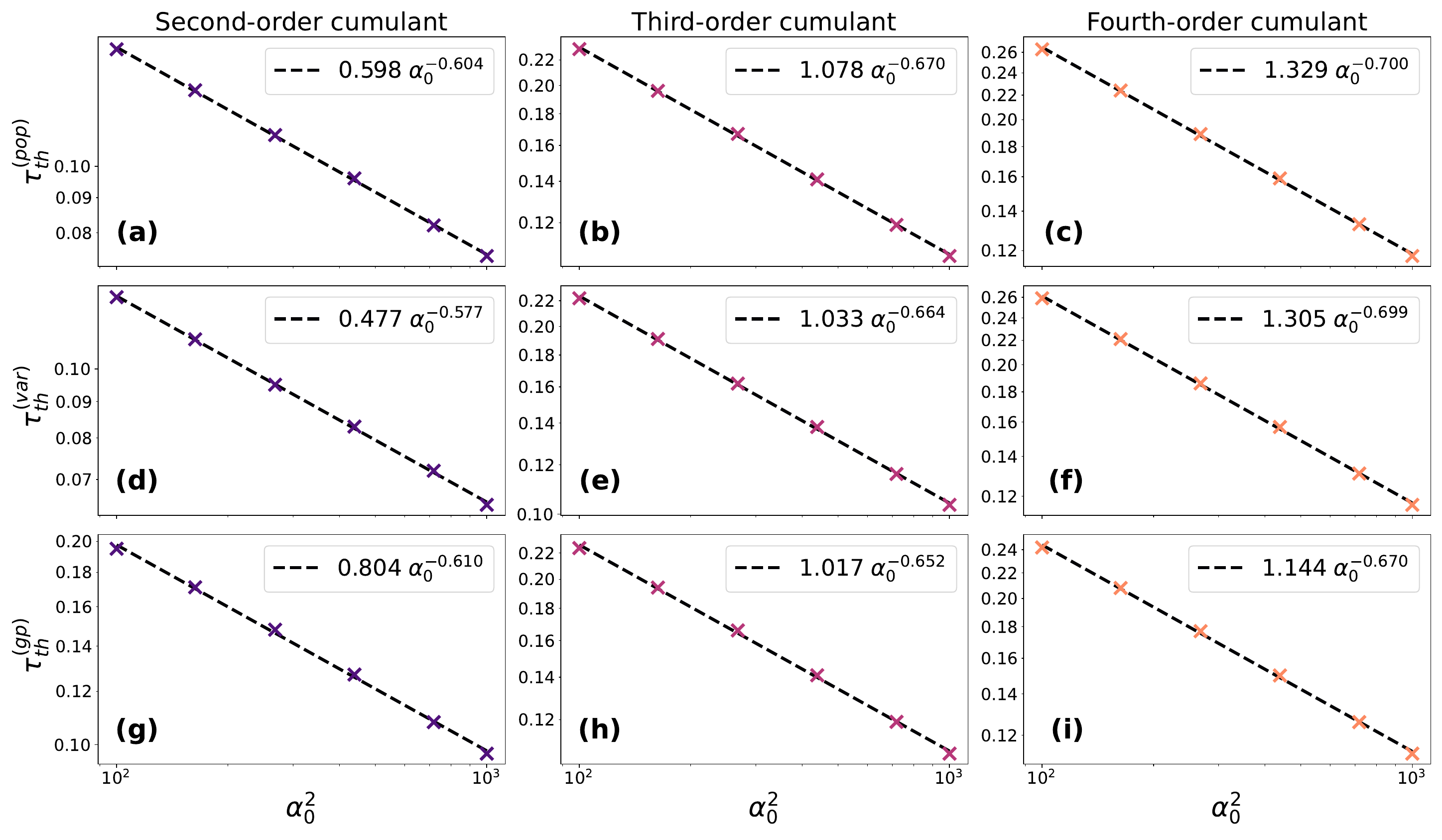}
\caption{ The cumulant time for signal-mode population \textbf{(a-c)}, pump mode variance along the momentum quadrature \textbf{(d-f)}, zero delay autocorrelation function for the pump mode \textbf{(g-i)} are all plotted on the log-log scale in the first, second, third and fourth rows respectively. We define the cumulant time for each quantity as the time at which the relative difference between the quantity of interest obtained from the cumulant expansion method and the full numerical solution is greater than 1$\%$ for the first time. All quantities corresponding to a particular order cumulant expansion are plotted as a function of $\alpha_{0}^{2}$ in one column. We notice the general threshold time decreases as a function of $\alpha_{0}^{2}$, which is consistent with the fact that the higher-order correlations appear faster in larger $\alpha_{0}^{2}$, and therefore the cumulant expansion breaks down sooner.}
\label{fig:fig10}
\end{figure*}
In this section, we analyze how the cumulant expansion method performs as a function of $\alpha_{0}^{2}$ for our system by comparing the data from the cumulant expansion method with that of the full numerical simulation data. Here, we focus on three particular quantities: the signal-mode population (equivalently the pump-mode population), the pump mode variance along the momentum quadrature and the pump mode zero delay autocorrelation function. In order to analyze the time interval over which the cumulant expansion method follows the full numerical simulation data, we define cumulant time for various quantities, which is the time at which the relative difference between the quantity of interest obtained from the cumulant expansion method and the full numerical solution is greater than 1$\%$ for the first time. In Fig. \ref{fig:fig10}, we plot the cumulant threshold times of the signal-mode population, pump-mode variance along the momentum quadrature and the pump mode zero delay autocorrelation function along row one, two and three respectively. The cumulant expansion data obtained from the second, third and the fourth order is plotted along columns one, two and three. For all quantities, we observe that the threshold time for each cumulant expansion follows a power-law with a negative exponent. Moreover, for a particular quantity, the magnitude of the power-law exponent increases with the order of the expansion. The negative exponent is in accordance with the observation that the correlations build up sooner (at smaller $\tau$) in larger system sizes and therefore the cumulant expansion breaks down sooner.  The larger exponent for higher order cumulant expansion shows that the time interval over which we obtain a better approximation of the dynamics with the $(n+1)^{st}$-order expansion is smaller compared to the $n^{th}$-order expansion.
\end{widetext}
\clearpage
\bibliography{Reference.bib}

\begin{thebibliography}{59}%
\makeatletter
\providecommand \@ifxundefined [1]{%
 \@ifx{#1\undefined}
}%
\providecommand \@ifnum [1]{%
 \ifnum #1\expandafter \@firstoftwo
 \else \expandafter \@secondoftwo
 \fi
}%
\providecommand \@ifx [1]{%
 \ifx #1\expandafter \@firstoftwo
 \else \expandafter \@secondoftwo
 \fi
}%
\providecommand \natexlab [1]{#1}%
\providecommand \enquote  [1]{``#1''}%
\providecommand \bibnamefont  [1]{#1}%
\providecommand \bibfnamefont [1]{#1}%
\providecommand \citenamefont [1]{#1}%
\providecommand \href@noop [0]{\@secondoftwo}%
\providecommand \href [0]{\begingroup \@sanitize@url \@href}%
\providecommand \@href[1]{\@@startlink{#1}\@@href}%
\providecommand \@@href[1]{\endgroup#1\@@endlink}%
\providecommand \@sanitize@url [0]{\catcode `\\12\catcode `\$12\catcode `\&12\catcode `\#12\catcode `\^12\catcode `\_12\catcode `\%12\relax}%
\providecommand \@@startlink[1]{}%
\providecommand \@@endlink[0]{}%
\providecommand \url  [0]{\begingroup\@sanitize@url \@url }%
\providecommand \@url [1]{\endgroup\@href {#1}{\urlprefix }}%
\providecommand \urlprefix  [0]{URL }%
\providecommand \Eprint [0]{\href }%
\providecommand \doibase [0]{https://doi.org/}%
\providecommand \selectlanguage [0]{\@gobble}%
\providecommand \bibinfo  [0]{\@secondoftwo}%
\providecommand \bibfield  [0]{\@secondoftwo}%
\providecommand \translation [1]{[#1]}%
\providecommand \BibitemOpen [0]{}%
\providecommand \bibitemStop [0]{}%
\providecommand \bibitemNoStop [0]{.\EOS\space}%
\providecommand \EOS [0]{\spacefactor3000\relax}%
\providecommand \BibitemShut  [1]{\csname bibitem#1\endcsname}%
\let\auto@bib@innerbib\@empty
\bibitem [{\citenamefont {Knill}\ \emph {et~al.}(2001)\citenamefont {Knill}, \citenamefont {Laflamme},\ and\ \citenamefont {Milburn}}]{Knill2001}%
  \BibitemOpen
  \bibfield  {author} {\bibinfo {author} {\bibfnamefont {E.}~\bibnamefont {Knill}}, \bibinfo {author} {\bibfnamefont {R.}~\bibnamefont {Laflamme}},\ and\ \bibinfo {author} {\bibfnamefont {G.~J.}\ \bibnamefont {Milburn}},\ }\bibfield  {title} {\bibinfo {title} {A scheme for efficient quantum computation with linear optics},\ }\href@noop {} {\bibfield  {journal} {\bibinfo  {journal} {nature}\ }\textbf {\bibinfo {volume} {409}},\ \bibinfo {pages} {46} (\bibinfo {year} {2001})}\BibitemShut {NoStop}%
\bibitem [{\citenamefont {Menicucci}\ \emph {et~al.}(2006)\citenamefont {Menicucci}, \citenamefont {Van~Loock}, \citenamefont {Gu}, \citenamefont {Weedbrook}, \citenamefont {Ralph},\ and\ \citenamefont {Nielsen}}]{Menicucci2006}%
  \BibitemOpen
  \bibfield  {author} {\bibinfo {author} {\bibfnamefont {N.~C.}\ \bibnamefont {Menicucci}}, \bibinfo {author} {\bibfnamefont {P.}~\bibnamefont {Van~Loock}}, \bibinfo {author} {\bibfnamefont {M.}~\bibnamefont {Gu}}, \bibinfo {author} {\bibfnamefont {C.}~\bibnamefont {Weedbrook}}, \bibinfo {author} {\bibfnamefont {T.~C.}\ \bibnamefont {Ralph}},\ and\ \bibinfo {author} {\bibfnamefont {M.~A.}\ \bibnamefont {Nielsen}},\ }\bibfield  {title} {\bibinfo {title} {Universal quantum computation with continuous-variable cluster states},\ }\href@noop {} {\bibfield  {journal} {\bibinfo  {journal} {Physical review letters}\ }\textbf {\bibinfo {volume} {97}},\ \bibinfo {pages} {110501} (\bibinfo {year} {2006})}\BibitemShut {NoStop}%
\bibitem [{\citenamefont {Asavanant}\ \emph {et~al.}(2019)\citenamefont {Asavanant}, \citenamefont {Shiozawa}, \citenamefont {Yokoyama}, \citenamefont {Charoensombutamon}, \citenamefont {Emura}, \citenamefont {Alexander}, \citenamefont {Takeda}, \citenamefont {Yoshikawa}, \citenamefont {Menicucci}, \citenamefont {Yonezawa} \emph {et~al.}}]{asavanant2019}%
  \BibitemOpen
  \bibfield  {author} {\bibinfo {author} {\bibfnamefont {W.}~\bibnamefont {Asavanant}}, \bibinfo {author} {\bibfnamefont {Y.}~\bibnamefont {Shiozawa}}, \bibinfo {author} {\bibfnamefont {S.}~\bibnamefont {Yokoyama}}, \bibinfo {author} {\bibfnamefont {B.}~\bibnamefont {Charoensombutamon}}, \bibinfo {author} {\bibfnamefont {H.}~\bibnamefont {Emura}}, \bibinfo {author} {\bibfnamefont {R.~N.}\ \bibnamefont {Alexander}}, \bibinfo {author} {\bibfnamefont {S.}~\bibnamefont {Takeda}}, \bibinfo {author} {\bibfnamefont {J.-i.}\ \bibnamefont {Yoshikawa}}, \bibinfo {author} {\bibfnamefont {N.~C.}\ \bibnamefont {Menicucci}}, \bibinfo {author} {\bibfnamefont {H.}~\bibnamefont {Yonezawa}}, \emph {et~al.},\ }\bibfield  {title} {\bibinfo {title} {Generation of time-domain-multiplexed two-dimensional cluster state},\ }\href@noop {} {\bibfield  {journal} {\bibinfo  {journal} {Science}\ }\textbf {\bibinfo {volume} {366}},\ \bibinfo {pages} {373} (\bibinfo {year} {2019})}\BibitemShut {NoStop}%
\bibitem [{\citenamefont {Zhong}\ \emph {et~al.}(2020)\citenamefont {Zhong}, \citenamefont {Wang}, \citenamefont {Deng}, \citenamefont {Chen}, \citenamefont {Peng}, \citenamefont {Luo}, \citenamefont {Qin}, \citenamefont {Wu}, \citenamefont {Ding}, \citenamefont {Hu} \emph {et~al.}}]{Zhong2020}%
  \BibitemOpen
  \bibfield  {author} {\bibinfo {author} {\bibfnamefont {H.-S.}\ \bibnamefont {Zhong}}, \bibinfo {author} {\bibfnamefont {H.}~\bibnamefont {Wang}}, \bibinfo {author} {\bibfnamefont {Y.-H.}\ \bibnamefont {Deng}}, \bibinfo {author} {\bibfnamefont {M.-C.}\ \bibnamefont {Chen}}, \bibinfo {author} {\bibfnamefont {L.-C.}\ \bibnamefont {Peng}}, \bibinfo {author} {\bibfnamefont {Y.-H.}\ \bibnamefont {Luo}}, \bibinfo {author} {\bibfnamefont {J.}~\bibnamefont {Qin}}, \bibinfo {author} {\bibfnamefont {D.}~\bibnamefont {Wu}}, \bibinfo {author} {\bibfnamefont {X.}~\bibnamefont {Ding}}, \bibinfo {author} {\bibfnamefont {Y.}~\bibnamefont {Hu}}, \emph {et~al.},\ }\bibfield  {title} {\bibinfo {title} {Quantum computational advantage using photons},\ }\href@noop {} {\bibfield  {journal} {\bibinfo  {journal} {Science}\ }\textbf {\bibinfo {volume} {370}},\ \bibinfo {pages} {1460} (\bibinfo {year} {2020})}\BibitemShut {NoStop}%
\bibitem [{\citenamefont {Arrazola}\ \emph {et~al.}(2021)\citenamefont {Arrazola}, \citenamefont {Bergholm}, \citenamefont {Br{\'a}dler}, \citenamefont {Bromley}, \citenamefont {Collins}, \citenamefont {Dhand}, \citenamefont {Fumagalli}, \citenamefont {Gerrits}, \citenamefont {Goussev}, \citenamefont {Helt} \emph {et~al.}}]{Arrazola2021}%
  \BibitemOpen
  \bibfield  {author} {\bibinfo {author} {\bibfnamefont {J.~M.}\ \bibnamefont {Arrazola}}, \bibinfo {author} {\bibfnamefont {V.}~\bibnamefont {Bergholm}}, \bibinfo {author} {\bibfnamefont {K.}~\bibnamefont {Br{\'a}dler}}, \bibinfo {author} {\bibfnamefont {T.~R.}\ \bibnamefont {Bromley}}, \bibinfo {author} {\bibfnamefont {M.~J.}\ \bibnamefont {Collins}}, \bibinfo {author} {\bibfnamefont {I.}~\bibnamefont {Dhand}}, \bibinfo {author} {\bibfnamefont {A.}~\bibnamefont {Fumagalli}}, \bibinfo {author} {\bibfnamefont {T.}~\bibnamefont {Gerrits}}, \bibinfo {author} {\bibfnamefont {A.}~\bibnamefont {Goussev}}, \bibinfo {author} {\bibfnamefont {L.~G.}\ \bibnamefont {Helt}}, \emph {et~al.},\ }\bibfield  {title} {\bibinfo {title} {Quantum circuits with many photons on a programmable nanophotonic chip},\ }\href@noop {} {\bibfield  {journal} {\bibinfo  {journal} {Nature}\ }\textbf {\bibinfo {volume} {591}},\ \bibinfo {pages} {54} (\bibinfo {year} {2021})}\BibitemShut {NoStop}%
\bibitem [{\citenamefont {Hamilton}\ \emph {et~al.}(2017)\citenamefont {Hamilton}, \citenamefont {Kruse}, \citenamefont {Sansoni}, \citenamefont {Barkhofen}, \citenamefont {Silberhorn},\ and\ \citenamefont {Jex}}]{hamilton2017gaussian}%
  \BibitemOpen
  \bibfield  {author} {\bibinfo {author} {\bibfnamefont {C.~S.}\ \bibnamefont {Hamilton}}, \bibinfo {author} {\bibfnamefont {R.}~\bibnamefont {Kruse}}, \bibinfo {author} {\bibfnamefont {L.}~\bibnamefont {Sansoni}}, \bibinfo {author} {\bibfnamefont {S.}~\bibnamefont {Barkhofen}}, \bibinfo {author} {\bibfnamefont {C.}~\bibnamefont {Silberhorn}},\ and\ \bibinfo {author} {\bibfnamefont {I.}~\bibnamefont {Jex}},\ }\bibfield  {title} {\bibinfo {title} {Gaussian boson sampling},\ }\href@noop {} {\bibfield  {journal} {\bibinfo  {journal} {Physical review letters}\ }\textbf {\bibinfo {volume} {119}},\ \bibinfo {pages} {170501} (\bibinfo {year} {2017})}\BibitemShut {NoStop}%
\bibitem [{\citenamefont {Kruse}\ \emph {et~al.}(2019)\citenamefont {Kruse}, \citenamefont {Hamilton}, \citenamefont {Sansoni}, \citenamefont {Barkhofen}, \citenamefont {Silberhorn},\ and\ \citenamefont {Jex}}]{kruse2019detailed}%
  \BibitemOpen
  \bibfield  {author} {\bibinfo {author} {\bibfnamefont {R.}~\bibnamefont {Kruse}}, \bibinfo {author} {\bibfnamefont {C.~S.}\ \bibnamefont {Hamilton}}, \bibinfo {author} {\bibfnamefont {L.}~\bibnamefont {Sansoni}}, \bibinfo {author} {\bibfnamefont {S.}~\bibnamefont {Barkhofen}}, \bibinfo {author} {\bibfnamefont {C.}~\bibnamefont {Silberhorn}},\ and\ \bibinfo {author} {\bibfnamefont {I.}~\bibnamefont {Jex}},\ }\bibfield  {title} {\bibinfo {title} {Detailed study of gaussian boson sampling},\ }\href@noop {} {\bibfield  {journal} {\bibinfo  {journal} {Physical Review A}\ }\textbf {\bibinfo {volume} {100}},\ \bibinfo {pages} {032326} (\bibinfo {year} {2019})}\BibitemShut {NoStop}%
\bibitem [{\citenamefont {Deshpande}\ \emph {et~al.}(2022)\citenamefont {Deshpande}, \citenamefont {Mehta}, \citenamefont {Vincent}, \citenamefont {Quesada}, \citenamefont {Hinsche}, \citenamefont {Ioannou}, \citenamefont {Madsen}, \citenamefont {Lavoie}, \citenamefont {Qi}, \citenamefont {Eisert} \emph {et~al.}}]{deshpande2022quantum}%
  \BibitemOpen
  \bibfield  {author} {\bibinfo {author} {\bibfnamefont {A.}~\bibnamefont {Deshpande}}, \bibinfo {author} {\bibfnamefont {A.}~\bibnamefont {Mehta}}, \bibinfo {author} {\bibfnamefont {T.}~\bibnamefont {Vincent}}, \bibinfo {author} {\bibfnamefont {N.}~\bibnamefont {Quesada}}, \bibinfo {author} {\bibfnamefont {M.}~\bibnamefont {Hinsche}}, \bibinfo {author} {\bibfnamefont {M.}~\bibnamefont {Ioannou}}, \bibinfo {author} {\bibfnamefont {L.}~\bibnamefont {Madsen}}, \bibinfo {author} {\bibfnamefont {J.}~\bibnamefont {Lavoie}}, \bibinfo {author} {\bibfnamefont {H.}~\bibnamefont {Qi}}, \bibinfo {author} {\bibfnamefont {J.}~\bibnamefont {Eisert}}, \emph {et~al.},\ }\bibfield  {title} {\bibinfo {title} {Quantum computational advantage via high-dimensional gaussian boson sampling},\ }\href@noop {} {\bibfield  {journal} {\bibinfo  {journal} {Science advances}\ }\textbf {\bibinfo {volume} {8}},\ \bibinfo {pages} {eabi7894} (\bibinfo {year} {2022})}\BibitemShut {NoStop}%
\bibitem [{\citenamefont {Grier}\ \emph {et~al.}(2022)\citenamefont {Grier}, \citenamefont {Brod}, \citenamefont {Arrazola}, \citenamefont {de~Andrade~Alonso},\ and\ \citenamefont {Quesada}}]{grier2022complexity}%
  \BibitemOpen
  \bibfield  {author} {\bibinfo {author} {\bibfnamefont {D.}~\bibnamefont {Grier}}, \bibinfo {author} {\bibfnamefont {D.~J.}\ \bibnamefont {Brod}}, \bibinfo {author} {\bibfnamefont {J.~M.}\ \bibnamefont {Arrazola}}, \bibinfo {author} {\bibfnamefont {M.~B.}\ \bibnamefont {de~Andrade~Alonso}},\ and\ \bibinfo {author} {\bibfnamefont {N.}~\bibnamefont {Quesada}},\ }\bibfield  {title} {\bibinfo {title} {The complexity of bipartite gaussian boson sampling},\ }\href@noop {} {\bibfield  {journal} {\bibinfo  {journal} {Quantum}\ }\textbf {\bibinfo {volume} {6}},\ \bibinfo {pages} {863} (\bibinfo {year} {2022})}\BibitemShut {NoStop}%
\bibitem [{\citenamefont {Pezz{\'e}}\ and\ \citenamefont {Smerzi}(2008)}]{Pezze2008}%
  \BibitemOpen
  \bibfield  {author} {\bibinfo {author} {\bibfnamefont {L.}~\bibnamefont {Pezz{\'e}}}\ and\ \bibinfo {author} {\bibfnamefont {A.}~\bibnamefont {Smerzi}},\ }\bibfield  {title} {\bibinfo {title} {Mach-zehnder interferometry at the heisenberg limit with coherent and squeezed-vacuum light},\ }\href@noop {} {\bibfield  {journal} {\bibinfo  {journal} {Physical review letters}\ }\textbf {\bibinfo {volume} {100}},\ \bibinfo {pages} {073601} (\bibinfo {year} {2008})}\BibitemShut {NoStop}%
\bibitem [{\citenamefont {Aasi}\ \emph {et~al.}(2013)\citenamefont {Aasi}, \citenamefont {Abadie}, \citenamefont {Abbott}, \citenamefont {Abbott}, \citenamefont {Abbott}, \citenamefont {Abernathy}, \citenamefont {Adams}, \citenamefont {Adams}, \citenamefont {Addesso}, \citenamefont {Adhikari} \emph {et~al.}}]{Aasi2013}%
  \BibitemOpen
  \bibfield  {author} {\bibinfo {author} {\bibfnamefont {J.}~\bibnamefont {Aasi}}, \bibinfo {author} {\bibfnamefont {J.}~\bibnamefont {Abadie}}, \bibinfo {author} {\bibfnamefont {B.}~\bibnamefont {Abbott}}, \bibinfo {author} {\bibfnamefont {R.}~\bibnamefont {Abbott}}, \bibinfo {author} {\bibfnamefont {T.}~\bibnamefont {Abbott}}, \bibinfo {author} {\bibfnamefont {M.}~\bibnamefont {Abernathy}}, \bibinfo {author} {\bibfnamefont {C.}~\bibnamefont {Adams}}, \bibinfo {author} {\bibfnamefont {T.}~\bibnamefont {Adams}}, \bibinfo {author} {\bibfnamefont {P.}~\bibnamefont {Addesso}}, \bibinfo {author} {\bibfnamefont {R.}~\bibnamefont {Adhikari}}, \emph {et~al.},\ }\bibfield  {title} {\bibinfo {title} {Enhanced sensitivity of the ligo gravitational wave detector by using squeezed states of light},\ }\href@noop {} {\bibfield  {journal} {\bibinfo  {journal} {Nature Photonics}\ }\textbf {\bibinfo {volume} {7}},\ \bibinfo {pages} {613} (\bibinfo {year} {2013})}\BibitemShut {NoStop}%
\bibitem [{\citenamefont {Laudenbach}\ \emph {et~al.}(2018)\citenamefont {Laudenbach}, \citenamefont {Pacher}, \citenamefont {Fung}, \citenamefont {Poppe}, \citenamefont {Peev}, \citenamefont {Schrenk}, \citenamefont {Hentschel}, \citenamefont {Walther},\ and\ \citenamefont {H{\"u}bel}}]{Laudenbach2018}%
  \BibitemOpen
  \bibfield  {author} {\bibinfo {author} {\bibfnamefont {F.}~\bibnamefont {Laudenbach}}, \bibinfo {author} {\bibfnamefont {C.}~\bibnamefont {Pacher}}, \bibinfo {author} {\bibfnamefont {C.-H.~F.}\ \bibnamefont {Fung}}, \bibinfo {author} {\bibfnamefont {A.}~\bibnamefont {Poppe}}, \bibinfo {author} {\bibfnamefont {M.}~\bibnamefont {Peev}}, \bibinfo {author} {\bibfnamefont {B.}~\bibnamefont {Schrenk}}, \bibinfo {author} {\bibfnamefont {M.}~\bibnamefont {Hentschel}}, \bibinfo {author} {\bibfnamefont {P.}~\bibnamefont {Walther}},\ and\ \bibinfo {author} {\bibfnamefont {H.}~\bibnamefont {H{\"u}bel}},\ }\bibfield  {title} {\bibinfo {title} {Continuous-variable quantum key distribution with gaussian modulation—the theory of practical implementations},\ }\href@noop {} {\bibfield  {journal} {\bibinfo  {journal} {Advanced Quantum Technologies}\ }\textbf {\bibinfo {volume} {1}},\ \bibinfo {pages} {1800011} (\bibinfo {year} {2018})}\BibitemShut {NoStop}%
\bibitem [{\citenamefont {Niset}\ \emph {et~al.}(2009)\citenamefont {Niset}, \citenamefont {Fiur{\'a}{\v{s}}ek},\ and\ \citenamefont {Cerf}}]{Niset2009}%
  \BibitemOpen
  \bibfield  {author} {\bibinfo {author} {\bibfnamefont {J.}~\bibnamefont {Niset}}, \bibinfo {author} {\bibfnamefont {J.}~\bibnamefont {Fiur{\'a}{\v{s}}ek}},\ and\ \bibinfo {author} {\bibfnamefont {N.~J.}\ \bibnamefont {Cerf}},\ }\bibfield  {title} {\bibinfo {title} {No-go theorem for gaussian quantum error correction},\ }\href@noop {} {\bibfield  {journal} {\bibinfo  {journal} {Physical review letters}\ }\textbf {\bibinfo {volume} {102}},\ \bibinfo {pages} {120501} (\bibinfo {year} {2009})}\BibitemShut {NoStop}%
\bibitem [{\citenamefont {Louisell}\ \emph {et~al.}(1961)\citenamefont {Louisell}, \citenamefont {Yariv},\ and\ \citenamefont {Siegman}}]{Louisell1961quantum}%
  \BibitemOpen
  \bibfield  {author} {\bibinfo {author} {\bibfnamefont {W.}~\bibnamefont {Louisell}}, \bibinfo {author} {\bibfnamefont {A.}~\bibnamefont {Yariv}},\ and\ \bibinfo {author} {\bibfnamefont {A.}~\bibnamefont {Siegman}},\ }\bibfield  {title} {\bibinfo {title} {Quantum fluctuations and noise in parametric processes. i.},\ }\href@noop {} {\bibfield  {journal} {\bibinfo  {journal} {Physical Review}\ }\textbf {\bibinfo {volume} {124}},\ \bibinfo {pages} {1646} (\bibinfo {year} {1961})}\BibitemShut {NoStop}%
\bibitem [{\citenamefont {Mollow}\ and\ \citenamefont {Glauber}(1967{\natexlab{a}})}]{Mollow1967_1}%
  \BibitemOpen
  \bibfield  {author} {\bibinfo {author} {\bibfnamefont {B.}~\bibnamefont {Mollow}}\ and\ \bibinfo {author} {\bibfnamefont {R.}~\bibnamefont {Glauber}},\ }\bibfield  {title} {\bibinfo {title} {Quantum theory of parametric amplification. i},\ }\href@noop {} {\bibfield  {journal} {\bibinfo  {journal} {Physical Review}\ }\textbf {\bibinfo {volume} {160}},\ \bibinfo {pages} {1076} (\bibinfo {year} {1967}{\natexlab{a}})}\BibitemShut {NoStop}%
\bibitem [{\citenamefont {Mollow}\ and\ \citenamefont {Glauber}(1967{\natexlab{b}})}]{Mollow1967_2}%
  \BibitemOpen
  \bibfield  {author} {\bibinfo {author} {\bibfnamefont {B.~R.}\ \bibnamefont {Mollow}}\ and\ \bibinfo {author} {\bibfnamefont {R.~J.}\ \bibnamefont {Glauber}},\ }\bibfield  {title} {\bibinfo {title} {Quantum theory of parametric amplification. ii},\ }\href@noop {} {\bibfield  {journal} {\bibinfo  {journal} {Phys. Rev.}\ }\textbf {\bibinfo {volume} {160}},\ \bibinfo {pages} {1097} (\bibinfo {year} {1967}{\natexlab{b}})}\BibitemShut {NoStop}%
\bibitem [{\citenamefont {Walls}(1969)}]{Walls1969topics_dissertation}%
  \BibitemOpen
  \bibfield  {author} {\bibinfo {author} {\bibfnamefont {D.}~\bibnamefont {Walls}},\ }\emph {\bibinfo {title} {Topics in non linear quantum optics}},\ \href@noop {} {Ph.D. thesis},\ \bibinfo  {school} {Harvard University} (\bibinfo {year} {1969})\BibitemShut {NoStop}%
\bibitem [{\citenamefont {Walls}\ and\ \citenamefont {Barakat}(1970)}]{Walls1970quantum}%
  \BibitemOpen
  \bibfield  {author} {\bibinfo {author} {\bibfnamefont {D.~F.}\ \bibnamefont {Walls}}\ and\ \bibinfo {author} {\bibfnamefont {R.}~\bibnamefont {Barakat}},\ }\bibfield  {title} {\bibinfo {title} {Quantum-mechanical amplification and frequency conversion with a trilinear hamiltonian},\ }\href@noop {} {\bibfield  {journal} {\bibinfo  {journal} {Physical Review A}\ }\textbf {\bibinfo {volume} {1}},\ \bibinfo {pages} {446} (\bibinfo {year} {1970})}\BibitemShut {NoStop}%
\bibitem [{\citenamefont {Walls}\ and\ \citenamefont {Tindle}(1972)}]{Walls1972nonlinear}%
  \BibitemOpen
  \bibfield  {author} {\bibinfo {author} {\bibfnamefont {D.}~\bibnamefont {Walls}}\ and\ \bibinfo {author} {\bibfnamefont {C.}~\bibnamefont {Tindle}},\ }\bibfield  {title} {\bibinfo {title} {Nonlinear quantum effects in optics},\ }\href@noop {} {\bibfield  {journal} {\bibinfo  {journal} {Journal of Physics A: General Physics}\ }\textbf {\bibinfo {volume} {5}},\ \bibinfo {pages} {534} (\bibinfo {year} {1972})}\BibitemShut {NoStop}%
\bibitem [{\citenamefont {Tucker}\ and\ \citenamefont {Walls}(1969)}]{Tucker1969quantum}%
  \BibitemOpen
  \bibfield  {author} {\bibinfo {author} {\bibfnamefont {J.}~\bibnamefont {Tucker}}\ and\ \bibinfo {author} {\bibfnamefont {D.~F.}\ \bibnamefont {Walls}},\ }\bibfield  {title} {\bibinfo {title} {Quantum theory of parametric frequency conversion},\ }\href@noop {} {\bibfield  {journal} {\bibinfo  {journal} {Annals of Physics}\ }\textbf {\bibinfo {volume} {52}},\ \bibinfo {pages} {1} (\bibinfo {year} {1969})}\BibitemShut {NoStop}%
\bibitem [{\citenamefont {Agrawal}\ and\ \citenamefont {Mehta}(1974)}]{Agrawal1974dynamics}%
  \BibitemOpen
  \bibfield  {author} {\bibinfo {author} {\bibfnamefont {G.}~\bibnamefont {Agrawal}}\ and\ \bibinfo {author} {\bibfnamefont {C.}~\bibnamefont {Mehta}},\ }\bibfield  {title} {\bibinfo {title} {Dynamics of parametric processes with a trilinear hamiltonian},\ }\href@noop {} {\bibfield  {journal} {\bibinfo  {journal} {Journal of Physics A: Mathematical, Nuclear and General}\ }\textbf {\bibinfo {volume} {7}},\ \bibinfo {pages} {607} (\bibinfo {year} {1974})}\BibitemShut {NoStop}%
\bibitem [{\citenamefont {Gambini}(1977)}]{Gambini1977parametric}%
  \BibitemOpen
  \bibfield  {author} {\bibinfo {author} {\bibfnamefont {R.}~\bibnamefont {Gambini}},\ }\bibfield  {title} {\bibinfo {title} {Parametric amplification with a trilinear hamiltonian},\ }\href@noop {} {\bibfield  {journal} {\bibinfo  {journal} {Physical Review A}\ }\textbf {\bibinfo {volume} {15}},\ \bibinfo {pages} {1157} (\bibinfo {year} {1977})}\BibitemShut {NoStop}%
\bibitem [{\citenamefont {Drobn{\`y}}\ and\ \citenamefont {Jex}(1992)}]{Drobny1992quantum}%
  \BibitemOpen
  \bibfield  {author} {\bibinfo {author} {\bibfnamefont {G.}~\bibnamefont {Drobn{\`y}}}\ and\ \bibinfo {author} {\bibfnamefont {I.}~\bibnamefont {Jex}},\ }\bibfield  {title} {\bibinfo {title} {Quantum properties of field modes in trilinear optical processes},\ }\href@noop {} {\bibfield  {journal} {\bibinfo  {journal} {Physical Review A}\ }\textbf {\bibinfo {volume} {46}},\ \bibinfo {pages} {499} (\bibinfo {year} {1992})}\BibitemShut {NoStop}%
\bibitem [{\citenamefont {Drobn{\`y}}\ \emph {et~al.}(1997)\citenamefont {Drobn{\`y}}, \citenamefont {Bandilla},\ and\ \citenamefont {Jex}}]{Drobny1997quantum}%
  \BibitemOpen
  \bibfield  {author} {\bibinfo {author} {\bibfnamefont {G.}~\bibnamefont {Drobn{\`y}}}, \bibinfo {author} {\bibfnamefont {A.}~\bibnamefont {Bandilla}},\ and\ \bibinfo {author} {\bibfnamefont {I.}~\bibnamefont {Jex}},\ }\bibfield  {title} {\bibinfo {title} {Quantum description of nonlinearly interacting oscillators via classical trajectories},\ }\href@noop {} {\bibfield  {journal} {\bibinfo  {journal} {Physical Review A}\ }\textbf {\bibinfo {volume} {55}},\ \bibinfo {pages} {78} (\bibinfo {year} {1997})}\BibitemShut {NoStop}%
\bibitem [{\citenamefont {Coelho}\ \emph {et~al.}(2009)\citenamefont {Coelho}, \citenamefont {Barbosa}, \citenamefont {Cassemiro}, \citenamefont {Villar}, \citenamefont {Martinelli},\ and\ \citenamefont {Nussenzveig}}]{Coelho2009three}%
  \BibitemOpen
  \bibfield  {author} {\bibinfo {author} {\bibfnamefont {A.}~\bibnamefont {Coelho}}, \bibinfo {author} {\bibfnamefont {F.}~\bibnamefont {Barbosa}}, \bibinfo {author} {\bibfnamefont {K.~N.}\ \bibnamefont {Cassemiro}}, \bibinfo {author} {\bibfnamefont {A.~d.~S.}\ \bibnamefont {Villar}}, \bibinfo {author} {\bibfnamefont {M.}~\bibnamefont {Martinelli}},\ and\ \bibinfo {author} {\bibfnamefont {P.}~\bibnamefont {Nussenzveig}},\ }\bibfield  {title} {\bibinfo {title} {Three-color entanglement},\ }\href@noop {} {\bibfield  {journal} {\bibinfo  {journal} {Science}\ }\textbf {\bibinfo {volume} {326}},\ \bibinfo {pages} {823} (\bibinfo {year} {2009})}\BibitemShut {NoStop}%
\bibitem [{\citenamefont {Bandilla}\ \emph {et~al.}(2000)\citenamefont {Bandilla}, \citenamefont {Drobn{\`y}},\ and\ \citenamefont {Jex}}]{Bandilla2000parametric}%
  \BibitemOpen
  \bibfield  {author} {\bibinfo {author} {\bibfnamefont {A.}~\bibnamefont {Bandilla}}, \bibinfo {author} {\bibfnamefont {G.}~\bibnamefont {Drobn{\`y}}},\ and\ \bibinfo {author} {\bibfnamefont {I.}~\bibnamefont {Jex}},\ }\bibfield  {title} {\bibinfo {title} {Parametric down-conversion and maximal pump depletion},\ }\href@noop {} {\bibfield  {journal} {\bibinfo  {journal} {Journal of Optics B: Quantum and Semiclassical Optics}\ }\textbf {\bibinfo {volume} {2}},\ \bibinfo {pages} {265} (\bibinfo {year} {2000})}\BibitemShut {NoStop}%
\bibitem [{\citenamefont {Xing}\ and\ \citenamefont {Ralph}(2023)}]{Xing2023pump}%
  \BibitemOpen
  \bibfield  {author} {\bibinfo {author} {\bibfnamefont {W.}~\bibnamefont {Xing}}\ and\ \bibinfo {author} {\bibfnamefont {T.}~\bibnamefont {Ralph}},\ }\bibfield  {title} {\bibinfo {title} {Pump depletion in optical parametric amplification},\ }\href@noop {} {\bibfield  {journal} {\bibinfo  {journal} {Physical Review A}\ }\textbf {\bibinfo {volume} {107}},\ \bibinfo {pages} {023712} (\bibinfo {year} {2023})}\BibitemShut {NoStop}%
\bibitem [{\citenamefont {Scharf}\ and\ \citenamefont {Walls}(1984)}]{Scharf1984effect}%
  \BibitemOpen
  \bibfield  {author} {\bibinfo {author} {\bibfnamefont {G.}~\bibnamefont {Scharf}}\ and\ \bibinfo {author} {\bibfnamefont {D.}~\bibnamefont {Walls}},\ }\bibfield  {title} {\bibinfo {title} {Effect of pump quantisation of squeezing in parametric amplifiers},\ }\href@noop {} {\bibfield  {journal} {\bibinfo  {journal} {Optics communications}\ }\textbf {\bibinfo {volume} {50}},\ \bibinfo {pages} {245} (\bibinfo {year} {1984})}\BibitemShut {NoStop}%
\bibitem [{\citenamefont {Milburn}\ and\ \citenamefont {Walls}(1981)}]{Milburn1981production}%
  \BibitemOpen
  \bibfield  {author} {\bibinfo {author} {\bibfnamefont {G.}~\bibnamefont {Milburn}}\ and\ \bibinfo {author} {\bibfnamefont {D.}~\bibnamefont {Walls}},\ }\bibfield  {title} {\bibinfo {title} {Production of squeezed states in a degenerate parametric amplifier},\ }\href@noop {} {\bibfield  {journal} {\bibinfo  {journal} {Optics Communications}\ }\textbf {\bibinfo {volume} {39}},\ \bibinfo {pages} {401} (\bibinfo {year} {1981})}\BibitemShut {NoStop}%
\bibitem [{\citenamefont {Hillery}\ and\ \citenamefont {Zubairy}(1984)}]{Hillery1984path}%
  \BibitemOpen
  \bibfield  {author} {\bibinfo {author} {\bibfnamefont {M.}~\bibnamefont {Hillery}}\ and\ \bibinfo {author} {\bibfnamefont {M.~S.}\ \bibnamefont {Zubairy}},\ }\bibfield  {title} {\bibinfo {title} {Path-integral approach to the quantum theory of the degenerate parametric amplifier},\ }\href@noop {} {\bibfield  {journal} {\bibinfo  {journal} {Physical Review A}\ }\textbf {\bibinfo {volume} {29}},\ \bibinfo {pages} {1275} (\bibinfo {year} {1984})}\BibitemShut {NoStop}%
\bibitem [{\citenamefont {Crouch}\ and\ \citenamefont {Braunstein}(1988)}]{Crouch1988limitations}%
  \BibitemOpen
  \bibfield  {author} {\bibinfo {author} {\bibfnamefont {D.~D.}\ \bibnamefont {Crouch}}\ and\ \bibinfo {author} {\bibfnamefont {S.~L.}\ \bibnamefont {Braunstein}},\ }\bibfield  {title} {\bibinfo {title} {Limitations to squeezing in a parametric amplifier due to pump quantum fluctuations},\ }\href@noop {} {\bibfield  {journal} {\bibinfo  {journal} {Physical Review A}\ }\textbf {\bibinfo {volume} {38}},\ \bibinfo {pages} {4696} (\bibinfo {year} {1988})}\BibitemShut {NoStop}%
\bibitem [{\citenamefont {Cryer-Jenkins}\ \emph {et~al.}(2023)\citenamefont {Cryer-Jenkins}, \citenamefont {Enzian}, \citenamefont {Freisem}, \citenamefont {Moroney}, \citenamefont {Price}, \citenamefont {Svela}, \citenamefont {Major},\ and\ \citenamefont {Vanner}}]{Cryer2023}%
  \BibitemOpen
  \bibfield  {author} {\bibinfo {author} {\bibfnamefont {E.~A.}\ \bibnamefont {Cryer-Jenkins}}, \bibinfo {author} {\bibfnamefont {G.}~\bibnamefont {Enzian}}, \bibinfo {author} {\bibfnamefont {L.}~\bibnamefont {Freisem}}, \bibinfo {author} {\bibfnamefont {N.}~\bibnamefont {Moroney}}, \bibinfo {author} {\bibfnamefont {J.~J.}\ \bibnamefont {Price}}, \bibinfo {author} {\bibfnamefont {A.~{\O}.}\ \bibnamefont {Svela}}, \bibinfo {author} {\bibfnamefont {K.~D.}\ \bibnamefont {Major}},\ and\ \bibinfo {author} {\bibfnamefont {M.~R.}\ \bibnamefont {Vanner}},\ }\bibfield  {title} {\bibinfo {title} {Second-order coherence across the brillouin lasing threshold},\ }\href {https://opg.optica.org/optica/abstract.cfm?URI=optica-10-11-1432} {\bibfield  {journal} {\bibinfo  {journal} {Optica}\ }\textbf {\bibinfo {volume} {10}},\ \bibinfo {pages} {1432} (\bibinfo {year} {2023})}\BibitemShut {NoStop}%
\bibitem [{\citenamefont {Nation}\ and\ \citenamefont {Blencowe}(2010)}]{Nation2010}%
  \BibitemOpen
  \bibfield  {author} {\bibinfo {author} {\bibfnamefont {P.~D.}\ \bibnamefont {Nation}}\ and\ \bibinfo {author} {\bibfnamefont {M.~P.}\ \bibnamefont {Blencowe}},\ }\bibfield  {title} {\bibinfo {title} {The trilinear hamiltonian: a zero-dimensional model of hawking radiation from a quantized source},\ }\href@noop {} {\bibfield  {journal} {\bibinfo  {journal} {New Journal of Physics}\ }\textbf {\bibinfo {volume} {12}},\ \bibinfo {pages} {095013} (\bibinfo {year} {2010})}\BibitemShut {NoStop}%
\bibitem [{\citenamefont {Levy}\ and\ \citenamefont {Kosloff}(2012)}]{Levy2012}%
  \BibitemOpen
  \bibfield  {author} {\bibinfo {author} {\bibfnamefont {A.}~\bibnamefont {Levy}}\ and\ \bibinfo {author} {\bibfnamefont {R.}~\bibnamefont {Kosloff}},\ }\bibfield  {title} {\bibinfo {title} {Quantum absorption refrigerator},\ }\href@noop {} {\bibfield  {journal} {\bibinfo  {journal} {Physical review letters}\ }\textbf {\bibinfo {volume} {108}},\ \bibinfo {pages} {070604} (\bibinfo {year} {2012})}\BibitemShut {NoStop}%
\bibitem [{\citenamefont {Correa}\ \emph {et~al.}(2014)\citenamefont {Correa}, \citenamefont {Palao}, \citenamefont {Alonso},\ and\ \citenamefont {Adesso}}]{Correa2014}%
  \BibitemOpen
  \bibfield  {author} {\bibinfo {author} {\bibfnamefont {L.~A.}\ \bibnamefont {Correa}}, \bibinfo {author} {\bibfnamefont {J.~P.}\ \bibnamefont {Palao}}, \bibinfo {author} {\bibfnamefont {D.}~\bibnamefont {Alonso}},\ and\ \bibinfo {author} {\bibfnamefont {G.}~\bibnamefont {Adesso}},\ }\bibfield  {title} {\bibinfo {title} {Quantum-enhanced absorption refrigerators},\ }\href@noop {} {\bibfield  {journal} {\bibinfo  {journal} {Scientific reports}\ }\textbf {\bibinfo {volume} {4}},\ \bibinfo {pages} {3949} (\bibinfo {year} {2014})}\BibitemShut {NoStop}%
\bibitem [{\citenamefont {Abdo}\ \emph {et~al.}(2013)\citenamefont {Abdo}, \citenamefont {Kamal},\ and\ \citenamefont {Devoret}}]{Abdo2013nondegenerate}%
  \BibitemOpen
  \bibfield  {author} {\bibinfo {author} {\bibfnamefont {B.}~\bibnamefont {Abdo}}, \bibinfo {author} {\bibfnamefont {A.}~\bibnamefont {Kamal}},\ and\ \bibinfo {author} {\bibfnamefont {M.}~\bibnamefont {Devoret}},\ }\bibfield  {title} {\bibinfo {title} {Nondegenerate three-wave mixing with the josephson ring modulator},\ }\href@noop {} {\bibfield  {journal} {\bibinfo  {journal} {Physical Review B}\ }\textbf {\bibinfo {volume} {87}},\ \bibinfo {pages} {014508} (\bibinfo {year} {2013})}\BibitemShut {NoStop}%
\bibitem [{\citenamefont {Drummond}\ and\ \citenamefont {Hillery}(2014)}]{Drummond2014quantum}%
  \BibitemOpen
  \bibfield  {author} {\bibinfo {author} {\bibfnamefont {P.~D.}\ \bibnamefont {Drummond}}\ and\ \bibinfo {author} {\bibfnamefont {M.}~\bibnamefont {Hillery}},\ }\href@noop {} {\emph {\bibinfo {title} {The quantum theory of nonlinear optics}}}\ (\bibinfo  {publisher} {Cambridge University Press},\ \bibinfo {year} {2014})\BibitemShut {NoStop}%
\bibitem [{\citenamefont {Allevi}\ \emph {et~al.}(2014)\citenamefont {Allevi}, \citenamefont {Jedrkiewicz}, \citenamefont {Brambilla}, \citenamefont {Gatti}, \citenamefont {Pe{\v{r}}ina~Jr}, \citenamefont {Haderka},\ and\ \citenamefont {Bondani}}]{Allevi2014}%
  \BibitemOpen
  \bibfield  {author} {\bibinfo {author} {\bibfnamefont {A.}~\bibnamefont {Allevi}}, \bibinfo {author} {\bibfnamefont {O.}~\bibnamefont {Jedrkiewicz}}, \bibinfo {author} {\bibfnamefont {E.}~\bibnamefont {Brambilla}}, \bibinfo {author} {\bibfnamefont {A.}~\bibnamefont {Gatti}}, \bibinfo {author} {\bibfnamefont {J.}~\bibnamefont {Pe{\v{r}}ina~Jr}}, \bibinfo {author} {\bibfnamefont {O.}~\bibnamefont {Haderka}},\ and\ \bibinfo {author} {\bibfnamefont {M.}~\bibnamefont {Bondani}},\ }\bibfield  {title} {\bibinfo {title} {Coherence properties of high-gain twin beams},\ }\href@noop {} {\bibfield  {journal} {\bibinfo  {journal} {Physical Review A}\ }\textbf {\bibinfo {volume} {90}},\ \bibinfo {pages} {063812} (\bibinfo {year} {2014})}\BibitemShut {NoStop}%
\bibitem [{\citenamefont {Pe{\v{r}}ina~Jr}\ \emph {et~al.}(2016)\citenamefont {Pe{\v{r}}ina~Jr}, \citenamefont {Haderka}, \citenamefont {Allevi},\ and\ \citenamefont {Bondani}}]{Pevrina2016}%
  \BibitemOpen
  \bibfield  {author} {\bibinfo {author} {\bibfnamefont {J.}~\bibnamefont {Pe{\v{r}}ina~Jr}}, \bibinfo {author} {\bibfnamefont {O.}~\bibnamefont {Haderka}}, \bibinfo {author} {\bibfnamefont {A.}~\bibnamefont {Allevi}},\ and\ \bibinfo {author} {\bibfnamefont {M.}~\bibnamefont {Bondani}},\ }\bibfield  {title} {\bibinfo {title} {Internal dynamics of intense twin beams and their coherence},\ }\href@noop {} {\bibfield  {journal} {\bibinfo  {journal} {Scientific Reports}\ }\textbf {\bibinfo {volume} {6}},\ \bibinfo {pages} {22320} (\bibinfo {year} {2016})}\BibitemShut {NoStop}%
\bibitem [{\citenamefont {Fl{\'o}rez}\ \emph {et~al.}(2020)\citenamefont {Fl{\'o}rez}, \citenamefont {Lundeen},\ and\ \citenamefont {Chekhova}}]{Florez2020pump}%
  \BibitemOpen
  \bibfield  {author} {\bibinfo {author} {\bibfnamefont {J.}~\bibnamefont {Fl{\'o}rez}}, \bibinfo {author} {\bibfnamefont {J.~S.}\ \bibnamefont {Lundeen}},\ and\ \bibinfo {author} {\bibfnamefont {M.~V.}\ \bibnamefont {Chekhova}},\ }\bibfield  {title} {\bibinfo {title} {Pump depletion in parametric down-conversion with low pump energies},\ }\href@noop {} {\bibfield  {journal} {\bibinfo  {journal} {Optics Letters}\ }\textbf {\bibinfo {volume} {45}},\ \bibinfo {pages} {4264} (\bibinfo {year} {2020})}\BibitemShut {NoStop}%
\bibitem [{\citenamefont {Ding}\ \emph {et~al.}(2018)\citenamefont {Ding}, \citenamefont {Maslennikov}, \citenamefont {Habl{\"u}tzel},\ and\ \citenamefont {Matsukevich}}]{Ding2018}%
  \BibitemOpen
  \bibfield  {author} {\bibinfo {author} {\bibfnamefont {S.}~\bibnamefont {Ding}}, \bibinfo {author} {\bibfnamefont {G.}~\bibnamefont {Maslennikov}}, \bibinfo {author} {\bibfnamefont {R.}~\bibnamefont {Habl{\"u}tzel}},\ and\ \bibinfo {author} {\bibfnamefont {D.}~\bibnamefont {Matsukevich}},\ }\bibfield  {title} {\bibinfo {title} {Quantum simulation with a trilinear hamiltonian},\ }\href@noop {} {\bibfield  {journal} {\bibinfo  {journal} {Physical Review Letters}\ }\textbf {\bibinfo {volume} {121}},\ \bibinfo {pages} {130502} (\bibinfo {year} {2018})}\BibitemShut {NoStop}%
\bibitem [{\citenamefont {Birrittella}\ \emph {et~al.}(2020)\citenamefont {Birrittella}, \citenamefont {Alsing},\ and\ \citenamefont {Gerry}}]{Birrittella2020phase}%
  \BibitemOpen
  \bibfield  {author} {\bibinfo {author} {\bibfnamefont {R.~J.}\ \bibnamefont {Birrittella}}, \bibinfo {author} {\bibfnamefont {P.~M.}\ \bibnamefont {Alsing}},\ and\ \bibinfo {author} {\bibfnamefont {C.~C.}\ \bibnamefont {Gerry}},\ }\bibfield  {title} {\bibinfo {title} {Phase effects in coherently stimulated down-conversion with a quantized pump field},\ }\href@noop {} {\bibfield  {journal} {\bibinfo  {journal} {Physical Review A}\ }\textbf {\bibinfo {volume} {101}},\ \bibinfo {pages} {013813} (\bibinfo {year} {2020})}\BibitemShut {NoStop}%
\bibitem [{\citenamefont {Agarwal}(2012)}]{Agarwal2012quantum}%
  \BibitemOpen
  \bibfield  {author} {\bibinfo {author} {\bibfnamefont {G.~S.}\ \bibnamefont {Agarwal}},\ }\href@noop {} {\emph {\bibinfo {title} {Quantum optics}}}\ (\bibinfo  {publisher} {Cambridge University Press},\ \bibinfo {year} {2012})\BibitemShut {NoStop}%
\bibitem [{\citenamefont {Virtanen}\ \emph {et~al.}(2020)\citenamefont {Virtanen}, \citenamefont {Gommers}, \citenamefont {Oliphant}, \citenamefont {Haberland}, \citenamefont {Reddy}, \citenamefont {Cournapeau}, \citenamefont {Burovski}, \citenamefont {Peterson}, \citenamefont {Weckesser}, \citenamefont {Bright}, \citenamefont {{van der Walt}}, \citenamefont {Brett}, \citenamefont {Wilson}, \citenamefont {Millman}, \citenamefont {Mayorov}, \citenamefont {Nelson}, \citenamefont {Jones}, \citenamefont {Kern}, \citenamefont {Larson}, \citenamefont {Carey}, \citenamefont {Polat}, \citenamefont {Feng}, \citenamefont {Moore}, \citenamefont {{VanderPlas}}, \citenamefont {Laxalde}, \citenamefont {Perktold}, \citenamefont {Cimrman}, \citenamefont {Henriksen}, \citenamefont {Quintero}, \citenamefont {Harris}, \citenamefont {Archibald}, \citenamefont {Ribeiro}, \citenamefont {Pedregosa}, \citenamefont {{van Mulbregt}},\ and\ \citenamefont {{SciPy 1.0 Contributors}}}]{2020SciPy-NMeth}%
  \BibitemOpen
  \bibfield  {author} {\bibinfo {author} {\bibfnamefont {P.}~\bibnamefont {Virtanen}}, \bibinfo {author} {\bibfnamefont {R.}~\bibnamefont {Gommers}}, \bibinfo {author} {\bibfnamefont {T.~E.}\ \bibnamefont {Oliphant}}, \bibinfo {author} {\bibfnamefont {M.}~\bibnamefont {Haberland}}, \bibinfo {author} {\bibfnamefont {T.}~\bibnamefont {Reddy}}, \bibinfo {author} {\bibfnamefont {D.}~\bibnamefont {Cournapeau}}, \bibinfo {author} {\bibfnamefont {E.}~\bibnamefont {Burovski}}, \bibinfo {author} {\bibfnamefont {P.}~\bibnamefont {Peterson}}, \bibinfo {author} {\bibfnamefont {W.}~\bibnamefont {Weckesser}}, \bibinfo {author} {\bibfnamefont {J.}~\bibnamefont {Bright}}, \bibinfo {author} {\bibfnamefont {S.~J.}\ \bibnamefont {{van der Walt}}}, \bibinfo {author} {\bibfnamefont {M.}~\bibnamefont {Brett}}, \bibinfo {author} {\bibfnamefont {J.}~\bibnamefont {Wilson}}, \bibinfo {author} {\bibfnamefont {K.~J.}\ \bibnamefont {Millman}}, \bibinfo {author} {\bibfnamefont {N.}~\bibnamefont {Mayorov}}, \bibinfo {author} {\bibfnamefont
  {A.~R.~J.}\ \bibnamefont {Nelson}}, \bibinfo {author} {\bibfnamefont {E.}~\bibnamefont {Jones}}, \bibinfo {author} {\bibfnamefont {R.}~\bibnamefont {Kern}}, \bibinfo {author} {\bibfnamefont {E.}~\bibnamefont {Larson}}, \bibinfo {author} {\bibfnamefont {C.~J.}\ \bibnamefont {Carey}}, \bibinfo {author} {\bibfnamefont {{\.I}.}~\bibnamefont {Polat}}, \bibinfo {author} {\bibfnamefont {Y.}~\bibnamefont {Feng}}, \bibinfo {author} {\bibfnamefont {E.~W.}\ \bibnamefont {Moore}}, \bibinfo {author} {\bibfnamefont {J.}~\bibnamefont {{VanderPlas}}}, \bibinfo {author} {\bibfnamefont {D.}~\bibnamefont {Laxalde}}, \bibinfo {author} {\bibfnamefont {J.}~\bibnamefont {Perktold}}, \bibinfo {author} {\bibfnamefont {R.}~\bibnamefont {Cimrman}}, \bibinfo {author} {\bibfnamefont {I.}~\bibnamefont {Henriksen}}, \bibinfo {author} {\bibfnamefont {E.~A.}\ \bibnamefont {Quintero}}, \bibinfo {author} {\bibfnamefont {C.~R.}\ \bibnamefont {Harris}}, \bibinfo {author} {\bibfnamefont {A.~M.}\ \bibnamefont {Archibald}}, \bibinfo {author}
  {\bibfnamefont {A.~H.}\ \bibnamefont {Ribeiro}}, \bibinfo {author} {\bibfnamefont {F.}~\bibnamefont {Pedregosa}}, \bibinfo {author} {\bibfnamefont {P.}~\bibnamefont {{van Mulbregt}}},\ and\ \bibinfo {author} {\bibnamefont {{SciPy 1.0 Contributors}}},\ }\bibfield  {title} {\bibinfo {title} {{{SciPy} 1.0: Fundamental Algorithms for Scientific Computing in Python}},\ }\href {https://doi.org/10.1038/s41592-019-0686-2} {\bibfield  {journal} {\bibinfo  {journal} {Nature Methods}\ }\textbf {\bibinfo {volume} {17}},\ \bibinfo {pages} {261} (\bibinfo {year} {2020})}\BibitemShut {NoStop}%
\bibitem [{\citenamefont {Al-Mohy}\ and\ \citenamefont {Higham}(2011)}]{Al2011computing}%
  \BibitemOpen
  \bibfield  {author} {\bibinfo {author} {\bibfnamefont {A.~H.}\ \bibnamefont {Al-Mohy}}\ and\ \bibinfo {author} {\bibfnamefont {N.~J.}\ \bibnamefont {Higham}},\ }\bibfield  {title} {\bibinfo {title} {Computing the action of the matrix exponential, with an application to exponential integrators},\ }\href@noop {} {\bibfield  {journal} {\bibinfo  {journal} {SIAM journal on scientific computing}\ }\textbf {\bibinfo {volume} {33}},\ \bibinfo {pages} {488} (\bibinfo {year} {2011})}\BibitemShut {NoStop}%
\bibitem [{\citenamefont {Higham}\ and\ \citenamefont {Al-Mohy}(2010)}]{Higham2010computing}%
  \BibitemOpen
  \bibfield  {author} {\bibinfo {author} {\bibfnamefont {N.~J.}\ \bibnamefont {Higham}}\ and\ \bibinfo {author} {\bibfnamefont {A.~H.}\ \bibnamefont {Al-Mohy}},\ }\bibfield  {title} {\bibinfo {title} {Computing matrix functions},\ }\href@noop {} {\bibfield  {journal} {\bibinfo  {journal} {Acta Numerica}\ }\textbf {\bibinfo {volume} {19}},\ \bibinfo {pages} {159} (\bibinfo {year} {2010})}\BibitemShut {NoStop}%
\bibitem [{\citenamefont {Kubo}(1962)}]{Kubo1962}%
  \BibitemOpen
  \bibfield  {author} {\bibinfo {author} {\bibfnamefont {R.}~\bibnamefont {Kubo}},\ }\bibfield  {title} {\bibinfo {title} {Generalized cumulant expansion method},\ }\href@noop {} {\bibfield  {journal} {\bibinfo  {journal} {Journal of the Physical Society of Japan}\ }\textbf {\bibinfo {volume} {17}},\ \bibinfo {pages} {1100} (\bibinfo {year} {1962})}\BibitemShut {NoStop}%
\bibitem [{\citenamefont {Fisher}\ and\ \citenamefont {Wishart}(1932)}]{fisher1932derivation}%
  \BibitemOpen
  \bibfield  {author} {\bibinfo {author} {\bibfnamefont {R.~A.}\ \bibnamefont {Fisher}}\ and\ \bibinfo {author} {\bibfnamefont {J.}~\bibnamefont {Wishart}},\ }\bibfield  {title} {\bibinfo {title} {{The Derivation of the Pattern Formulae of Two-Way Partitions from those of Simpler Patterns}},\ }\href@noop {} {\bibfield  {journal} {\bibinfo  {journal} {Proceedings of the London Mathematical Society}\ }\textbf {\bibinfo {volume} {s2-33}},\ \bibinfo {pages} {195} (\bibinfo {year} {1932})}\BibitemShut {NoStop}%
\bibitem [{\citenamefont {Ursell}(1927)}]{ursell_1927}%
  \BibitemOpen
  \bibfield  {author} {\bibinfo {author} {\bibfnamefont {H.~D.}\ \bibnamefont {Ursell}},\ }\bibfield  {title} {\bibinfo {title} {The evaluation of gibbs' phase-integral for imperfect gases},\ }\href@noop {} {\bibfield  {journal} {\bibinfo  {journal} {Mathematical Proceedings of the Cambridge Philosophical Society}\ }\textbf {\bibinfo {volume} {23}},\ \bibinfo {pages} {685–697} (\bibinfo {year} {1927})}\BibitemShut {NoStop}%
\bibitem [{\citenamefont {Duneau}\ \emph {et~al.}(1973)\citenamefont {Duneau}, \citenamefont {Iagolnitzer},\ and\ \citenamefont {Souillard}}]{duneau1973decrease}%
  \BibitemOpen
  \bibfield  {author} {\bibinfo {author} {\bibfnamefont {M.}~\bibnamefont {Duneau}}, \bibinfo {author} {\bibfnamefont {D.}~\bibnamefont {Iagolnitzer}},\ and\ \bibinfo {author} {\bibfnamefont {B.}~\bibnamefont {Souillard}},\ }\bibfield  {title} {\bibinfo {title} {Decrease properties of truncated correlation functions and analyticity properties for classical lattices and continuous systems},\ }\href@noop {} {\bibfield  {journal} {\bibinfo  {journal} {Communications in Mathematical Physics}\ }\textbf {\bibinfo {volume} {31}},\ \bibinfo {pages} {191} (\bibinfo {year} {1973})}\BibitemShut {NoStop}%
\bibitem [{\citenamefont {Plankensteiner}\ \emph {et~al.}(2022)\citenamefont {Plankensteiner}, \citenamefont {Hotter},\ and\ \citenamefont {Ritsch}}]{Plankensteiner2022}%
  \BibitemOpen
  \bibfield  {author} {\bibinfo {author} {\bibfnamefont {D.}~\bibnamefont {Plankensteiner}}, \bibinfo {author} {\bibfnamefont {C.}~\bibnamefont {Hotter}},\ and\ \bibinfo {author} {\bibfnamefont {H.}~\bibnamefont {Ritsch}},\ }\bibfield  {title} {\bibinfo {title} {Quantumcumulants. jl: A julia framework for generalized mean-field equations in open quantum systems},\ }\href@noop {} {\bibfield  {journal} {\bibinfo  {journal} {Quantum}\ }\textbf {\bibinfo {volume} {6}},\ \bibinfo {pages} {617} (\bibinfo {year} {2022})}\BibitemShut {NoStop}%
\bibitem [{{\relax DLMF}()}]{NIST:DLMF}%
  \BibitemOpen
  {\relax DLMF},\ \href {https://dlmf.nist.gov/} {\bibinfo {title} {{\it NIST Digital Library of Mathematical Functions}}},\ \bibinfo {howpublished} {\url{https://dlmf.nist.gov/}, Release 1.1.11 of 2023-09-15},\ \bibinfo {note} {f.~W.~J. Olver, A.~B. {Olde Daalhuis}, D.~W. Lozier, B.~I. Schneider, R.~F. Boisvert, C.~W. Clark, B.~R. Miller, B.~V. Saunders, H.~S. Cohl, and M.~A. McClain, eds.}\BibitemShut {Stop}%
\bibitem [{\citenamefont {Armstrong}\ \emph {et~al.}(1962)\citenamefont {Armstrong}, \citenamefont {Bloembergen}, \citenamefont {Ducuing},\ and\ \citenamefont {Pershan}}]{Armstrong1962interactions}%
  \BibitemOpen
  \bibfield  {author} {\bibinfo {author} {\bibfnamefont {J.}~\bibnamefont {Armstrong}}, \bibinfo {author} {\bibfnamefont {N.}~\bibnamefont {Bloembergen}}, \bibinfo {author} {\bibfnamefont {J.}~\bibnamefont {Ducuing}},\ and\ \bibinfo {author} {\bibfnamefont {P.~S.}\ \bibnamefont {Pershan}},\ }\bibfield  {title} {\bibinfo {title} {Interactions between light waves in a nonlinear dielectric},\ }\href@noop {} {\bibfield  {journal} {\bibinfo  {journal} {Physical review}\ }\textbf {\bibinfo {volume} {127}},\ \bibinfo {pages} {1918} (\bibinfo {year} {1962})}\BibitemShut {NoStop}%
\bibitem [{\citenamefont {Yanagimoto}\ \emph {et~al.}(2022)\citenamefont {Yanagimoto}, \citenamefont {Ng}, \citenamefont {Yamamura}, \citenamefont {Onodera}, \citenamefont {Wright}, \citenamefont {Jankowski}, \citenamefont {Fejer}, \citenamefont {McMahon},\ and\ \citenamefont {Mabuchi}}]{Yanagimoto2022onset}%
  \BibitemOpen
  \bibfield  {author} {\bibinfo {author} {\bibfnamefont {R.}~\bibnamefont {Yanagimoto}}, \bibinfo {author} {\bibfnamefont {E.}~\bibnamefont {Ng}}, \bibinfo {author} {\bibfnamefont {A.}~\bibnamefont {Yamamura}}, \bibinfo {author} {\bibfnamefont {T.}~\bibnamefont {Onodera}}, \bibinfo {author} {\bibfnamefont {L.~G.}\ \bibnamefont {Wright}}, \bibinfo {author} {\bibfnamefont {M.}~\bibnamefont {Jankowski}}, \bibinfo {author} {\bibfnamefont {M.}~\bibnamefont {Fejer}}, \bibinfo {author} {\bibfnamefont {P.~L.}\ \bibnamefont {McMahon}},\ and\ \bibinfo {author} {\bibfnamefont {H.}~\bibnamefont {Mabuchi}},\ }\bibfield  {title} {\bibinfo {title} {Onset of non-gaussian quantum physics in pulsed squeezing with mesoscopic fields},\ }\href@noop {} {\bibfield  {journal} {\bibinfo  {journal} {Optica}\ }\textbf {\bibinfo {volume} {9}},\ \bibinfo {pages} {379} (\bibinfo {year} {2022})}\BibitemShut {NoStop}%
\bibitem [{\citenamefont {Yanagimoto}\ \emph {et~al.}(2024)\citenamefont {Yanagimoto}, \citenamefont {Ng}, \citenamefont {Jankowski}, \citenamefont {Nehra}, \citenamefont {McKenna}, \citenamefont {Onodera}, \citenamefont {Wright}, \citenamefont {Hamerly}, \citenamefont {Marandi}, \citenamefont {Fejer},\ and\ \citenamefont {Mabuchi}}]{Yanagimoto:24}%
  \BibitemOpen
  \bibfield  {author} {\bibinfo {author} {\bibfnamefont {R.}~\bibnamefont {Yanagimoto}}, \bibinfo {author} {\bibfnamefont {E.}~\bibnamefont {Ng}}, \bibinfo {author} {\bibfnamefont {M.}~\bibnamefont {Jankowski}}, \bibinfo {author} {\bibfnamefont {R.}~\bibnamefont {Nehra}}, \bibinfo {author} {\bibfnamefont {T.~P.}\ \bibnamefont {McKenna}}, \bibinfo {author} {\bibfnamefont {T.}~\bibnamefont {Onodera}}, \bibinfo {author} {\bibfnamefont {L.~G.}\ \bibnamefont {Wright}}, \bibinfo {author} {\bibfnamefont {R.}~\bibnamefont {Hamerly}}, \bibinfo {author} {\bibfnamefont {A.}~\bibnamefont {Marandi}}, \bibinfo {author} {\bibfnamefont {M.~M.}\ \bibnamefont {Fejer}},\ and\ \bibinfo {author} {\bibfnamefont {H.}~\bibnamefont {Mabuchi}},\ }\bibfield  {title} {\bibinfo {title} {Mesoscopic ultrafast nonlinear optics—the emergence of multimode quantum non-gaussian physics},\ }\href {https://doi.org/10.1364/OPTICA.514075} {\bibfield  {journal} {\bibinfo  {journal} {Optica}\ }\textbf {\bibinfo {volume} {11}},\ \bibinfo {pages}
  {896} (\bibinfo {year} {2024})}\BibitemShut {NoStop}%
\bibitem [{\citenamefont {Serafini}(2023)}]{Serafini2023quantum}%
  \BibitemOpen
  \bibfield  {author} {\bibinfo {author} {\bibfnamefont {A.}~\bibnamefont {Serafini}},\ }\href@noop {} {\emph {\bibinfo {title} {Quantum continuous variables: a primer of theoretical methods}}}\ (\bibinfo  {publisher} {CRC press},\ \bibinfo {year} {2023})\BibitemShut {NoStop}%
\bibitem [{\citenamefont {Agust{\'\i}}\ \emph {et~al.}(2020)\citenamefont {Agust{\'\i}}, \citenamefont {Chang}, \citenamefont {Quijandr{\'\i}a}, \citenamefont {Johansson}, \citenamefont {Wilson},\ and\ \citenamefont {Sab{\'\i}n}}]{Agusti2020tripartite}%
  \BibitemOpen
  \bibfield  {author} {\bibinfo {author} {\bibfnamefont {A.}~\bibnamefont {Agust{\'\i}}}, \bibinfo {author} {\bibfnamefont {C.~S.}\ \bibnamefont {Chang}}, \bibinfo {author} {\bibfnamefont {F.}~\bibnamefont {Quijandr{\'\i}a}}, \bibinfo {author} {\bibfnamefont {G.}~\bibnamefont {Johansson}}, \bibinfo {author} {\bibfnamefont {C.~M.}\ \bibnamefont {Wilson}},\ and\ \bibinfo {author} {\bibfnamefont {C.}~\bibnamefont {Sab{\'\i}n}},\ }\bibfield  {title} {\bibinfo {title} {Tripartite genuine non-gaussian entanglement in three-mode spontaneous parametric down-conversion},\ }\href@noop {} {\bibfield  {journal} {\bibinfo  {journal} {Physical Review Letters}\ }\textbf {\bibinfo {volume} {125}},\ \bibinfo {pages} {020502} (\bibinfo {year} {2020})}\BibitemShut {NoStop}%
\bibitem [{\citenamefont {Shchukin}\ and\ \citenamefont {Vogel}(2005)}]{Shchukin2005inseparability}%
  \BibitemOpen
  \bibfield  {author} {\bibinfo {author} {\bibfnamefont {E.}~\bibnamefont {Shchukin}}\ and\ \bibinfo {author} {\bibfnamefont {W.}~\bibnamefont {Vogel}},\ }\bibfield  {title} {\bibinfo {title} {Inseparability criteria for continuous bipartite quantum states},\ }\href@noop {} {\bibfield  {journal} {\bibinfo  {journal} {Physical review letters}\ }\textbf {\bibinfo {volume} {95}},\ \bibinfo {pages} {230502} (\bibinfo {year} {2005})}\BibitemShut {NoStop}%
\bibitem [{\citenamefont {Steinlechner}\ \emph {et~al.}(2013)\citenamefont {Steinlechner}, \citenamefont {Ast}, \citenamefont {Kr{\"u}ger}, \citenamefont {Singh}, \citenamefont {Eberle}, \citenamefont {H{\"a}ndchen},\ and\ \citenamefont {Schnabel}}]{Steinlechner2013absorption}%
  \BibitemOpen
  \bibfield  {author} {\bibinfo {author} {\bibfnamefont {J.}~\bibnamefont {Steinlechner}}, \bibinfo {author} {\bibfnamefont {S.}~\bibnamefont {Ast}}, \bibinfo {author} {\bibfnamefont {C.}~\bibnamefont {Kr{\"u}ger}}, \bibinfo {author} {\bibfnamefont {A.~P.}\ \bibnamefont {Singh}}, \bibinfo {author} {\bibfnamefont {T.}~\bibnamefont {Eberle}}, \bibinfo {author} {\bibfnamefont {V.}~\bibnamefont {H{\"a}ndchen}},\ and\ \bibinfo {author} {\bibfnamefont {R.}~\bibnamefont {Schnabel}},\ }\bibfield  {title} {\bibinfo {title} {Absorption measurements of periodically poled potassium titanyl phosphate (ppktp) at 775 nm and 1550 nm},\ }\href@noop {} {\bibfield  {journal} {\bibinfo  {journal} {Sensors}\ }\textbf {\bibinfo {volume} {13}},\ \bibinfo {pages} {565} (\bibinfo {year} {2013})}\BibitemShut {NoStop}%
\end{thebibliography}%
\end{document}